\documentclass[a4paper,11pt]{article}
\usepackage{jcappub} 
\usepackage[normalem]{ulem}
\usepackage[utf8]{inputenc}
\usepackage{float}
\usepackage{siunitx}
\usepackage{graphicx}
\usepackage{gensymb}
\usepackage{cancel,soul,ulem}
\usepackage{amsmath, cases}
\usepackage[usenames]{color}
\usepackage{amssymb}
\usepackage[utf8]{inputenc}
\usepackage{amsmath, amssymb, amsthm, graphicx, epsfig, fancyhdr,epsfig, slashed}
\usepackage{mathrsfs}
\let\oldcdot\cdot
\usepackage{breqn}
\let\cdot\oldcdot
\usepackage{enumerate}
\usepackage{epsfig}  
\usepackage{subcaption} 
\usepackage{graphicx}   
\usepackage{slashed}       
\usepackage{tikz}
\usepackage{url}
\usepackage{color,soul} 
\usepackage{multirow}
\usepackage{comment}
\usepackage{mathtools}
\usepackage{bm}
\usepackage{hyperref}
\usepackage[all]{hypcap}
\hypersetup{  
    colorlinks=true,
    linkcolor=blue,
    filecolor=red,      
    urlcolor=cyan,
    citecolor=green,
    }

\newcommand{\kbh}{k_{\rm BH}}

\newcommand{\w}{w_{\phi}}

\newcommand{\Min}{M_{\rm in}}

\def\be{\begin{equation}}
\def\ee{\end{equation}}
\def\beq{\begin{equation}\begin{aligned}}
\def\eeq{\end{aligned}\end{equation}}

\begin{document}

\title{Gravitational Wave Signatures of Primordial Black Hole Reheating in Upcoming Interferometry Missions}

\author[a]{Debarun Paul,}
\emailAdd{debarun31paul@gmail.com}

\author[a]{Md Riajul Haque}
\emailAdd{riaj.0009@gmail.com}

\author[a]{and Supratik Pal}
\emailAdd{supratik@isical.ac.in}

\affiliation[a]{\,Physics and Applied Mathematics Unit, Indian Statistical Institute, 203 B.T. Road, Kolkata 700108, India}

\smallskip

\abstract{
We investigate the prospects of detecting a stochastic gravitational wave (GW) background from the primordial black hole (PBH) reheating epoch. If PBHs form during a non-standard cosmological phase prior to the radiation-dominated era, they can dominate the Universe’s energy density before evaporating via Hawking radiation. Such PBHs can generate induced GWs that may fall within the detectable range of future interferometry missions:  (i) through isocurvature perturbations arising from the inhomogeneous spatial distribution of PBHs, and  
(ii) through the amplification of adiabatic perturbations triggered by the abrupt transition from PBH domination to radiation domination.  We assess the detection prospects of such GW spectra using the signal-to-noise ratio, Fisher forecast analysis, and Markov chain Monte Carlo analysis with mock data from LISA and ET. Our findings reveal that ET exhibits superior sensitivity to both isocurvature- and adiabatic-induced GWs, covering a wide PBH mass range of $\Min \in (0.5-4\times 10^7)$ g. However, we find that the relative uncertainties associated with the parameter of the isocurvature source are quite high. LISA, by contrast, is mostly sensitive to the adiabatic source, with $\Min \in (2\times10^4-5\times 10^8)$ g. The combined effect of adiabatic and isocurvature sources on ET and LISA provides a multi-stage window into the post-inflationary Universe by constraining PBH mass, energy fraction, and the background equation of state.
}

\maketitle
\newpage
\section{Introduction}
\label{sec:intro}
Primordial Black Holes (PBHs) can form from high-density fluctuations in the early Universe \cite{Hawking:1974rv,Carr:1974nx,Hawking:1975vcx}, offering insights into dark matter, cosmology, and gravitational wave (GW) sources. Unlike astrophysical black holes, which originate from the final stages of stellar evolution, PBHs can span a wide range of masses depending on the formation epoch and the underlying cosmology. The study of PBHs has received considerable interest in recent years, as they offer novel eyes to look into various aspects of the early Universe, including inflationary dynamics, dark matter, baryogenesis, and non-standard reheating scenarios. Recent observations of GWs by LIGO-Virgo collaboration~\cite{LIGOScientific:2016aoc,LIGOScientific:2016sjg,LIGOScientific:2017bnn,LIGOScientific:2017vox,LIGOScientific:2017ycc,LIGOScientific:2017vwq}, along with the growing interest in exploring dark matter candidates and early Universe cosmology, have made the studies of PBHs a timely one, as they could potentially account for a fraction of dark matter or leave imprints on stochastic GW backgrounds. Thus, any signature of  PBHs in upcoming detectors could provide a powerful probe of the primordial Universe and dark matter, potentially unravelling new physics beyond the standard cosmological paradigm.

The formation of PBHs can be linked to large curvature perturbations generated during inflation. When these perturbations exceed a critical threshold, they can collapse gravitationally to form PBHs upon horizon reentry. PBHs are formed when local density contrast surpass a threshold value \(\delta_c\), with \(\frac{\delta \rho}{\rho} \gtrsim \delta_c \sim 1\), indicating gravitational collapse. Various mechanisms have been proposed to generate such fluctuations, including quantum fluctuations during single-field \cite{PhysRevD.48.543,PhysRevD.50.7173,Yokoyama:1998pt,Saito:2008em,Garcia-Bellido:2017mdw} and multi-field \cite{Yokoyama:1995ex,Randall:1995dj,Garcia-Bellido:1996mdl,Pi:2017gih} inflation, collapse of cosmic string loops \cite{MacGibbon:1990zk,Jenkins:2020ctp,Helfer:2018qgv,Matsuda:2005ez,Lake:2009nq}, domain wall collapse \cite{Rubin:2000dq,Rubin:2001yw}, and bubble collisions during phase transitions \cite{KodamaPTP1979}, and some more fancy mechanisms \cite{Lu:2024xnb,Lu:2024zwa,Flores:2024lng,Flores:2023zpf,Flores:2024eyy,Ballesteros:2024hhq}. While the exact formation mechanism for PBHs remain a topic of further investigation, 
an alternative, observation-driven, route to study PBH evolution is to perform a model-independent analysis and to investigate how the parameters can be constrained by future interferometry missions. 

While the standard reheating scenario assumes that the inflaton field decays into radiation, thereby initiating the radiation-dominated era \cite{Drewes:2017fmn,Garcia:2020eof,Haque:2020zco,Garcia:2020wiy,Haque:2022kez,Clery:2021bwz,Clery:2022wib,Co:2022bgh,Ahmed:2022tfm,Haque:2023zhb}, the PBH reheating scenario  presents an alternative, and interesting, possibility. In this framework, PBHs form during a non-standard cosmological epoch characterized by a background equation of state parameter $\w$ different from radiation. Depending on the properties of the inflaton potential and the dynamics of the post-inflationary era, the value of $\w$ can range between $0$ (matter-like domination) and $1$ (stiff fluid domination). In this scenarion, due to their non-relativistic nature, PBHs behave like pressureless matter and can dominate the Universe’s energy density before evaporating via Hawking radiation. This evaporation process injects energy into the cosmic plasma, triggering the reheating of the Universe and marking the onset of the radiation-dominated era \cite{Hidalgo:2011fj,Martin:2019nuw, Hooper:2019gtx, Hooper:2020evu, Hooper:2020otu, Bernal:2020bjf, Cheek:2021cfe, Cheek:2022mmy, Mazde:2022sdx,RiajulHaque:2023cqe, Calabrese:2023key, Barman:2024slw}. Such a scenario naturally connects the properties of PBHs to the dynamics of the reheating epoch, offering a novel observational window into the post-inflationary Universe.

A remarkable feature of the PBH reheating scenario is the generation of stochastic GW backgrounds. Two distinct mechanisms contribute to the production of GWs in this context. First, on the small scales where the PBHs can be treated as individual particles, the inhomogeneous spatial distribution of PBHs induces isocurvature density perturbations, which act as secondary sources of GWs during the PBH-dominated era \cite{Papanikolaou:2020qtd, Domenech:2020ssp, Dalianis:2020gup, Domenech:2021ztg, Domenech:2021wkk, Papanikolaou:2021uhe, Bhaumik:2022pil, Bhaumik:2022zdd,Papanikolaou:2022chm,Bhaumik:2024qzd,Domenech:2024wao,Gross:2024wkl}. These induced GWs originate from the Poissonian nature of the PBH number density fluctuations. On the other hand, on large scales PBHs behave as pressure-less fluid and the sudden transition from PBH domination to radiation domination, driven by PBH evaporation, amplifies adiabatic curvature perturbations and generates an additional GW signal~\cite{Inomata:2020lmk, White:2021hwi, Bhaumik:2022pil, Bhaumik:2022zdd, Bhaumik:2023wmw,Bhaumik:2024qzd,Domenech:2024wao}. The distinction between these two sources are purely phenomenological, reflecting differences in scales rather than any physical separation, although both are produced due to sudden changes in the background equation of state from PBH domination to radiation dominated epoch. 
The spectrum of these GWs carries distinctive features that encode information about the PBH parameters, the background equation of state, and the duration of the PBH-dominated era.

The detection of these GW spectra would offer a unique probe of the PBH reheating scenario and the associated non-standard cosmological dynamics~\cite{Bertone:2019irm}. Future interferometry missions, such as the Laser Interferometer Space Antenna (LISA)~\cite{amaroseoane2017laser,Baker:2019nia} and the Einstein Telescope (ET)~\cite{Punturo_2010,Hild:2010id}, are particularly well-suited to detect the GW spectra associated with PBH reheating \footnote{Recently, an analysis from the LISA Cosmology Working Group was conducted on forecasting LISA's capabilities to constrain the sources of scalar-induced gravitational waves~\cite{LISACosmologyWorkingGroup:2025vdz}. However, they did not examine in detail the sources associated with PBH reheating.

}. LISA, operating in the millihertz frequency range, is sensitive to GWs generated by heavier PBHs with masses around ($10^5$–$10^8$) g, while ET, probing higher frequencies, can detect the spectra from lighter PBHs with masses around ($1$–$10^7$) g. The complementary sensitivity ranges of these missions provide a unique opportunity to explore a wide range of PBH masses and the associated reheating dynamics.

In this work, we systematically investigate the GW signatures of PBH reheating and their detection prospects in future GW observatories. We outline the theoretical framework for PBH formation during non-standard reheating, the associated GW production mechanisms, and the expected spectral features. Our analysis employs a combination of signal-to-noise ratio calculations, Fisher forecast analysis, and Markov chain Monte Carlo (MCMC) analysis using mock data, leading to a quantitative estimation of the prospects of detecting these spectra. By linking the GWs amplitude and frequency to the PBH parameters — such as the formation mass, energy fraction and the background equation of state — we demonstrate how future GW observations can shed light on the physics of PBH reheating and the early Universe's non-standard evolution.

\textit{This paper is organized as follows}: In Section \ref{sec:PBH}, we provide a brief overview of the primordial black hole reheating scenario. Section \ref{sec:detection_prospects} outlines the detection prospects of the GW spectra in future interferometry missions. Section \ref{sec:GW_PBH} discusses the production of GW from PBH reheating, focusing on two distinct sources: isocurvature-induced GWs and adiabatic-induced GWs. We present the detection prospects and estimations for uncertainties associated with PBH and background parameters for both LISA and ET. In Section \ref{sec:full_spec}, we present a combined analysis of both GW sources, highlighting the complementary roles of LISA and ET in exploring the PBH reheating parameters. Finally, Section \ref{sec:conclusion} summarizes our findings and discusses the implications of future GW observations in probing the non-standard cosmological epoch associated with PBH reheating.

\section{Brief overview of the primordial black hole reheating}
\label{sec:PBH}

After inflation ends, the inflaton field oscillates around the minimum of its potential. Depending on the shape of the potential at the minima, the equation of state (EoS) \( \w \) of the scalar field can differ from that of radiation (\( \w = 1/3 \)), and the duration of reheating depends on how quickly the inflaton decays into standard model particles. In the standard reheating scenario, the Universe enters the radiation-dominated (RD) era as the oscillating inflaton field decays, with its energy dominating the Universe before it decays \cite{Garcia:2020eof,Haque:2020zco,Garcia:2020wiy,Haque:2022kez,Clery:2021bwz,Clery:2022wib}. 

However, in the scenario we are considering a different background evolution, where PBHs form during the inflaton-dominated phase. We consider this as a formation in some non-standard phase with a background equation of state parameter $\w$.
The time-dependent oscillations of the inflaton field can be expressed as
\begin{equation} \label{Eq:phi}
\phi(t) = \phi_0(t)\, \mathcal{P}(t)\,,
\end{equation}
where $\phi_0(t)$ denotes the oscillation amplitude and $\mathcal{P}(t)$ captures the periodic behaviour. The amplitude $\phi_0(t)$ evolves due to redshift and decay, assuming that the oscillation time scale is much shorter than the time scales associated with redshift and decay. Averaging over a single oscillation yields the relation $\langle\dot{\phi}^2\rangle \simeq \langle\phi\, dV_\phi/d\phi\rangle$, which implies that the inflaton energy density can be approximated as $\rho_\phi = \langle \dot{\phi}^2/2 + V(\phi) \rangle \sim V(\phi_0)$. Under these assumptions, the averaged equation of state (EoS) of the inflaton is given by~\cite{Garcia:2020eof,Bernal:2019mhf}
\be
w_\phi = \frac{P_\phi}{\rho_\phi} = \frac{\langle \dot{\phi}^2/2 - V(\phi) \rangle}{\langle \dot{\phi}^2/2 + V(\phi) \rangle} \simeq \frac{n - 2}{n + 2}\,, \label{eq: were n}
\ee
where we assume that the inflaton potential near its minimum takes the form $V(\phi)\sim \phi^{2n}$. For a potential that behaves as $V(\phi) \sim \phi^6$ near its minimum, the inflaton exhibits an average background equation of state $w_\phi = 0.5$. In scenarios where $w_\phi < 1/3$, we assume that the inflaton decays during the PBH-dominated era that is taken care of by the decay width.  On the other hand, when $w_\phi > 1/3$, the inflaton must likewise remain stable at least until the PBH domination epoch to maintain consistency with the underlying cosmological framework. Important to note that, after formation, PBHs evolve like non-relativistic matter, causing their energy density to dilute more slowly than the total energy density, since the effective EoS satisfies $0<\w\leq 1$ \footnote{We exclude exact matter domination ($\w=0$) because PBH formation in an early matter-dominated phase would take much longer to form an apparent horizon \cite{Escriva:2020tak} and preventing PBHs from becoming dominant, what we assume here.}. As a result, PBHs could dominate the Universe before decaying, provided their initial abundance is large enough \cite{Hidalgo:2011fj,Martin:2019nuw, Hooper:2019gtx, Hooper:2020evu,RiajulHaque:2023cqe}. We thus envision a PBH-dominated era sandwiched between the $\w$ dominated phase and the standard radiation-dominated era. 

\begin{figure}
    \centering
    \includegraphics[scale=0.35]{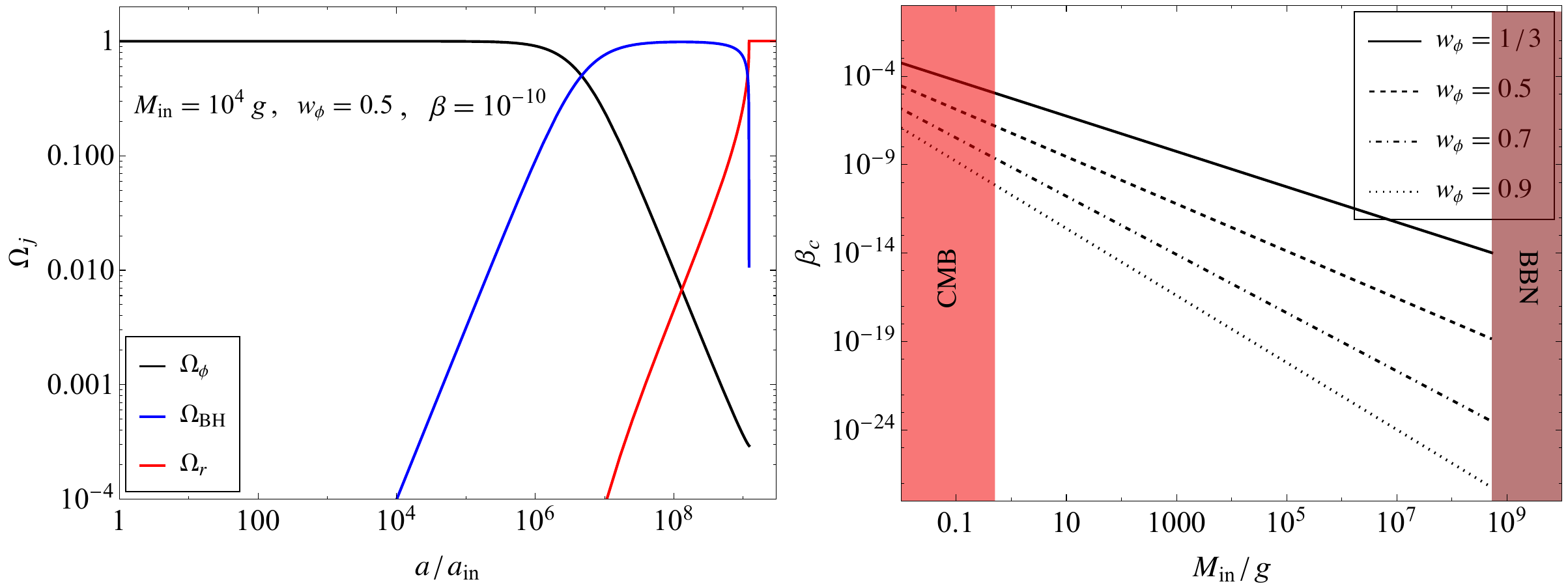}
    \caption{\it \textbf{Left panel:}  The evolution of the normalized energy densities, $ \Omega_j = \rho_j / 3 M_{\rm p}^2 H^2 $, as a function of the scale factor is shown. This figure illustrates the PBH reheating scenario, where PBH evaporation completes after it has dominated the entire background. \textbf{Right panel:} $\beta_c$ as a function of  $M_{\rm in}$ for various values of  $\w$, with shaded regions indicating the excluded ranges: the red region represents the CMB constraint on the formation mass of PBHs, and the brown region corresponds to the BBN constraint.}
    \label{fig:betac}
\end{figure}

The evolution of the PBHs is typically described by two key quantities: the formation mass ($M_{\rm in}$) and the energy fraction ($\beta$), which represent the portion of the total energy density that goes into creating PBHs. Assuming ultralight PBHs form during the post-inflationary era in  a background of $w_\phi$, the Boltzmann equations for the various energy components can be written as:
\begin{equation}
\frac{\mathrm{d} \rho_{\phi}}{\mathrm{d} a} + 3(1+\w) \frac{\rho_{\phi}}{a} = - \frac{\Gamma_\phi \rho_\phi (1 + \w)}{aH} \,,
\end{equation}
\begin{equation}
\frac{\mathrm{d} \rho_{\rm r}}{\mathrm{d} a} + 4 \frac{\rho_{\rm r}}{a} = -\frac{\rho_{\rm BH}}{M_{\rm BH}} \frac{\mathrm{d} M_{\rm BH}}{\mathrm{d} a} + \frac{\Gamma_\phi \rho_\phi (1 + \w)}{aH} \simeq   -\frac{\rho_{\rm BH}}{M_{\rm BH}} \frac{\mathrm{d} M_{\rm BH}}{\mathrm{d} a}\,,
\end{equation}
\begin{equation}
\frac{\mathrm{d} \rho_{\rm BH}}{\mathrm{d} a} + 3 \frac{\rho_{\rm BH}}{a} = \frac{\rho_{\rm BH}}{M_{\rm BH}} \frac{\mathrm{d} M_{\rm BH}}{\mathrm{d} a}\,,
\end{equation}
\begin{equation}
\frac{\mathrm{d} M_{\rm BH}}{\mathrm{d} a} = -\epsilon \frac{M_{\rm p}^4}{M_{\rm BH}^2} \frac{1}{aH} \,. \label{eq: pbh mass}
\end{equation}
Here, $ \epsilon \equiv 3.8\, \left( \frac{\pi}{480} \right) g_{\ast} (T_{\rm BH})$, and $g_{\ast} (T_{\rm BH})$ represents the number of relativistic degrees of freedom. The factor $3.8$ accounts for the greybody factor \cite{Arbey:2019mbc, Cheek:2021odj, Baldes:2020nuv}. $\rho_\phi, \rho_{\rm r}$ and $\rho_{\rm BH}$ represent the energy densities of inflaton, radiation and PBHs, respectively. Note that in the change of the inflaton energy density, we neglect the decay term compared to the Universe's expansion ($ \Gamma_\phi \ll H$), assuming that lifetime of the inflaton is much longer than the timescale required for the Universe to reach PBH-domination. As a result, the dominant source of radiation energy density will be the evaporation of PBHs. The change in the PBH energy density is influenced by two processes: Hawking evaporation and dilution due to cosmic expansion. The mass of a PBH at any time can be determined by solving Eq.~\eqref{eq: pbh mass} as
\begin{equation}
M_{\rm BH} = M_{\rm in} \left( 1 - \Gamma_{\rm BH} (t - t_{\rm in}) \right)^{1/3} ,
\end{equation}
where $\Gamma_{\rm BH} = \frac{3 \epsilon M_{\rm p}^4}{M_{\rm in}^3}$, and  $t_{\rm in}$  and $a_{\rm in}$ are the formation time of the PBH and the scale factor associated with the formation, respectively. The lifetime of a PBH is given by $t_{\rm ev} = 1 / \Gamma_{\rm BH}$. We assume that PBHs form during a $\w$ dominated phase due to the gravitational collapse of density fluctuations. Therefore, the formation mass of the PBHs ($M_{\rm in}$) can be expressed as:
\begin{equation} \label{eq:Min}
M_{\rm in} = \gamma \frac{4}{3} \pi \frac{ \rho_\phi (a_{\rm in})}{H_{\rm in}^3} = \frac{4 \pi \gamma M_{\rm p}^2}{H_{\rm in}} \,,
\end{equation}
where  $\gamma \equiv \w^{3/2}$ is the collapse efficiency \cite{Carr:1974nx} \footnote{At this point, it is worth mentioning that a more precise method exists for estimating the formation mass of PBHs, which is sensitive to both the $\w$ of the cosmological background fluid and the profile of the scalar fluctuations \cite{Musco:2012au, Musco:2008hv, Hawke:2002rf, Niemeyer:1997mt, Escriva:2021pmf, Escriva:2019nsa, Escriva:2020tak, Escriva:2021aeh}. However, a comprehensive analytical study to fully understand the values of $\gamma$ in the context of PBH formation during post-inflationary era has yet to be conducted. Therefore, we have opted to use the analytical relation for collapse efficiency as suggested by Carr and Hawking.}. In Fig.~\ref{fig:betac}, we illustrate the evolution of various energy components as a function of the normalized scale factor $ a/a_{\rm in}$. From the figure, it is evident that there exists a critical value of $\beta$, denoted $\beta_c$, above which the PBH energy density exceeds the inflaton energy density before the PBHs evaporate, with reheating occurring primarily through PBH evaporation. We refer to this scenario as PBH reheating. Consequently, $\beta_c$ turns out to be
\begin{equation}
\beta_{\rm c}= \left[ \frac{\epsilon}{2\pi(1+\w)\gamma}\frac{M_{\rm p}^2}{M_{\rm in}^2}\right]^{\frac{2\w}{1+\w}}\simeq \left(\frac{2.8\times 10^{-6}}{\w^{3/4}\sqrt{(1+\w)}}\right)^{\frac{4\w}{1+\w}}
\left(\frac{ 1\,\rm g}{\Min}\right)^{\frac{4\w}{1+\w}}.
\label{eq:betac}
\end{equation}
It can be found from the above Eq.~\eqref{eq:betac} that  $\beta_{\rm c} \simeq 5.6\times 10^{-6}\left(\frac{1~\rm{g}}{\Min}\right)$ for $\w=1/3$, 
and $\beta_{\rm c} \simeq 6.1\times 10^{-8}\left(\frac{1~\rm{g}}{M_{\rm in}}\right)^{\frac{4}{3}}$ for $\w=1/2$.
In the right panel of Fig.~\ref{fig:betac}, how the critical parameter $\beta_{\rm c}$ varies with the background EoS is depicted. The figure clearly illustrates that PBH-domination becomes easier, as the value of $\w$ increases, resulting in a shift of $\beta_{\rm c}$ to lower values.

The end of the evaporation marks the point of reheating, and the radiation temperature corresponding to this evaporation point is given by: \footnote{Note that for $\beta < \beta_{\rm c}$, unlike the case where PBHs dominate the background before they evaporate, PBHs are formed and evaporate during the inflaton-dominated phase, without ever becoming the dominant energy component of the Universe. In this scenario, reheating through PBH evaporation can only occur if the inflaton's equation of state corresponds to a stiff fluid, \textit{i.e.}, $\w>1/3$, and the production rate due to inflaton decay or scattering remains subdominant, such that $\Gamma_\phi \rho_\phi (1 + \w) < - \frac{\rho_{\rm BH}}{M_{\rm BH}} \frac{dM_{\rm BH}}{dt}$ (for more details, see Ref. \cite{RiajulHaque:2023cqe,Haque:2024cdh}). We exclude this type of PBH reheating scenario throughout our analysis.}
\begin{equation}\label{eq:Teva}
T_{\rm ev}= \left(\frac{360\,\epsilon^2}{\pi^2\,g_{\ast}(T_{\rm ev})}\right)^{1/4}\left(\frac{M_{\rm p}}{M_{\rm in}}\right)^{3/2}M_{\rm p}=2.7\times 10^{10}\left(\frac{1\,\rm{g}}{M_{\rm in}}\right)^{3/2}\,,
\end{equation} 
where we assume $g_{\ast}(T_{\rm ev})=106.75$ and $g_{\ast}(T_{\rm BH})=108$.
To ensure successful Big Bang Nucleosynthesis (BBN), the Universe must be radiation-dominated during BBN. This requires all PBHs to evaporate before BBN, \textit{i.e.}, $T_{\rm ev} \geq T_{\rm BBN} = 4 \, \text{MeV}$ \cite{Kawasaki:1999na,Kawasaki:2000en,Hasegawa:2019jsa}. This sets an upper limit on $M_{\rm in}$ as determined by Eq.~\eqref{eq:Teva}. On the other hand, data from Planck-2018, along with BICEP2/$Keck$ \cite{BICEP:2021xfz,BICEP2:2015nss}, 
provides an upper bound on the tensor-to-scalar ratio ($r_{0.05}<0.036$), which in turn constrains the Hubble parameter during inflation, assuming de Sitter like inflation. This constraint is given by $ H_{\rm inf} < 4.8 \times 10^{13}\, \text{GeV}$, providing a lower bound on $M_{\rm in}$ according to Eq.~\eqref{eq:Min}. Therefore, from both BBN and cosmic microwave background (CMB) observations, one can obtain an upper and lower bound on $M_{\rm in}$, respectively, given by: 
\begin{eqnarray}
0.5 \, \text{g} \left( \frac{\gamma}{0.2} \right) \lesssim M_{\rm in} \lesssim 4.8 \times 10^8 \, \text{g} \left( \frac{g_{\ast}(T_{\rm ev})}{106.75} \right)^{-1/6} \left( \frac{g_{\ast}(T_{\rm BH})}{108} \right)\,.
\end{eqnarray}


Ultralight PBHs can contribute to the generation of GWs in several ways. For instance, large curvature perturbations which is associated with the  PBH formation can induce GWs \cite{Baumann:2007zm, Espinosa:2018eve, Domenech:2019quo, Ragavendra:2020sop, Inomata:2023zup, Franciolini:2023pbf, Firouzjahi:2023lzg, Maity:2024odg}, stochastic GW background can be generated by gravitons produced from the evaporation of PBHs \cite{Fujita:2014hha}, induced GWs through fluctuations in the PBH number density, which are isocurvature in nature \cite{Domenech:2020ssp, Papanikolaou:2020qtd, Domenech:2021wkk, Papanikolaou:2022chm, Bhaumik:2022pil, Bhaumik:2022zdd, Papanikolaou:2022hkg, Papanikolaou:2024kjb, Domenech:2024wao, Bhaumik:2024qzd,Gross:2024wkl}. Additionally, the sharp transition from matter-domination (PBH-domination) to radiation-domination, driven by PBH evaporation, enhances induced GWs, which are adiabatic primordial curvature perturbations \cite{Inomata:2019ivs, Inomata:2020lmk, Bhaumik:2022pil, Bhaumik:2022zdd}. Since ultralight PBHs form near the end of inflation, the secondary GWs induced by scalar perturbations from PBH formation typically occur at very high frequencies, close to the wave number associated with end of inflation. The stochastic GW background resulting from PBH evaporation into gravitons also contributes to high-frequency GWs. Thus, we focus on two relevant sources within the detectable range: one related to induced GWs from isocurvature perturbations arising from PBH density fluctuations, and the other from adiabatic fluctuations, which become significant due to the sudden transition from PBH domination to radiation domination.
Our ultimate goal is to link the amplitude and frequency dependency  of these GWs to PBH and background parameters ($\w,\beta, \Min$), so that any future detection of stochastic GW background can provide insights into both the early PBH domination and the phase preceding it.

\section{Steps to investigate detection prospects in interferometry missions}
\label{sec:detection_prospects}

Before discussing specific GW sources and their detectability with future GWs missions like ET and LISA, let us first discuss very briefly the steps we have followed in investigating the detection prospects of such signals in interferometry missions, in order to make the present article  self-content from the point of view of the potential readers.

\subsection{Signal-to-noise ratio for different detectors}
\label{subsec:noise_SNR}
Noise plays a key role in any observational data, so evaluating the signal-to-noise ratio (SNR) is essential for determining the detectable prospects of the signal. SNR for the GW missions can be calculated as~\cite{Thrane:2013oya,Caprini:2015zlo}
\begin{eqnarray}
\label{eq:snr}
    \text{SNR}\;\equiv\; \sqrt{\tau_{\rm obs}\;\int_{f_{\rm min}}^{f_{\rm max}} df \left(\frac{\Omega_{\rm{GW}}(f,\{\theta\})h^2}{\Omega_{\rm GW}^{\rm noise}(f)h^2}\right)^2},
\end{eqnarray}
where $\Omega_{\rm{GW}}(f,\{\theta\})h^2$ represents the GW energy density predicted by a specific theory, with $\{\theta\}(\equiv \w,\beta, \Min$ in our case) being the model parameters. $\Omega_{\rm GW}^{\rm noise}(f)h^2$ contains noise characteristic of the detector, operating within the frequency-range between $[f_{\rm min},f_{\rm max}]$ and $\tau_{\rm obs}$ denotes the duration of observation. In this analysis, we have considered ET and LISA as the representative missions, for which the noise spectra for instrumental noise are described in App.~\ref{app:noise}. In Table-\ref{tab:detector_spec},  the frequency-range and observation time for the detectors are presented. For both detectors, we have set the observational threshold for the SNR to be $1$.
\begin{table}[!ht]
    \centering
    \renewcommand{\arraystretch}{1.2}
    \begin{tabular}{|c|c|c|c|}
    \hline
    \hline
       \textit{Detectors}  & \textit{Frequency range} & $\tau_{\rm obs}$\\
    \hline
       LISA  & $\left[10^{-4}-1\right]$ Hz & $4$ years\\
       ET  & $\left[1-10^4\right]$ Hz & $5$ years\\
    \hline
    \hline
    \end{tabular}
    \caption{\it Specification of the detectors under consideration.}
    \label{tab:detector_spec}
\end{table}

\subsection{Statistical analysis with likelihood function}
\label{subsec:likelihood}
A common approach in this context is to divide the frequency range of GW detectors into logarithmically spaced bins of equal weight~\cite{Gowling:2021gcy}, ensuring that each bin contains a well-defined contribution to the overall signal. In each bins, there exist 
\begin{eqnarray}\label{eq:frequencybin}
    n_b \equiv \left[(f_b - f_{b-1})\tau_{\rm obs}\right],
\end{eqnarray}
with $b\in [0,N_b]$. We have set $N_b=100$ in our analysis for all the detectors~\footnote{$N_b=\mathcal{O}(100)$ ensures the Gaussian approximation~\cite{Gowling:2021gcy}.}. The square brackets represent an integer value of $n_b$. Throughout our analysis, we have considered a Gaussian Likelihood function ($\mathscr{L}(\theta)$) for $\Omega_{\rm sig}(f_b,\{\theta\})$. For the mathematical convenience, instead of maximising $\mathscr{L}$, we generally maximise its logarithm, defined as
\begin{eqnarray}\label{eq:loglikelihood}
    \mathcal{L} (\theta) \equiv ln\,(\mathscr{L} (\theta)).
\end{eqnarray}
It is generally called the log-likelihood function or $\chi^2$-distribution.

\subsubsection*{Fisher forecast analysis:}
Although SNR analysis asses the parameter space where is the detection is robust, it does not provide precise measurements of the parameters being probed. Fisher matrix analysis estimates the uncertainties associated with the parameter, offering insight into the precision of the parameter during measurement. We adopted the following form of the Gaussian likelihood function~\cite{Dodelson:2003ft} to perform the Fisher analysis:
\begin{eqnarray}
\label{eq:likelihood_fisher}
    \mathscr{L} (\theta) = \prod_{b=1}^{N_b} \sqrt{\frac{n_b}{2\pi\Omega_{\rm n}(f_b)^2}}\,\, {\rm exp}\left( - \frac{n_b\left(\Omega_{\rm sig}(f_b,\{\theta\}) - \Omega_{\rm fid}(f_b)\right)^2}{\Omega_{\rm n}(f_b)^2}\right).
\end{eqnarray}
where  $\Omega_{\rm fid}(f_b)\equiv \Omega_{\rm sig}(f_b,\{\theta_{\rm fid})\}$ with $\{\theta_{\rm fid}\}$ indicates the fiducial values of the model parameters.
In terms of log-likelihood, the Fisher matrix can be expressed as~\cite{Dodelson:2003ft}
\begin{eqnarray}
    F_{ij} = \left\langle-\,\frac{\partial^2 \mathcal{L} (\theta)}{\partial \theta_i \partial \theta_j}\right\rangle,
\end{eqnarray}
where angular brackets indicate the ensemble average over observational data \textit{i.e.}, the data that would be collected if the true Universe were described by the given model, around which the derivative is evaluated. Under the Gaussian approximation, the Fisher matrix can be expressed as~\cite{Dodelson:2003ft}
\begin{eqnarray}
    F_{ij} = \tau_{\rm obs}\sum_{b=1}^{N_b}\frac{2\Delta f_b}{\Omega_{\rm n}^2}\frac{\partial \Omega_{\rm sig}}{\partial \theta_i}\frac{\partial \Omega_{\rm sig}}{\partial \theta_j}\, ,
\end{eqnarray}
where $\Delta f_b$ is the frequency binning, defined as $f_b - f_{b-1}$.
The reciprocal of the Fisher matrix, $\left[C_{ij}\right] \;\equiv\; \left[F_{ij}\right]^{-1}$, provides the covariance matrix, in which the square roots of the diagonal elements ($C_{ii}$) exhibit the uncertainties associated with the model parameters ($\theta_i$).

\subsubsection*{Markov chain Monte Carlo analysis:}
As discussed in the previous section, the Fisher matrix provides insights only into the uncertainties for a given fiducial value. To obtain a more comprehensive inference of the parameters, we perform a rigorous MCMC analysis for the GW detectors ET and LISA. In this context, we employ the same form of the likelihood function as defined in Eq.~\eqref{eq:likelihood_fisher}, however, we replace the fiducial model $\Omega_{\rm fid} (f_b)$ with mock data $\Omega_{\rm mock} (f_b)$. This substitution is essential in the MCMC framework, where the likelihood function must quantify the agreement between model predictions and (mock) observational data, rather than a predefined fiducial model. To generate the mock catalogues, we have chosen a set of parameters such that \textit{i)} SNR $>1$ and \textit{ii)} relative uncertainties on the parameters (obtained from Fisher analysis) less than $1$ \textit{i.e.} $\Delta \theta/\theta<1$. To make it more realistic, we have sampled the chosen set of parameters from a normal distribution, incorporating the noise spectra associated with the detectors~\cite{Ferreira:2022jcd}, ensuring that the mock catalogue is different depending on the spectra and the detectors. The desired posterior distributions of the parameters are expressed as the product of likelihood and priors on the parameters. The prior ranges are tabulated in Table-\ref{tab:prior}. To perform the whole analysis, we have used the publicly available sampler code \texttt{emcee}~\cite{2013PASP..125..306F}, whereas \texttt{GetDist}~\cite{Lewis:2019xzd} is used to plot the posterior distribution.

\begin{table}[!ht]
    \centering
    \renewcommand{\arraystretch}{1.2}
    \begin{tabular}{c|c}
    \hline
    \hline
        \textit{Parameter} & \textit{Prior}\\
        \hline
        $\w$ & Flat, 0.1 $\rightarrow$ 0.99 \\
        $\log_{10}(\beta)$ &  Flat, $-22$ $\rightarrow$ $0$\\
        $\log_{10}(\Min/1\,{\rm g})$ &  Flat, $-0.3$ $\rightarrow$ $8.62$\\
    \hline
    \hline
    \end{tabular}
    \caption{\it Priors on the all three parameters for ET and LISA.}
    \label{tab:prior}
\end{table}

\section{Gravitational waves from PBH reheating}
\label{sec:GW_PBH}
As mentioned earlier, two promising sources of gravitational waves are expected to arise in the PBH reheating scenario: one associated with isocurvature fluctuations and the other with adiabatic fluctuations. In this section, we will focus on each of these sources individually and investigate the detection prospects and possible constraints on  the associated PBH parameters.  We will make use of the statistical techniques as discussed in the previous section, for the two upcoming interferometer missions, namely, LISA and ET, and evaluate their potential detection prospects of the PBH parameters.

The observed quantity, GW spectral energy density, can be defined, in terms of oscillation averaged tensor power spectrum ($\overline{\mathcal{P}_h}$), as 
\begin{eqnarray}
\label{eq:omega_GW}
    \Omega_{\rm GW}(k,\eta) = \frac{k^2}{12\mathcal{H}^2}\overline{\mathcal{P}_h(k,\eta)}.
\end{eqnarray}
Here, \(\mathcal{H}\) is the Hubble parameter defined in terms of the conformal time ($\eta$), \textit{i.e.}, $\mathcal{H} \equiv \frac{da}{d\eta} \frac{1}{a}$.
This spectra is calculated at the conformal time during the RD epoch when GWs are well inside the horizon and propagate freely. Hence, at the present time, GW spectra can be calculated as~\cite{2018PhRvD..97j3528A}
\begin{eqnarray}
\label{eq:GW_present}
    \Omega_{\rm GW}^{(0)}(k) \,{ h}^2\simeq c_{\rm g}\, \Omega_{\rm r}^{(0)}  { h}^2\, \Omega_{\rm GW} (k,\eta)\simeq1.62\times 10^{-5} \,\Omega_{\rm GW} (k,\eta).
\end{eqnarray}
The parameter $ c_g$ is given by $c_g \equiv \left( \frac{g_{\ast k}}{g_{\ast 0}} \right) \left( \frac{g_{\ast\mathrm{s}0}}{g_{\ast\mathrm{s}k}} \right)^{4/3}$, where $ (g_{\ast k}, g_{\ast\mathrm{s}k})$ and  $(g_{\ast 0}, g_{\ast \mathrm{s}0})$ denote the effective number of relativistic degrees of freedom associated with the energy density of radiation and entropy, respectively, at the time when the wave number $k$ re-enters the Hubble radius and at the present epoch. Additionally, $ \Omega_{\rm r}^{(0)} h^2 \simeq 4.18 \times 10^{-5}$ represents the radiation energy density today.
Now $\overline{\mathcal{P}_h (k,\eta)}$ can be expressed, in terms of curvature power spectrum ($\mathcal{P}_{\Phi}$), as~\cite{2018PhRvD..97l3532K}
\begin{eqnarray}
\label{eq:av_tensorPS}
    \overline{\mathcal{P}_{h}(k,\eta)}=8 \int_{0}^{\infty} dv \int_{|1-v|}^{1+v} du \left(\frac{\left(1+v^2-u^2\right)^2-4 v^2}{4 u v}\right)^2 {\cal P}_{\Phi}(u k){\cal P}_{\Phi}(v k)\overline{I^2}(x,u,v)    \,,
    \label{eq:TensorPowerSpectrum}
\end{eqnarray}
with $x\equiv k\eta$. The kernel function, $I(x,u,v)$ carries the complete information of time-evolution in $\mathcal{P}_h$. Now due to the sudden transition from PBH-domination to RD, the contribution, coming right after the end of PBH-domination, is the dominant one~\cite{Inomata:2019ivs}. Thus, $\mathcal{P}_{\Phi}$ can be estimated at the end of PBH-domination. 
We now present the form of the gravitational wave spectrum for two different sources after solving the momentum integral in  $u$ and $v$, along with their detection prospects in light of upcoming interferometry missions.
\subsection{Gravitational waves induced by isocurvature fluctuations}
\label{subsec:density_fluctuation}
For $\beta>\beta_{\rm c}$, PBHs indirectly serve as a significant source of induced gravitational waves. The initial inhomogeneous distribution of PBHs creates isocurvature fluctuations, which, by the time PBHs dominate, transform into adiabatic perturbations. It can be assumed that the PBH-distribution to be a Poisson spectrum for the density fluctuation and we define the initial isocurvature perturbation as $\frac{\delta \rho_{\rm BH}(\mathbf{x})}{\rho_{\rm BH}}\equiv S_i(\mathbf{x})$. If $d$ be the mean separation between the PBHs, the two-point correlation for $S_i$ in the real-space, can be expressed as~\cite{Papanikolaou:2020qtd}
\begin{eqnarray}
\label{eq:density_contrast1}
    \left\langle \frac{\delta \rho_{\rm BH}(\mathbf{x})}{\rho_{\rm BH}} \frac{\delta \rho_{\rm BH}(\mathbf{\Tilde{x}})}{\rho_{\rm BH}} \right\rangle = \frac{4\pi}{3} \left(\frac{d}{a}\right)^3 \delta(\mathbf{x}-\mathbf{\Tilde{x}})\,.
\end{eqnarray}
In Fourier space this correlation has the following form~\cite{Papanikolaou:2020qtd,Domenech:2020ssp} 
\begin{eqnarray}
    \label{eq:density_contrast_fourier}
    \left\langle S_i(k) S_i(\Tilde{k}) \right\rangle = \frac{2\pi^2}{k^3}\,\mathcal{P}_{\rm BH,i}(k)\,\delta(k+\Tilde{k}),
\end{eqnarray}
where the initial power spectra ($\mathcal{P}_{\rm BH,i} (k)$) can be defined as
\begin{eqnarray}
    \label{eq:scale-inv_PS}
    \mathcal{P}_{\rm BH,i} (k) \equiv \frac{k^3}{2\pi} \frac{4\pi}{3} \left(\frac{d}{a}\right)^3 = \frac{2}{3\pi} \left(\frac{k}{k_{\rm UV}}\right)^3.
\end{eqnarray}
Hence, the initial power spectrum is directly related with the ultraviolet (UV) cutoff scale ($k_{\rm UV}$) which can be expressed as~\cite{Domenech:2020ssp} 
\begin{eqnarray}
    \label{eq:kUV}
    k_{\rm UV} \simeq 1.1 \times 10^4\, {\rm Hz}\left(\frac{g_{\ast, s}(T_{\rm ev})}{106.75}\right)^{-1/3} \left(\frac{g_{\ast}(T_{\rm ev})}{106.75}\right)^{1/4} \left(\frac{g_{\ast}(T_{\rm BH})}{108}\right)^{1/6}\left(\frac{M_{\rm in}}{10^4 \, \rm g}\right)^{-5/6}.
\end{eqnarray}
The evolution of the Newtonian potential ($\Phi$), can be related with initial isocurvature perturbation as
\begin{eqnarray}
\label{eq:PhiEq}
    \Phi'' +  3 \mathcal{H}(1+c_s^2)\Phi' + \left((1+3c_s^2)\mathcal{H}^2+2\mathcal{H}' +c_s^2 k^2\right)\Phi = \frac{1}{2}a^2c_s^2\rho_{\rm BH} S_i \,,
\end{eqnarray}
with  $c_s$ representing the sound speed. The prime ($'$) in the expression indicates the derivative with respect to $\eta$. This isocurvature perturbation acts as a source term in the tensor power spectrum $\mathcal{P}_h(k)$, ultimately inducing second-order GWs. The spectral energy density of these GWs, particularly near the resonant peak, is given by~\cite{Domenech:2024wao}:
\begin{eqnarray}
\label{eq:OGWres}
    \Omega_{\rm GW, res}(k)\approx \Omega_{\rm GW, res}^{\rm peak}
    \left(\frac{k}{k_{\rm UV}}\right)^{11/3}
    \Theta^{\rm (iso)}_{\rm UV}(k),
\end{eqnarray}
where the amplitude near the peak can be expressed as 
\begin{eqnarray}
\label{eq:OGWresPeak}
   \Omega_{\rm GW, res}^{\rm peak} = C^4(\w)\frac{c_s^{7/3}(c_s^2-1)^2}{576 \times 6^{1/3}\pi}
    \left(\frac{k_{\rm BH}}{k_{\rm UV}}\right)^8
    \left(\frac{k_{\rm UV}}{k_{\rm ev}}\right)^{17/3},
\end{eqnarray}
with $C(\w) = \frac{9}{20} \alpha_{\rm fit}^{-\frac{1}{3\w}}\left(3+\frac{1-3\w}{1+3\w}\right)^{-\frac{1}{3\w}}$, considering $\alpha_{\rm fit}\approx 0.135$. The PBH domination scale ($k_{\rm BH}$) is defined as following 
\begin{eqnarray}
\label{eq:kBH}
    k_{\rm BH} &=& \sqrt{2}\gamma^{1/3}\beta^{\frac{1+\w}{6\w}}k_{\rm UV}.
\end{eqnarray}
The cutoff function, $\Theta^{\rm (iso)}_{\rm UV}(k)$, is defined in App.~\ref{app:theta_UV}. The infrared (IR) region belongs to relatively lower frequencies where $k\ll k_{\rm UV}$. The prime contribution to this regime can be expressed as~\cite{Domenech:2024wao}
\begin{eqnarray}
\label{eq:OGWIR}
    \Omega_{\rm GW, IR}(k) &=& C^4(\w) \frac{c_s^4 }{120 \pi ^2}\left(\frac{2}{3}\right)^{\frac{1}{3}} 
    \left(\frac{k_{\rm BH}}{k_{\rm UV}}\right)^8
    \left(\frac{k_{\rm UV}}{k_{\rm ev}}\right)^{\frac{14}{3}}
    \left(\frac{k}{k_{\rm UV}}\right) \\
    &\simeq& 9.03 \times 10^{24}\, C^4(\w) \,\beta^{\frac{4(1+\w)} {3\w}}
    \left(\frac{\gamma}{0.2}\right)^{\frac{8}{3}}
    \left(\frac{M_{\rm in}}{10^4\text{g}}\right)^{\frac{28}{9}}\left(\frac{k}{k_{\rm UV}}\right).
\end{eqnarray}
In addition, a transition frequency $ f_{\rm T}$ marks the boundary between the infrared (IR) tail of the GW spectrum and the region near the resonant peak \cite{Domenech:2020ssp}. This frequency is determined by equating the IR and resonant components of the GW spectrum:  
$\Omega_{\rm GW,IR} (f_{\rm T}) = \Omega_{\rm GW,res} (f_{\rm T})$.
The transition frequency can be expressed as:
\begin{eqnarray} 
f_{\rm T} = \frac{k_{\rm UV}}{2\pi} \left[\left(10^{-6} \frac{M_{\rm in}}{10^4 \, {\rm g}}\right)^{-2/3}\right]^{3/8}\,,
\end{eqnarray}    
which depends solely on the PBH formation mass $M_{\rm in}$.  
Thus, the complete isocurvature-induced GW spectrum is given by:  
\begin{eqnarray}
\label{eq:GW_iso}
    \Omega_{\rm GW, iso}(f) =
    \begin{cases}
        \Omega_{\rm GW,res}(f), & f \geq f_{\rm T}, \\
        \Omega_{\rm GW,IR}(f), & f < f_{\rm T}.
    \end{cases}
\end{eqnarray}

Finally, the present-day isocurvature-induced GW spectrum, $ \Omega_{\rm GW,iso}^{(0)}(f)\, h^2$, can be obtained using Eq.~\eqref{eq:GW_present}. 
\begin{figure*}[!ht]
    \centering
    \begin{subfigure}{.32\textwidth}
    \includegraphics[width=\textwidth]{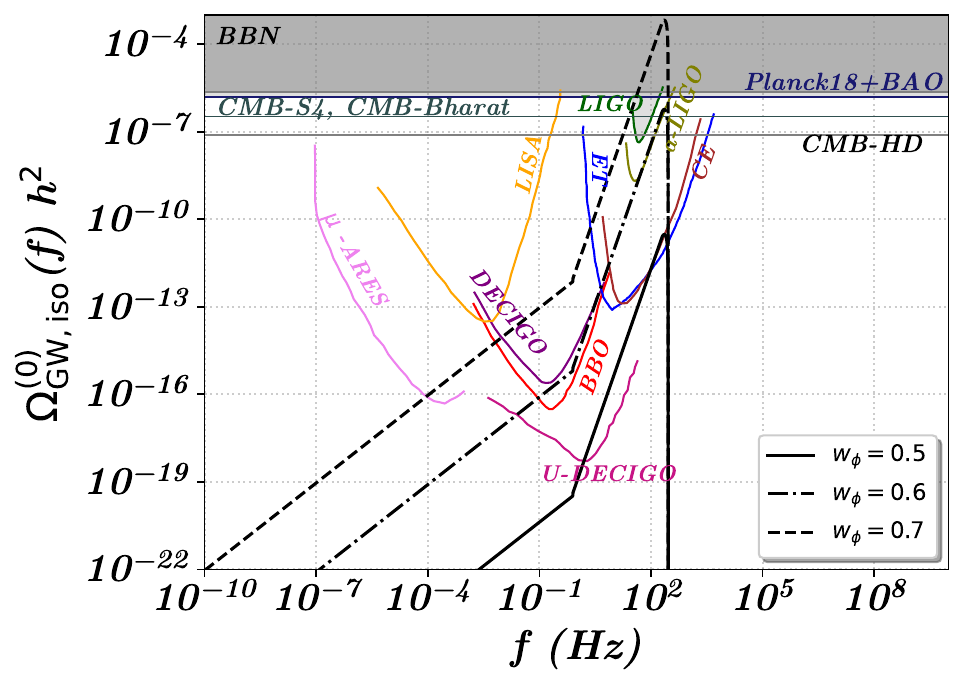}
    \caption{\it $\beta=10^{-10}$ and $\Min = 10^5$ g}
    \label{fig:GWiso_wvarry}
    \end{subfigure}
    \hfill
    \begin{subfigure}{.32\textwidth}
    \includegraphics[width=\textwidth]{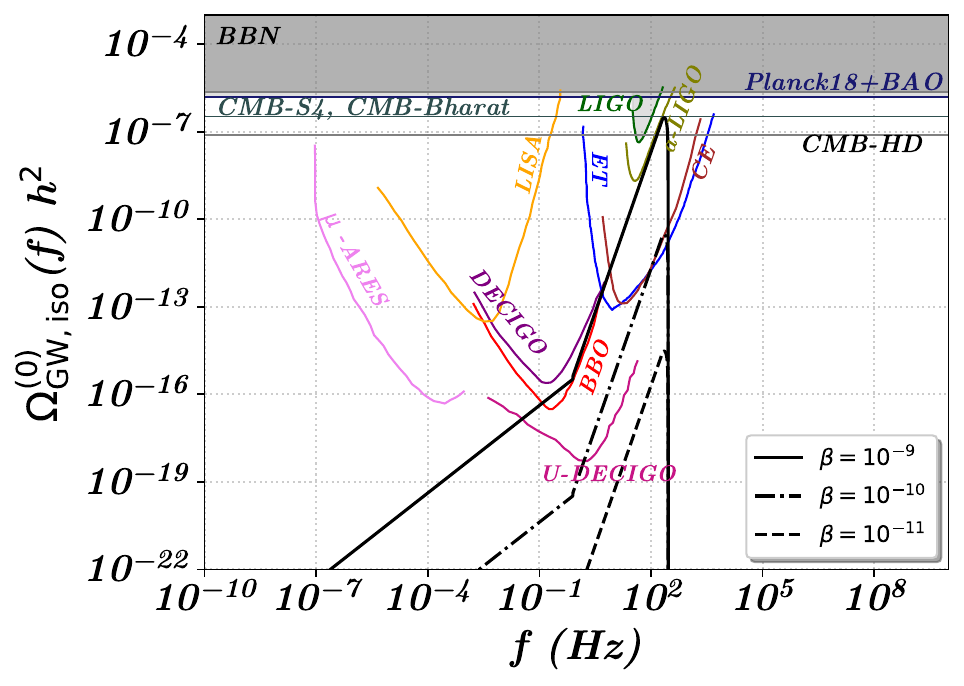}
    \caption{\it $\w=0.5$ and $\Min = 10^5$ g}
    \label{fig:GWiso_betavarry}   
    \end{subfigure}
    \hfill
    \begin{subfigure}{.32\textwidth}
    \includegraphics[width=\textwidth]{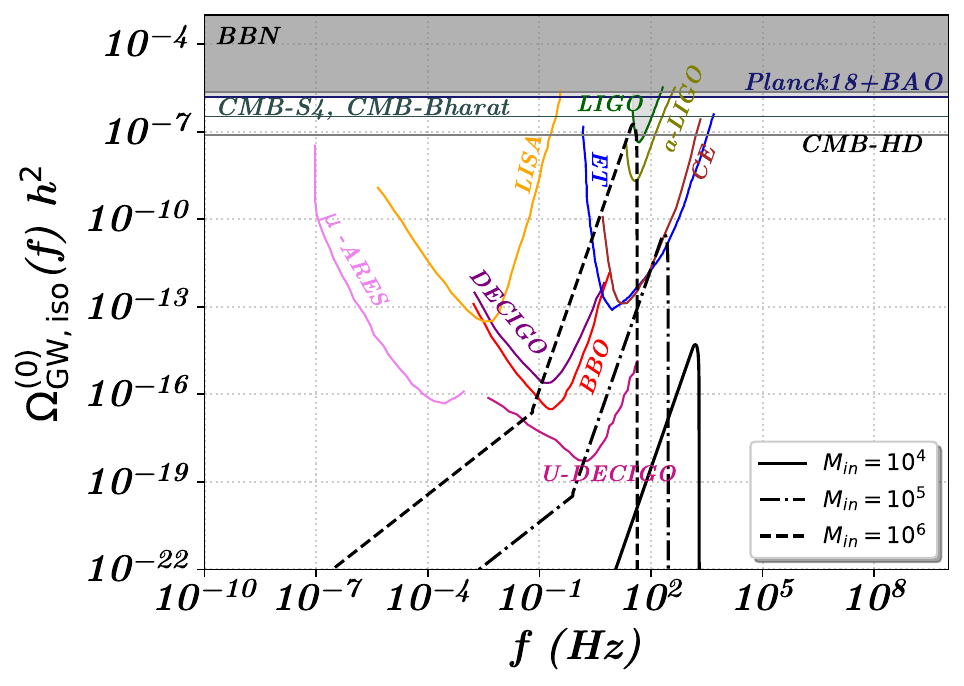}
    \caption{\it $\w=0.5$ and $\beta = 10^{-10}$}
    \label{fig:GWiso_mvarry}   
    \end{subfigure}
    \caption{\it Induced gravitational wave spectrum at the present epoch due to \textbf{isocurvature fluctuations}. Each plot represents the variation of a single parameter while the other two are kept fixed.}
    \label{fig:GWiso}
\end{figure*}

In Fig.~\ref{fig:GWiso}, we illustrate how each parameter affects the GW spectrum while keeping others fixed. The spectral slope near the peak,  $\Omega_{\rm GW,res} \propto k^{11/3}$ (\textit{i.e.}, Eq.~\eqref{eq:OGWres}), remains independent of $\w$, with the peak frequency solely determined by $M_{\rm in}$ (\textit{i.e.}, the peak appear at the wave number $k_{\rm UV}$). Lighter PBHs form and evaporate earlier, shifting GW peaks to higher frequencies (Fig.~\ref{fig:GWiso_mvarry}). For a fixed $M_{\rm in}$, the peak frequency remains unchanged, but its amplitude increases with $\w$ (Fig.~\ref{fig:GWiso_wvarry}) and $\beta$ (Fig.~\ref{fig:GWiso_betavarry}).  

This behaviour can be explained using Eq.~\eqref{eq:OGWresPeak}, as the background equation of state $\w$, where PBHs form, influences the GW amplitude through $C^4(\w)$ and  $\beta^{\frac{4(1+\w)}{3\w}}$. This $\beta$ dependency arises from the ratio $(\kbh/k_{\rm UV})^8$ in Eq.~\eqref{eq:OGWresPeak}, making the amplitude highly sensitive to $\w$.  The transfer function term, $C_4(\w)$, enhances the GW amplitude for lower $\w$ by reducing the gravitational potential decay, while $\beta^{\frac{4(1+\w)}{3\w}}$ has the opposite effect, increasing the amplitude for stiffer $\w$ and suppressing it for softer $\w$. This can be easily understood from the fact that a stiffer equation of state ($\w > 1/3$) prolongs the PBH-dominated phase, enhancing the GW signal, while a softer one ($\w < 1/3$) shortens it, leading to suppression.

However, the isocurvature-induced peak alone cannot break the degeneracy between $\w$ and $\beta$, as both influence the amplitude. To resolve this, we consider induced gravitational waves from inflationary adiabatic fluctuations, which we explore in the next section.

\subsubsection{Statistical analysis in the light of the detectors}
\label{subsubsec:parameter_estimation_density_fluctuation}

Fig.~\ref{fig:GWiso} shows that the GW spectra from the isocurvature source are well within the detection capabilities of ET and LISA. Therefore, we proceed with the statistical analysis of this source for the two detectors, as outlined in Sec.~\ref{sec:detection_prospects}.
\subsubsection*{Evaluation of signal-to-noise ratio:}
\begin{figure*}[!ht]
    \centering
    \begin{subfigure}{.45\textwidth}
    \includegraphics[width=\textwidth]{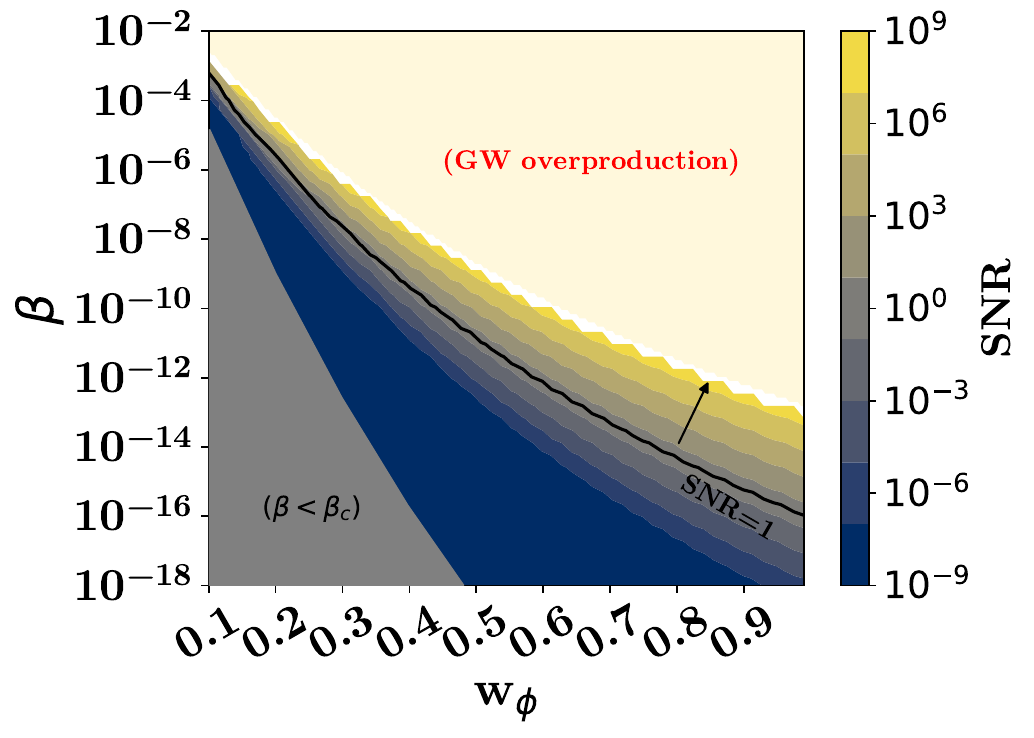}
    \caption{\it $\Min=10^5$ g}
    \label{fig:SNR_iso_ET_m5}
    \end{subfigure}
    \hfill
    \begin{subfigure}{.45\textwidth}
    \includegraphics[width=\textwidth]{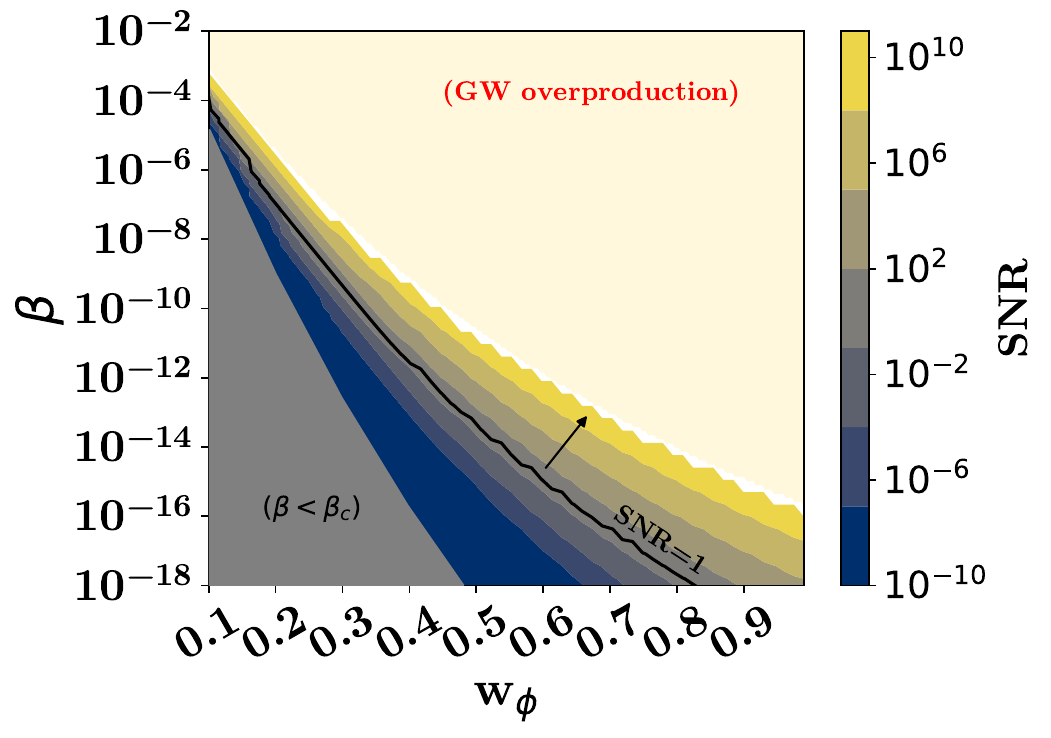}
    \caption{\it $\Min=10^7$ g}
    \label{fig:SNR_iso_ET_m7}   
    \end{subfigure}
    \caption{\it SNR plots for \textbf{ET} with two different values of $\Min$ considering \textbf{isocurvature fluctuation}. The black solid line represents ${\rm SNR}=1$ with the arrow indicating the region ${\rm SNR}>1$. GW overproduction, derived from $\Delta N_{\rm eff}$ limit at BBN, and $\beta<\beta_c$ are indicated by light orange and grey regions, respectively.}
    \label{fig:SNR_iso_ET}
\end{figure*}

\begin{figure*}[!ht]
    \centering
    \begin{subfigure}{.45\textwidth}
    \includegraphics[width=\textwidth]{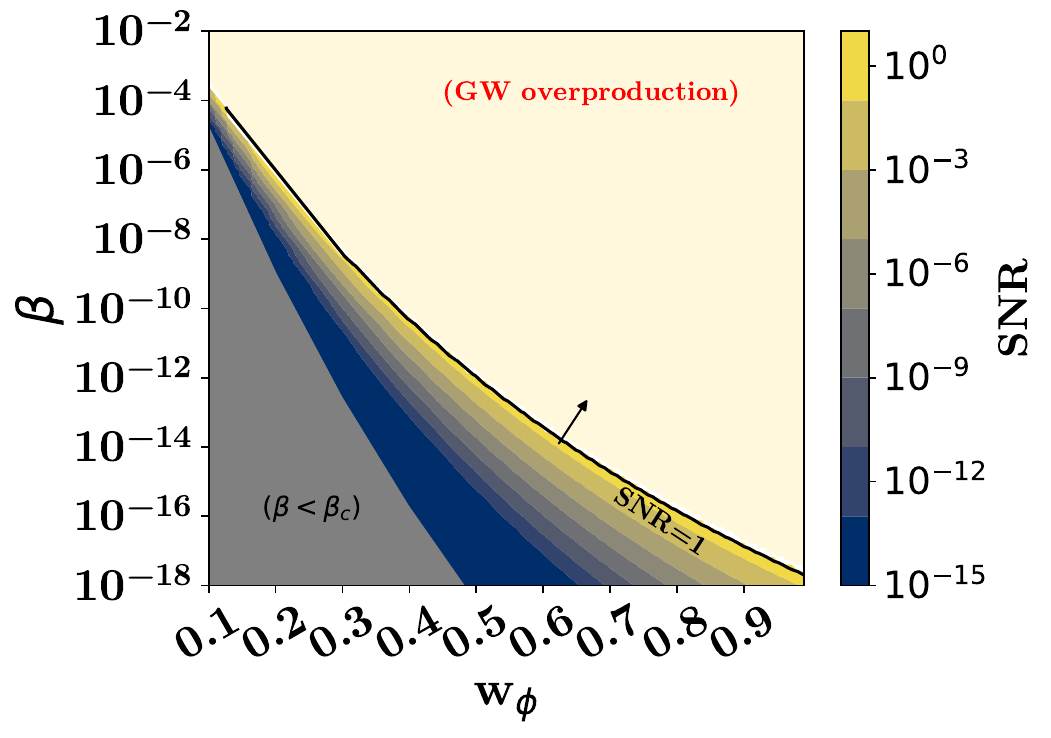}
    \caption{\it $\Min=2\times10^8$ g}
    \label{fig:SNR_iso_LISA_m9}
    \end{subfigure}
    \begin{subfigure}{.45\textwidth}
    \includegraphics[width=\textwidth]{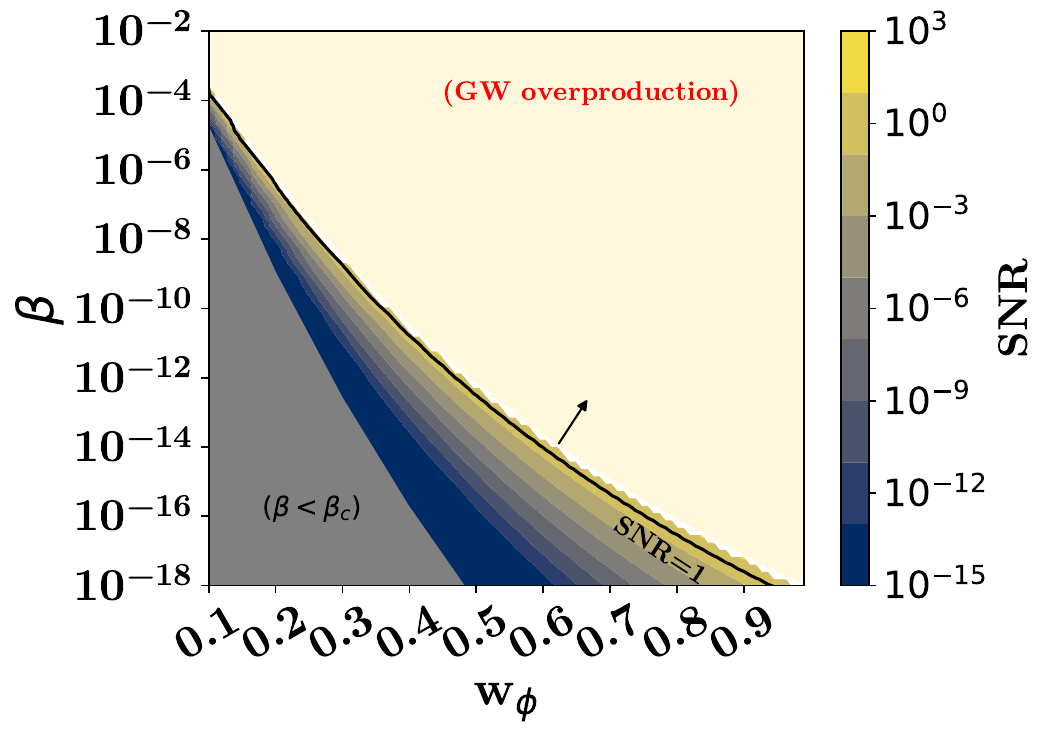}
    \caption{\it $\Min=4.8\times10^8$ g}
    \label{fig:SNR_iso_LISA_m10}   
    \end{subfigure}
    \caption{\it Illustration of SNR is same as Fig.~\ref{fig:SNR_iso_ET}, but for \textbf{LISA}.}
    \label{fig:SNR_iso_LISA}
\end{figure*}

Fig.~\ref{fig:SNR_iso_ET} and Fig.~\ref{fig:SNR_iso_LISA} depict the SNR in the $\w$-$\beta$ plane for different values of $\Min$, corresponding to the future GW missions ET and LISA, respectively. The black solid line represents SNR = 1, while the regions where $\rm{SNR} > 1$ are indicated by black arrows in both cases. Additionally, the region where $\beta < \beta_c$ and the constraints from GW overproduction—derived from the upper bound on $\Delta N_{\rm eff}$ from BBN, with $\Delta N_{\rm eff} = 0.28$ based on Planck legacy data~\cite{Planck:2018vyg}, are highlighted in grey and light orange, respectively. In both scenarios, the SNR increases with $\beta$ for a fixed $\w$ and vice versa, a trend discussed in detail in the previous section. 

\begin{figure}[!ht]
    \centering
    \includegraphics[scale=0.5]{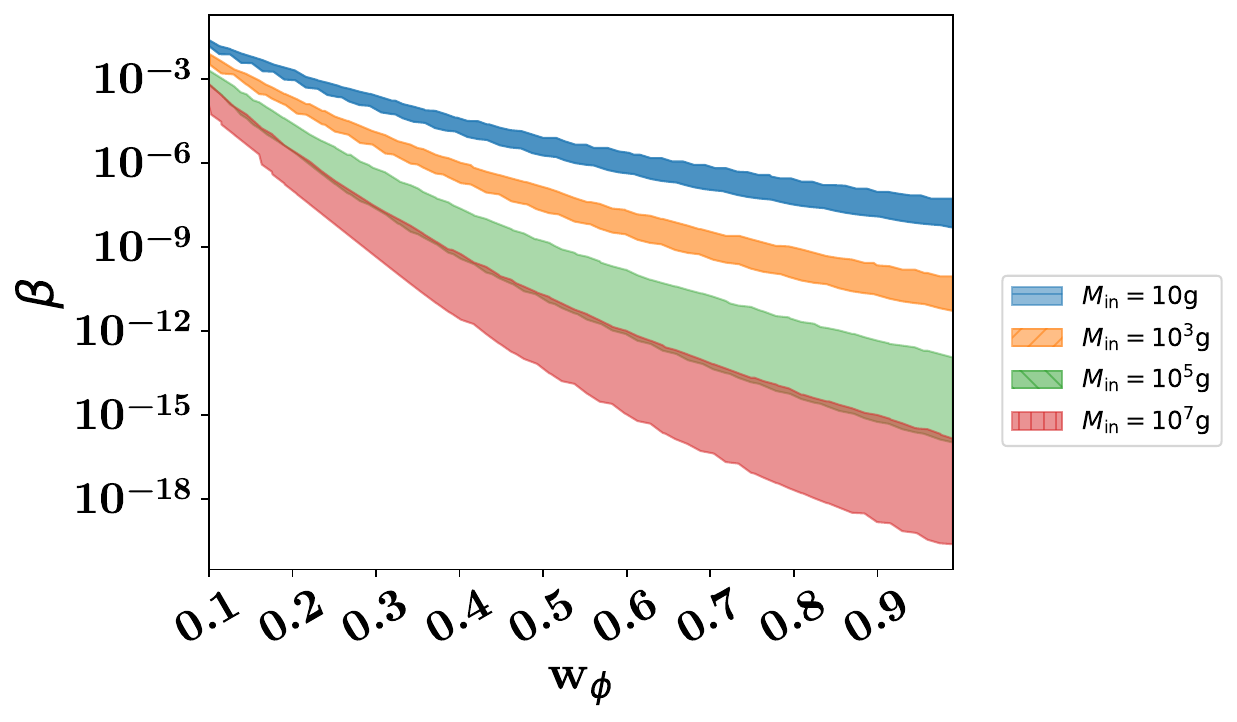}
    \caption{\it Illustration of extent of parameter space for various $\Min$ where ${\rm SNR}\geq1$, considering \textbf{isocurvature induced GWs}, for ET. The upper limit comes from the GW overproduction, derived from $\Delta N_{\rm eff}$ constraints at BBN.}
    \label{fig:snr_iso_et_all_mass}
\end{figure}

In Fig.~\ref{fig:snr_iso_et_all_mass}, we show the range of parameters in order to achieve an $\rm SNR >1$ for the interferometry mission ET in the $\w-\beta$  plane for different values of $M_{\rm in}$. One interesting outcome of this analysis is that as the mass increases from a very low value, say $1 \,\rm g$, the region where $\rm SNR >1$ in the $\w-\beta$ plane expands. However, after reaching an optimal value around $10^{6} \,\rm g$, the region starts to shrink. Beyond $5\times 10^{7} \,\rm g$, no region in the $\w-\beta$ plane remains where the SNR exceeds one. Such behaviour can be easily understood by comparing the peak frequency of the induced GW spectrum from the isocurvature source with the noise curve of the ET (see, for instance, Fig. \ref{fig:noise} in the appendix). The peak frequency of this GWs spectrum is associated with $k_{\rm UV}$ and is given by: $f_{\text{peak}} \sim 1.7 \times 10^3 \, (M_{\rm in}/10^{4})^{-5/6} \text{ Hz} $. Since  ET is most sensitive to frequencies in the range of approximately $1-4 \times 10^3$ Hz, the peak frequency of the spectrum, which lies within the sensitivity for the mass range $10^{3} - 10^{7}$ g, determines the extent of the high-SNR region. Thus, any formation mass outside this range — either above or below — leads to a lower detection probability. This trend is also evident in Fig.~\ref{fig:snr_iso_et_all_mass}, where for masses decreasing below $ 10^3$ g, the parameter space shrinks, with the minimum mass providing  $\rm SNR > 1$ being around $1$ g. Moreover, from the noise sensitivity curve, we see that the lowest value of $ \Omega_{\rm GW}^{\rm noise} h^2$ occurs around 10 Hz, which corresponds to a formation mass of approximately $10^6$ g. This aligns with our findings, as we observe the largest compatible parameter space in the $\w-\beta$ plane around this mass scale.

On the other hand, for LISA, which is sensitive to frequencies in the range of $10^{-5} - 0.5$ Hz, we find that only PBH masses above $10^{8}$ g are compatible with its detection. However, the SNR values remain very close to 1, even for the maximum allowed PBH formation mass of $\sim 5 \times 10^8$ g (calculated by considering the evaporation point at the time of BBN), which is also clear from Fig. \ref{fig:SNR_iso_LISA}. This SNR-based analysis clearly indicates that, in terms of detection possibilities, the isocurvature source is more favourably detected by ET than LISA.

\subsubsection*{Fisher forecast analysis:}
As discussed earlier, the Fisher analysis provides insights into parameter uncertainties for a given source (see Sec.~\ref{subsec:likelihood} for the analysis setup). Here, we use the Fisher forecast to estimate the uncertainties associated with the isocurvature fluctuation parameters, $\w$ and $\beta$, for both ET and LISA. Note that in this analysis, we do not account for any uncertainties in the formation mass, $M_{\rm in}$. Since Fisher forecast analysis depends on the choice of fiducial, we have chosen the fiducials from the range with SNR $\geq 1$ and outlined the range of parameters where the relative uncertainties associated with the parameters remain below $10$~\footnote{Typically, precise parameter estimation requires the relative uncertainties on the parameters to be less than $1$. However, for isocurvature fluctuation, the is no region satisfying the condition $\Delta \w/\w\,(\Delta\beta/\beta)<1$.}.
\begin{figure*}[!ht]
    \centering
    \begin{subfigure}{.31\textwidth}
    \includegraphics[width=\textwidth]{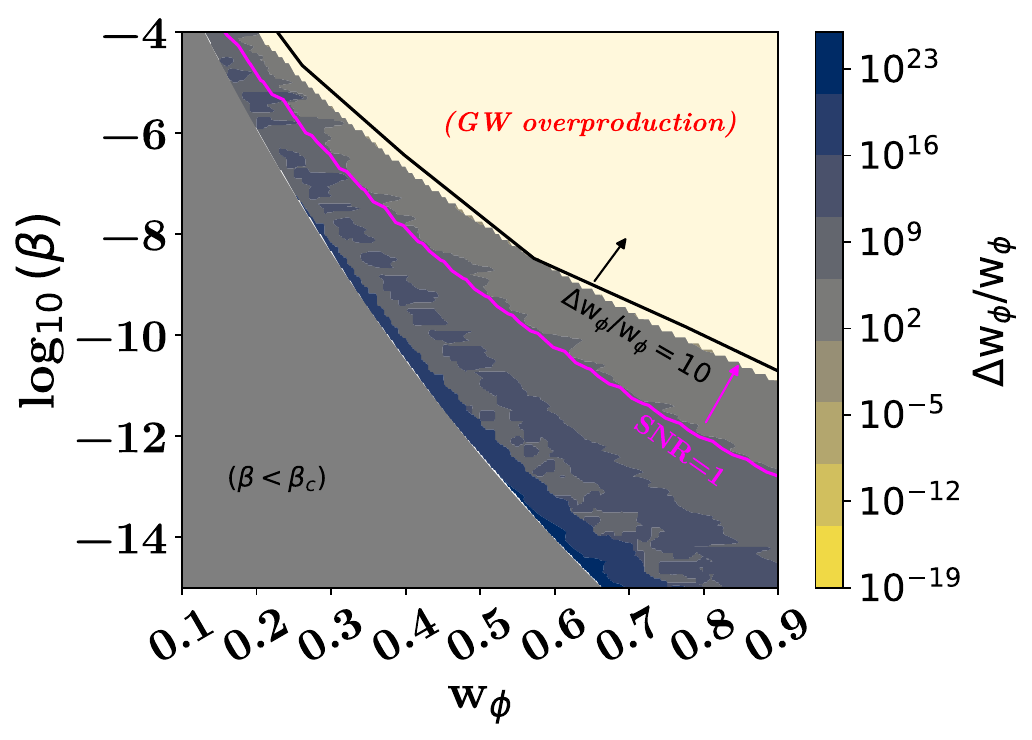}
    \caption{\it $\Min=10^4$ g}
    \label{fig:fisher_iso_ET_deltaw_m4}
    \end{subfigure}
    \begin{subfigure}{.31\textwidth}
    \includegraphics[width=\textwidth]{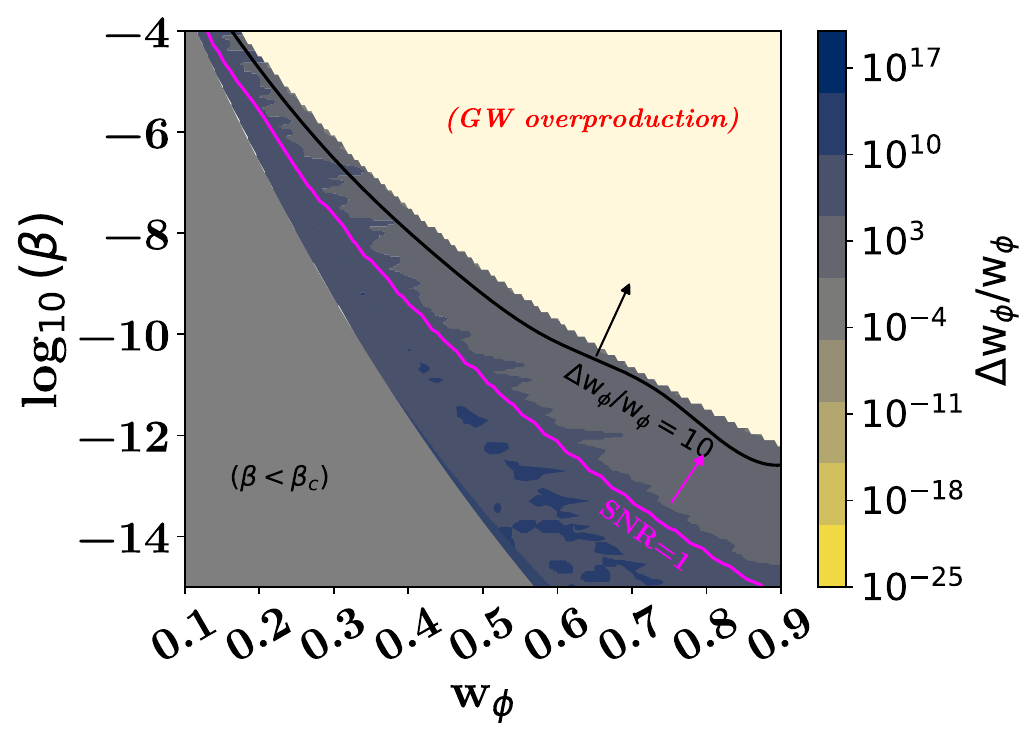}
    \caption{\it $\Min=10^5$ g}
    \label{fig:fisher_iso_ET_deltaw_m5}   
    \end{subfigure}
    \begin{subfigure}{.31\textwidth}
    \includegraphics[width=\textwidth]{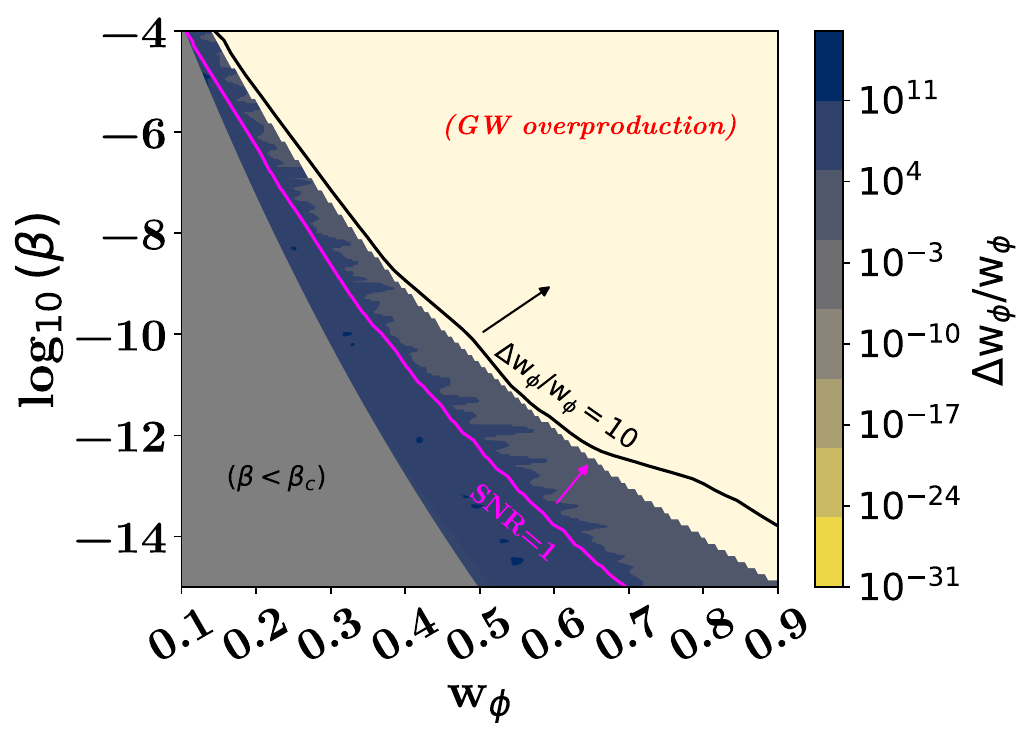}
    \caption{\it $\Min=10^6$ g}
    \label{fig:fisher_iso_ET_deltaw_m6}   
    \end{subfigure}
    \caption{\it Illustration of relative uncertainties on $\w$, calculated by Fisher analysis, considering \textbf{isocurvature fluctuation}, for \textbf{ET}. Black (magenta) solid line indicates ${\rm SNR}=1$ ($\Delta \w/\w=10$). The black (magenta) arrow indicates where ${\rm SNR}>1$ ($\Delta \w/\w<10$).}
    \label{fig:fisher_iso_ET_deltaw}
\end{figure*}
One of the key findings of the Fisher analysis is that in most of the parameter space where ${\rm SNR} > 1$, the relative uncertainties $\Delta \w/\w$ and $\Delta \beta/\beta$ are both large (\textit{i.e.}, $\Delta \w/\w, \Delta \beta/\beta > 10$), as evident in Fig.~\ref{fig:fisher_iso_ET_deltaw} and Fig.~\ref{fig:fisher_iso_ET_deltabeta}. We find that only for masses around $\Min = 10^5$ g, the SNR $>1$ region is compatible with the uncertainty constraint $\Delta \w/\w < 10$, though $\Delta \beta/\beta$ remains greater than 10.  

\begin{figure*}[!ht]
    \centering
    \begin{subfigure}{.45\textwidth}
    \includegraphics[width=\textwidth]{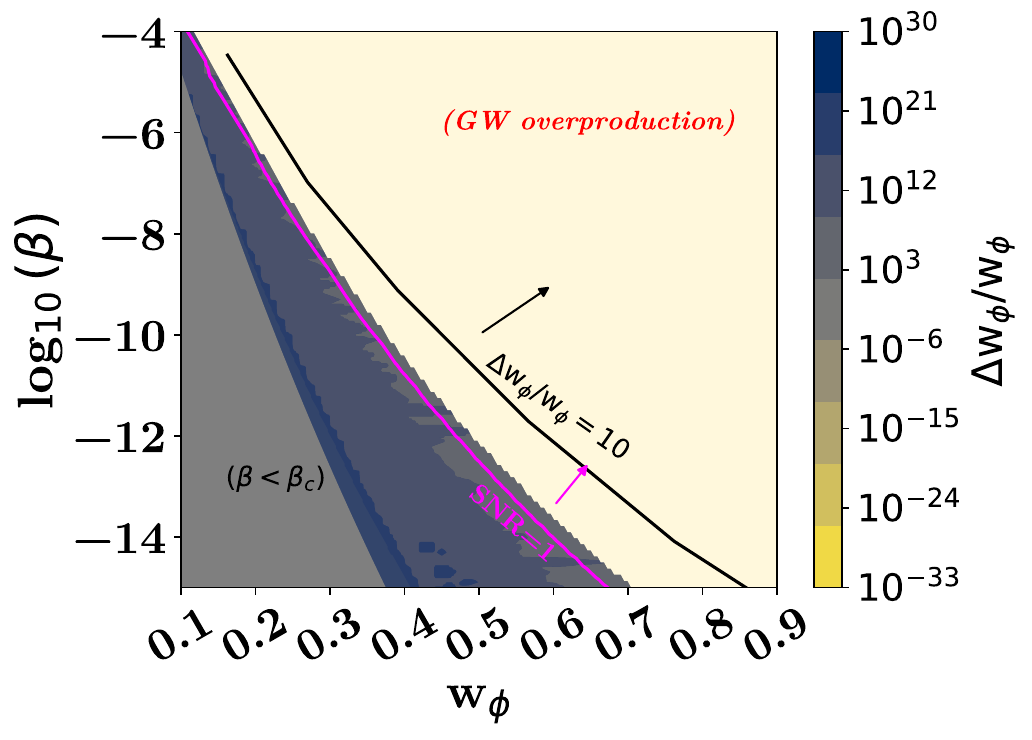}
    \caption{\it Relative uncertainties on $\w$}
    \label{fig:fisher_iso_LISA_deltaw}
    \end{subfigure}
    \begin{subfigure}{.45\textwidth}
    \includegraphics[width=\textwidth]{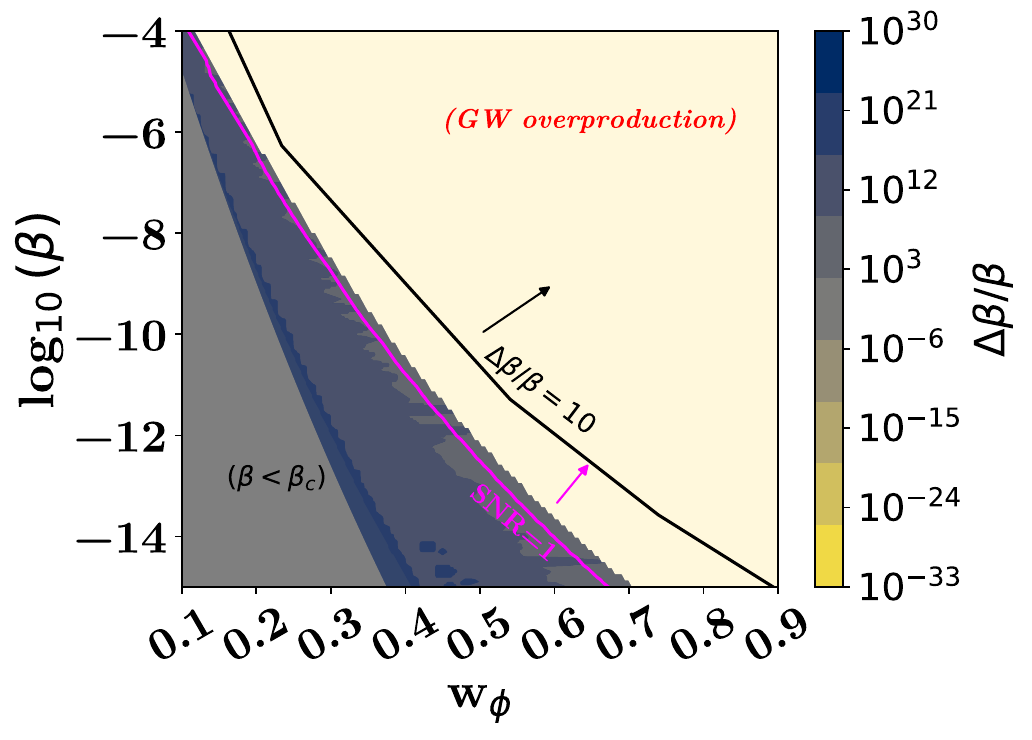}
    \caption{\it Relative uncertainties on $\beta$}
    \label{fig:fisher_iso_LISA_deltabeta}   
    \end{subfigure}
    \caption{\it Illustration of relative uncertainties on $\w$ and $\beta$, calculated by Fisher analysis, considering \textbf{isocurvature fluctuation}, for \textbf{LISA}. The analysis is performed for $\Min=4.8\times10^8$ g. Rest is same as Fig.~\ref{fig:fisher_iso_ET_deltaw}.}
    \label{fig:fisher_iso_LISA}
\end{figure*}

Thus, the Fisher analysis suggests that while ET has a high detection potential, providing high SNR values for isocurvature fluctuations, it is less effective in precisely constraining the associated parameters. On the other hand, for LISA, both the SNR and Fisher analyses indicate poor detection prospects, as shown in Fig.~\ref{fig:fisher_iso_LISA}.  

In summary, considering both SNR and Fisher analyses, we find that LISA has limited detection potential, while ET shows high SNR values but with large parameter uncertainties in most cases. This leads us to conclude that neither detector can convincingly achieve precise measurements of the parameters associated with isocurvature fluctuations, thus probing early physics considering such a source is difficult. Due to this limitation, in the next section, we explore another source of induced GWs, associated with adiabatic fluctuations originating from inflation.

\begin{figure*}[!ht]
    \centering
    \begin{subfigure}{.31\textwidth}
    \includegraphics[width=\textwidth]{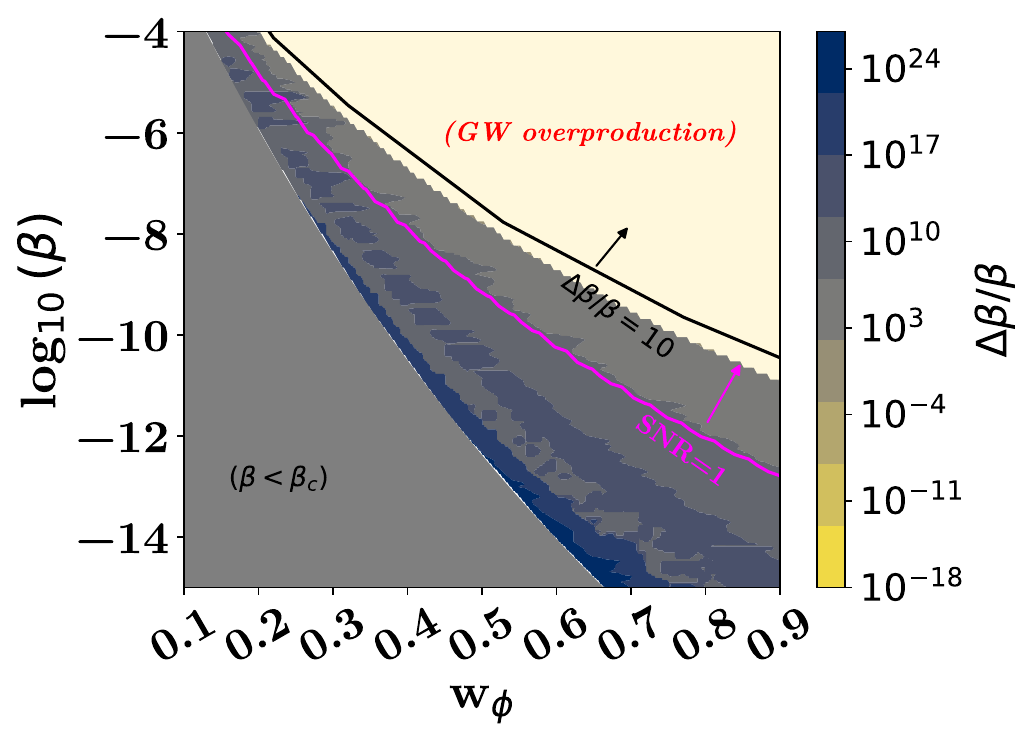}
    \caption{\it $\Min=10^4$ g}
    \label{fig:fisher_iso_ET_deltabeta_m4}
    \end{subfigure}
    \begin{subfigure}{.31\textwidth}
    \includegraphics[width=\textwidth]{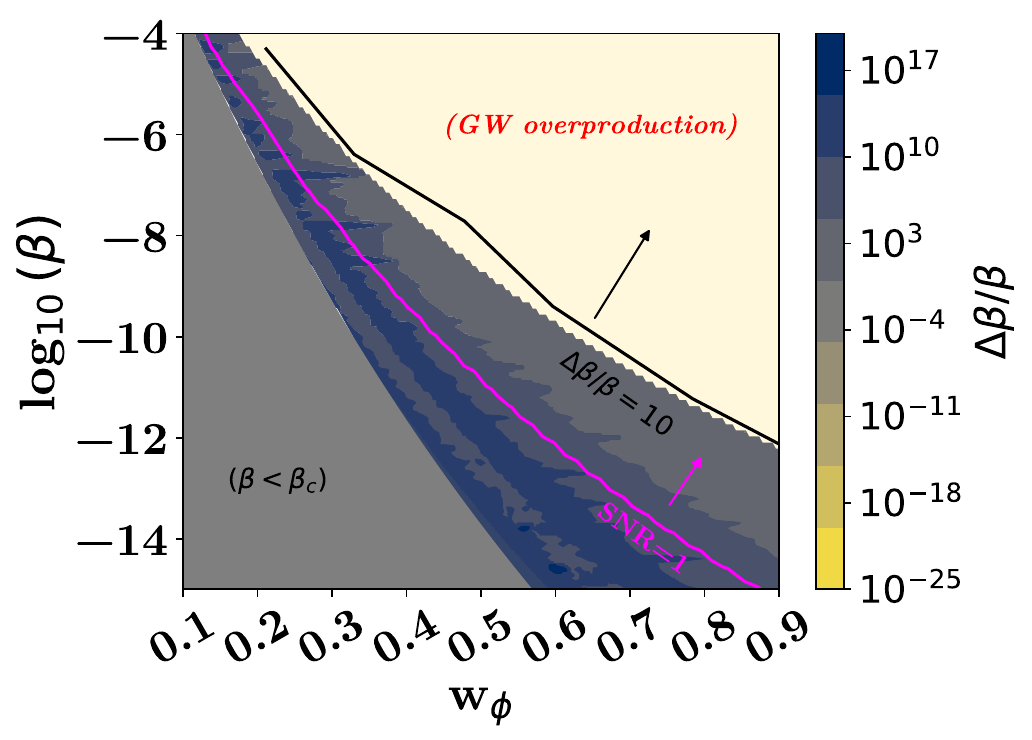}
    \caption{\it $\Min=10^5$ g}
    \label{fig:fisher_iso_ET_deltabeta_m5}   
    \end{subfigure}
    \begin{subfigure}{.31\textwidth}
    \includegraphics[width=\textwidth]{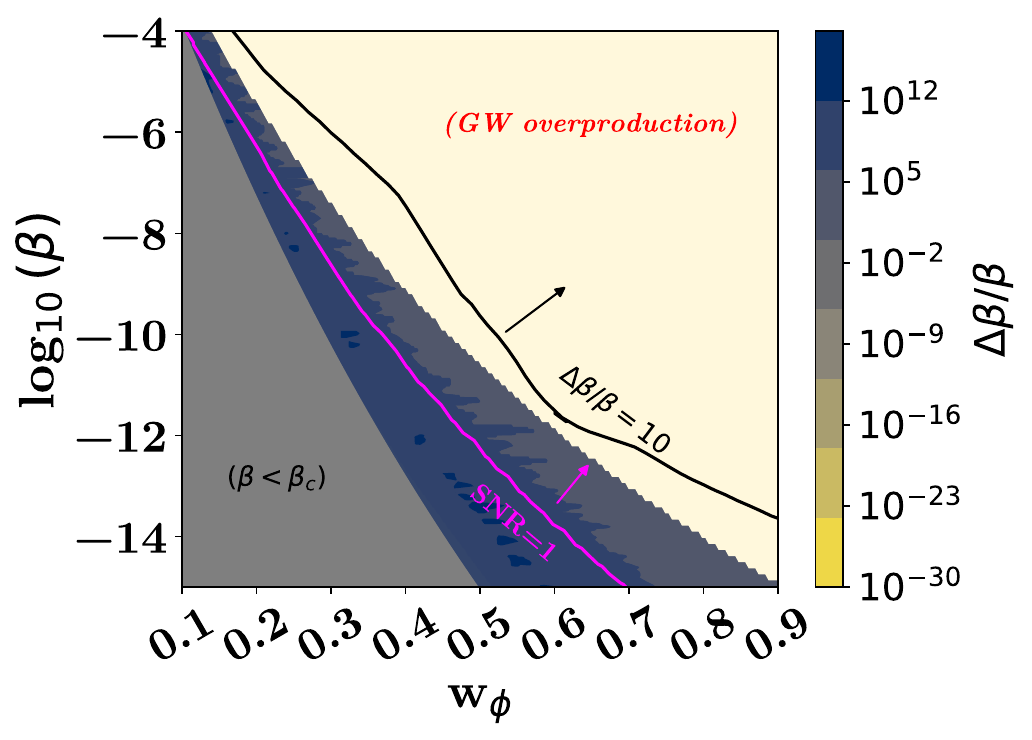}
    \caption{\it $\Min=10^6$ g}
    \label{fig:fisher_iso_ET_deltabeta_m6}   
    \end{subfigure}
    \caption{\it Representation is same as Fig.~\ref{fig:fisher_iso_ET_deltaw}. However, it is the illustration of relative uncertainties on $\beta$, for \textbf{ET}.}
    \label{fig:fisher_iso_ET_deltabeta}
\end{figure*}

\subsection{Gravitational waves induced by adiabatic fluctuations}
\label{subsec:adiabatic_induced_GW}
During the PBH-domination, the gravitational potential remains constant at the subhorizon scales. After PBH-domination, the gravitational potential oscillates due to radiation pressure in the RD era. During the transition to radiation domination, but before these oscillations begin, the potential on subhorizon scales decays, depending on the transition duration. Since PBH evaporation is nearly instantaneous, the transition is too short for significant decay before oscillations start. As a result, strong GWs are generated following the sudden PBH reheating scenario, even if the primordial curvature perturbation power spectrum remains nearly scale-invariant. We follow Ref.~\cite{Domenech:2024wao} in our study for this type of adiabatic induced GWs production. For the adiabatic initial conditions, $\mathcal{P}_{\Phi}$ is related to the initial curvature power spectrum as 
\begin{eqnarray}
    \mathcal{P}_{\Phi_0}(k) = A_s \left(\frac{k}{k_{\rm *}}\right)^{n_s-1}  \Theta(k_{\rm UV}-k) \Theta(k-k_{\rm IR})\,,
    \label{eq:ScalarPS}
\end{eqnarray}
with $k_{\rm UV}$ and $k_{\rm IR}$ being the UV and IR cutoff scale, respectively. To avoid complications arising from non-linear evolution, the primordial power spectrum is typically truncated at a high-wavenumber UV cutoff, $k_{\rm UV}$, which excludes modes that become non-linear by the end of the matter-dominated era. Conversely, an infrared (IR) cutoff, $k_{\rm IR}$, is imposed to eliminate large-scale modes that could conflict with CMB observations. These cutoffs ensure that the analysis remains within the regime of linear perturbation theory and observational consistency. To compute the induced GWs from these adiabatic perturbations, we insert the scalar power spectrum $\mathcal{P}_{\Phi_0} (k)$ and the transfer function of the adiabatic mode $\Phi(k)$ into the convolution integral that determines the tensor power spectrum (Eq. \eqref{eq:av_tensorPS}). Unlike the isocurvature case, the adiabatic spectrum depends on both the spectral index $n_s$ and the equation-of-state (EoS) parameter $w_\phi$, which controls the transfer function. For modes that enter the horizon during the pre PBH domination (stiff or soft) era, the adiabatic transfer function can be approximated by an effective power-law:
\begin{equation}
\Phi_{\text{ad}}(k \gg k_{\text{BH}}) = A_\Phi(w_\phi) \left(\frac{k}{k_{\rm BH}}\right)^{n(w_\phi)}\,.
\end{equation}
The effective exponent $n(w_\phi)$ depends on $w_\phi$, given piecewise:
\begin{equation} 
\label{eq:nw}
    n(\w)\simeq
        \begin{cases}
        b/2 - 2 & \quad \w\lesssim 1/5\\
        -1.83 - 0.285 b + 0.790 b^2 & \quad 1/5<\w<2/3\\
        -(2+b) & \quad \w\gtrsim2/3
    \end{cases} \, ,
\end{equation}
with $b\equiv (1-3\w)/(1+3\w)$ and for $k\ll \kbh$, $n(\w)=0$. The amplitude $A_\Phi(w_\phi)$ is determined using a numerical fit given by the following expression
\begin{equation} 
\label{eq:a_phi}
    A_{\Phi}(\w)=
        \begin{cases}
        \frac{2}{3b} \left(7.76+18.3\,b+12.5\,b^2 \right) & \quad \w< 1/5 \text{ or, } \w>2/3\\
        15.6 +39.2\, b + 21.3\, b^2 & \quad 1/5<\w<2/3
    \end{cases} \, .
\end{equation}
Schematically, this parametrization gives rise to the following integrals over momentum
\begin{eqnarray} 
\label{eq:TensorPS_Adi_Gen}
    \overline{\mathcal{P}_{h,\rm RD}}(k,\eta,\bar{x}\gg1) \propto & \int_{0}^{\infty} dv \int_{|1-v|}^{1+v} du \left(4 v^2-\left(1+v^2-u^2\right)^2\right)^2 (u v)^{n_{\rm eff}(\w)} \, \overline{\mathcal{I}^2_{\rm osc}}(\bar{x},u,v)\, , \nonumber\\
\end{eqnarray}
with $\bar{x}\equiv k\bar{\eta}$ and $\mathcal{I}^2_{\rm osc}$ being the oscillatory kernel function, used to estimate the kernel function at the end of PBH-domination, as mentioned in Eq.~\eqref{eq:av_tensorPS}. Here the effective spectral index ($n_{\rm eff}$) is related to the scalar spectral index ($n_s$) and the background cosmology via $\w$. Combining this with the scalar power spectrum, the effective spectral index in the GW source integral becomes:~\cite{Domenech:2024wao}
\begin{eqnarray}
\label{eq:neff}
    n_{\rm eff}(\w) &=-\frac{5}{3} + n_s + 2 n(\w).
\end{eqnarray}
This governs the slope of the induced GW spectrum at large $k$, and is negative for $n_s \lesssim 1$.
For $k\gg \kbh$, we obtain the resonant contribution for the GW spectrum, which can be written as 
\begin{eqnarray}
    \label{eq:OmegaGW_res_adi}
    \Omega_{\rm GW, res}^{\rm ad}(k)
    =& \Omega_{\rm GW, res}^{\rm ad, peak}
    \left(\frac{k}{k_{\rm UV}}\right)^{2 n_{\rm eff}(\w)+7} \Theta^{\rm (ad)}_{\rm UV}(k) \,,
\end{eqnarray}
where at $k=k_{\rm UV}$, the amplitude is given by
\begin{eqnarray}
    \label{eq:OmegaGWres_Peak_adi}
   \Omega_{\rm GW, res}^{\rm ad, peak} \simeq 9.72\times 10^{32} \frac{ 3^{2 n(\w)+n_s}}{4^{3 n(\w)+n_s}} A_s^2 A_\Phi(\w)^4 \gamma ^{-4 n(\w)/3}  \beta ^{-\frac{2 n(\w) (\w +1)}{3 \w }}\left(\frac{g_H}{108}\right)^{-\frac{17}{9}} \left(\frac{\Min}{10^4\text{g}}\right)^{\frac{34}{9}}\,.\nonumber\\
\end{eqnarray}
The details of the cutoff function, $\Theta^{\rm (ad)}_{\rm UV} (k)$, is discussed in App.~\ref{app:theta_UV}. At the proximity of $\kbh$ and at $k\ll \kbh$, we can define the intermediate and IR part of the GW spectrum as follows:
\begin{align}    
    \Omega_{\rm GW, mid}^{\rm ad}(k)&=& -\frac{ A_s^2 A_\Phi(\w)^4 c_s^4}{768 (1+2n_{\rm eff})}\left(\frac{3}{2}\right)^{\frac{2}{3}} \xi_1^{1+2n_{\rm eff}}
    \left(\frac{\kbh}{k_{\rm UV}}\right)^{2 n_s-\frac{7}{3}}
    \left(\frac{k_{\rm UV}}{k_{\rm ev}}\right)^{\frac{14}{3}}
    \left(\frac{k}{k_{\rm UV}}\right)^{5},\label{eq:OmegaGW_mid_adi}\\
    \Omega_{\rm GW, IR}^{\rm ad}(k)&=&\frac{ A_s^2 c_s^4 (3 \w +5)^4 }{10^4  (6 n_s+5) (\w +1)^4}
    \left(\frac{3}{2}\right)^{\frac{2}{3}}
    \left(\frac{\xi_2 \kbh}{k_{\rm UV}}\right)^{\frac{5}{3} + 2n_s}
    \left(\frac{k_{\rm UV}}{k_{\rm ev}}\right)^{\frac{14}{3}}
    \left(\frac{k}{k_{\rm UV}}\right) \label{eq:OmegaGW_IR_adi}.
\end{align}
Here, both $\xi_1$ and $\xi_2$ are the functions of $\w$. However, roughly $\xi_1$ is a $\mathcal{O}(1)$ parameter~\cite{Domenech:2024wao}. $\xi_2$ can be determined as follows:
\begin{equation} 
\label{eq:xi2}
    \xi_2 = 
        \begin{cases}
        \left(\frac{7.76+18.3b+12.5b^2}{(3\w+5)/(5\w+5)}\right)^{(1+3\w)/(3+3\w)} & \quad \w> 1/3\\
        \left(\frac{7.76+18.3b+12.5b^2}{(3\w+5)/(5\w+5)}\right)^{(2+6\w)/(3+15\w)} & \quad \w< 1/3
    \end{cases} \, .
\end{equation}
Now, the complete spectrum for the adiabatic contribution of the GW spectra ($\Omega_{\rm GW, ad}(k)$), is simply the sum of the three parts. At the present epoch, the adiabatic contribution of the GW spectra ($ \Omega_{\rm GW, ad}^{(0)}(f) h^2$) can be estimated using Eq.~\eqref{eq:GW_present}. 

\begin{figure*}[!ht]
    \centering
    \begin{subfigure}{.32\textwidth}
    \includegraphics[width=\textwidth]{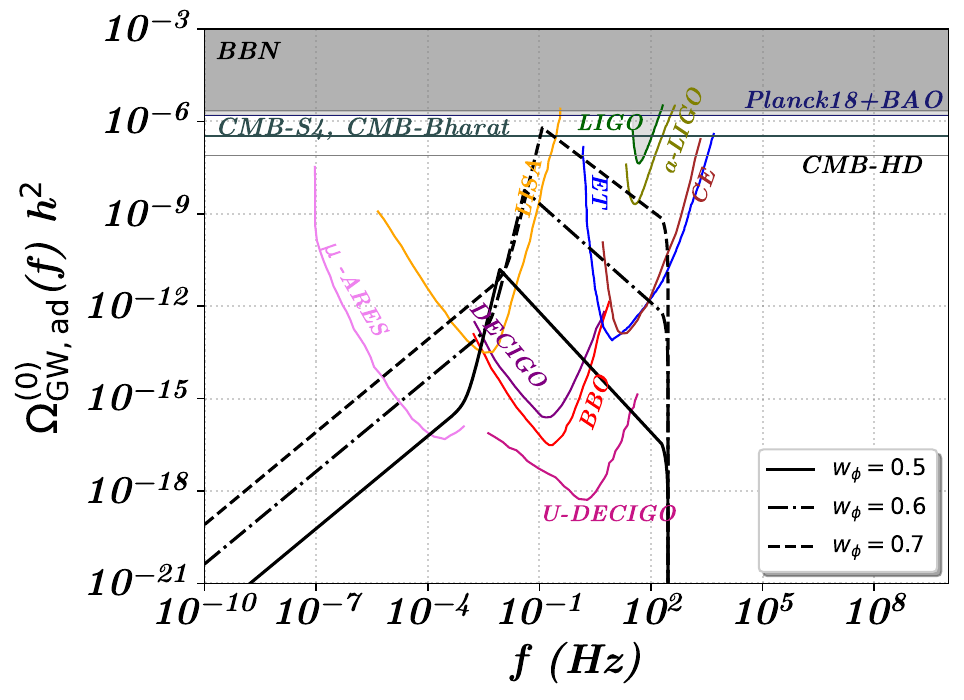}
    \caption{\it $\beta=10^{-10}$, $M_{\rm in}=10^5$ g}
    \label{fig:GW_sudden_w_fixed}
    \end{subfigure}
    \hfill
    \begin{subfigure}{.32\textwidth}
    \includegraphics[width=\textwidth]{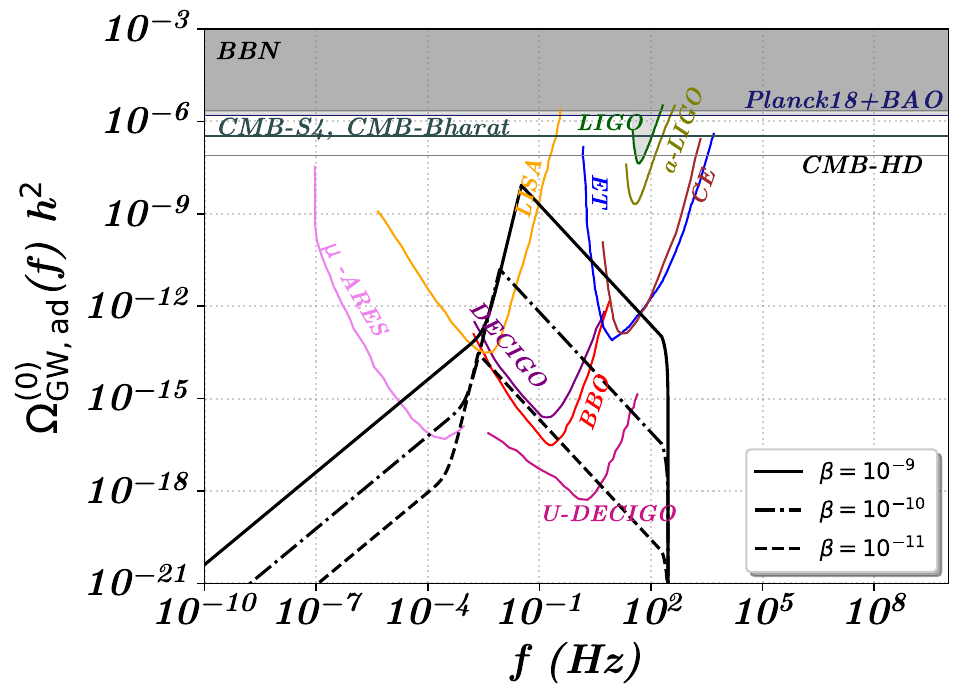}
    \caption{\it $\w=0.5$, $M_{\rm in}=10^5$ g}
    \label{fig:GW_sudden_beta_fixed}   
    \end{subfigure}
    \hfill
    \begin{subfigure}{.32\textwidth}
    \includegraphics[width=\textwidth]{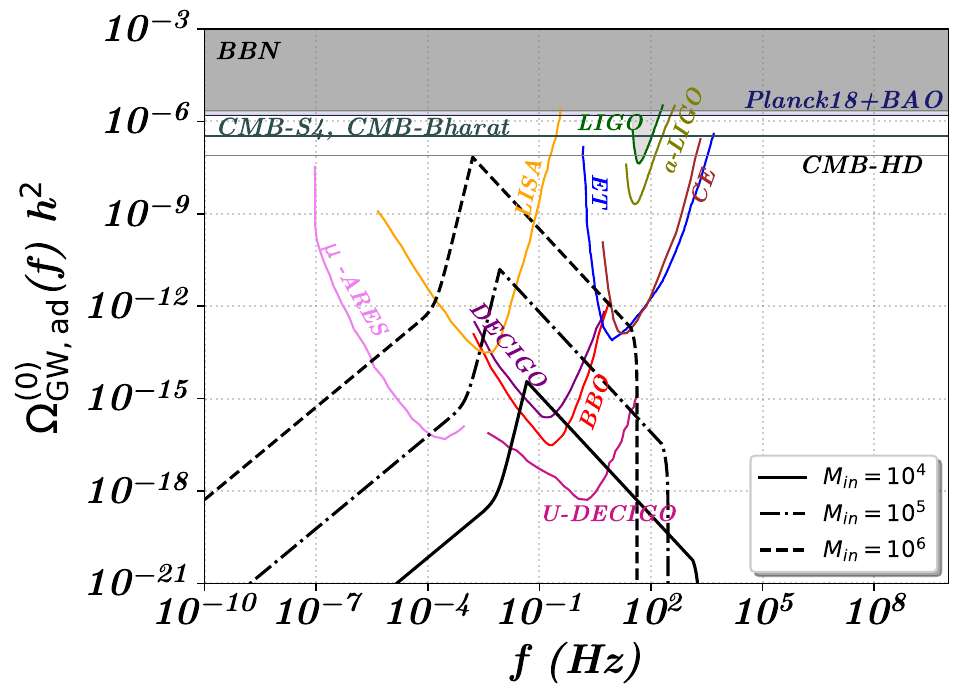}
    \caption{\it $\w=0.5$, $\beta=10^{-10}$}
    \label{fig:GW_sudden_mass_fixed}   
    \end{subfigure}
    \caption{\it Induced gravitational wave spectrum at the present epoch due to \textbf{adiabatic fluctuation}.  Each plot represents the variation of a single parameter while the other two are kept fixed.}
    \label{fig:GW_sudden_PBH}
\end{figure*}
Fig.~\ref{fig:GW_sudden_PBH} illustrates the impact of each of the three parameters on the GWs spectrum while keeping the others fixed, similar to our analysis for isocurvature fluctuations. Compared to the isocurvature case, the resonant part of the spectrum is broader and shifted to lower frequencies, as its peak is roughly associated with $ k_{\rm BH}$. In contrast, the isocurvature peak lies at $k_{\rm UV}$, closer to the PBH formation wavenumber $k_{\rm in}$. From Fig.~\ref{fig:GW_sudden_PBH}, it is evident that increasing $\w$ or $\beta$ shifts the peak frequency to the right, \textit{i.e.}, toward higher values. This can be understood as follows: a stiff EoS $\w$ causes a greater redshift of the background relative to PBHs, which behave like dust. As a result, PBH-domination occurs earlier, pushing $k_{\rm BH}$ closer to the formation wave number $ k_{\rm in}$. Since the peak of the spectrum lies around $k_{\rm BH}$, an increase in $\w$ naturally leads to a shift in the peak frequency towards higher frequency. A similar effect occurs for higher $\beta$, as a larger $\beta$ facilitates earlier PBH-domination. Consequently, in both cases, the peak frequency shifts to higher values.

The behaviour of the GW spectra across different region (resonant part, intermediate part, IR part) can be understood using Eqs.~\eqref{eq:OmegaGW_res_adi}, \eqref{eq:OmegaGW_mid_adi} and \eqref{eq:OmegaGW_IR_adi}. At the resonant part of the GW spectra, the spectral dependence follows $k^{2n_{\rm eff}(\w)+7}$, where $n_{\rm eff}(\w)$ has a dependence on $n_s$, in contrast to the isocurvature fluctuation. For $n_s\lesssim 1$, the exponent $(2n_{\rm eff}(\w)+7)$ becomes negative, explaining the negative slope of the GW spectrum at this region. Furthermore, GW spectra, at the intermediate and IR regimes, scales as $k^5$ and $k$, respectively. This transition is evident in Fig.~\ref{fig:GW_sudden_PBH}, where the spectrum exhibits a sharp variation with $f$ at the intermediate part, compared to IR part.

\subsubsection{Statistical analysis in the light of the detectors}
\label{subsubsec:parameter_estimation_induced}
Similar to the isocurvature scenario, in this section, we proceed with the statistical analysis of the spectrum for ET and LISA, as they appear more promising for detecting the adiabatic spectrum, as evident from Fig.~\ref{fig:GW_sudden_PBH}. LISA shows comparatively better prospects for detecting the adiabatic than the isocurvature source, at least based on what is observed in the Fig.~\ref{fig:GW_sudden_PBH}.
\subsubsection*{Evaluation of signal-to-noise ratio:}
\begin{figure*}[!ht]
    \centering
    \begin{subfigure}{.47\textwidth}
    \includegraphics[width=\textwidth]{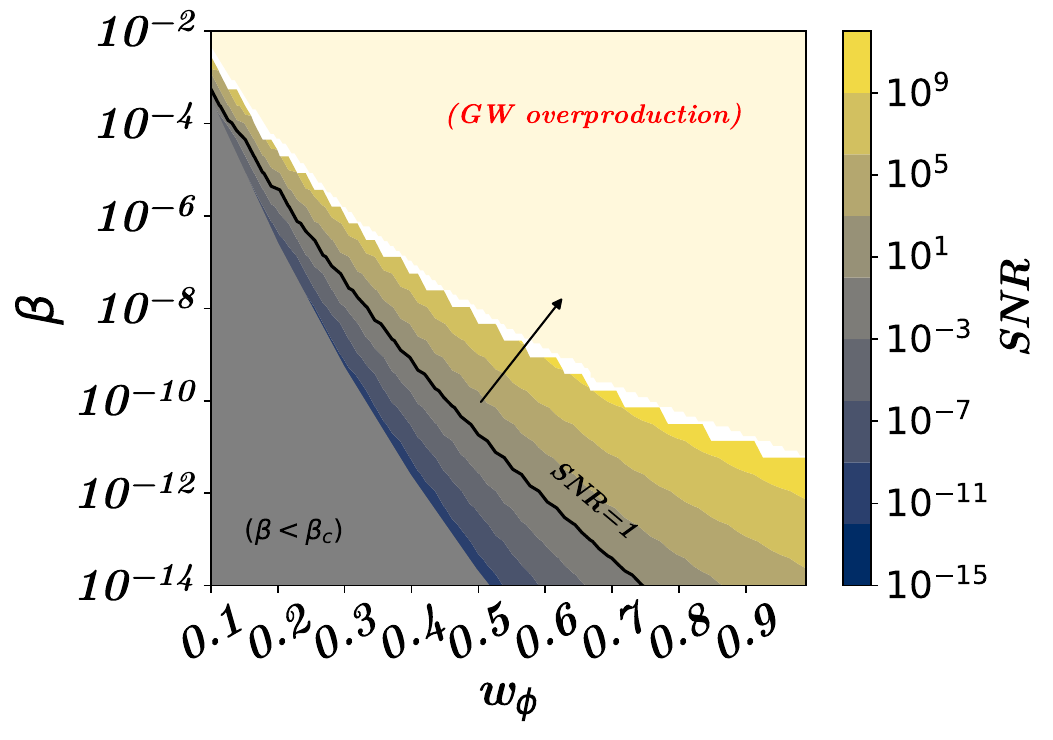}
    \caption{\it $M_{\rm in}=10^5$ g}
    \label{fig:SNR_sudden_ET_m5}
    \end{subfigure}
    \hfill
    \begin{subfigure}{.47\textwidth}
    \includegraphics[width=\textwidth]{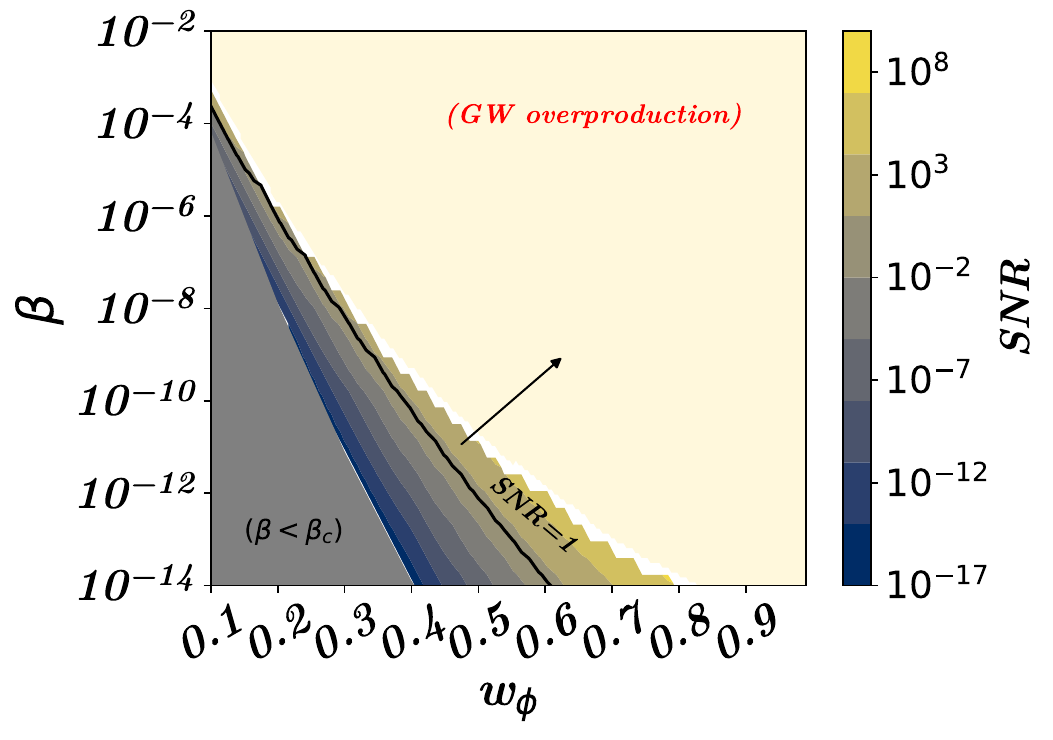}
    \caption{\it $M_{\rm in}=10^7$ g}
    \label{fig:SNR_sudden_ET_m7}   
    \end{subfigure}
    \caption{\it SNR plots for \textbf{ET} with two different values of $\Min$ considering \textbf{adiabatic fluctuation}. Rest of the demonstration is same as Fig.~\ref{fig:SNR_iso_ET}.}
    \label{fig:SNR_sudden_ET}
\end{figure*}

\begin{figure*}[!ht]
    \centering
    \begin{subfigure}{.47\textwidth}
    \includegraphics[width=\textwidth]{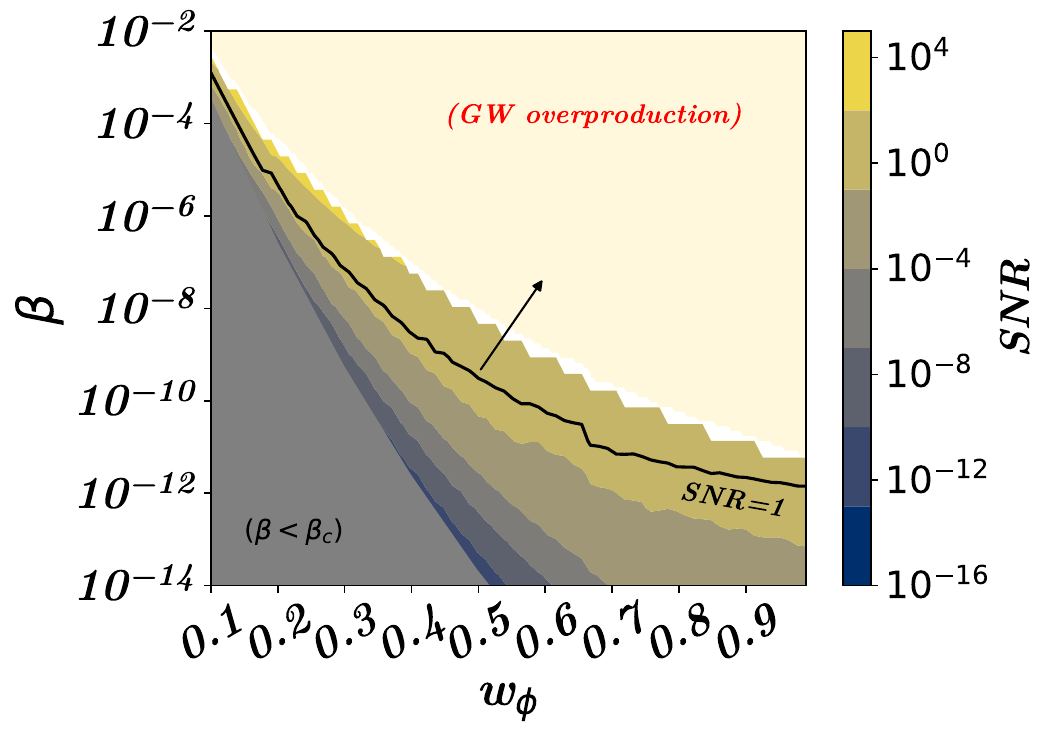}
    \caption{\it $M_{\rm in}=10^5$ g}
    \label{fig:SNR_sudden_LISA_m5}
    \end{subfigure}
    \hfill
    \begin{subfigure}{.47\textwidth}
    \includegraphics[width=\textwidth]{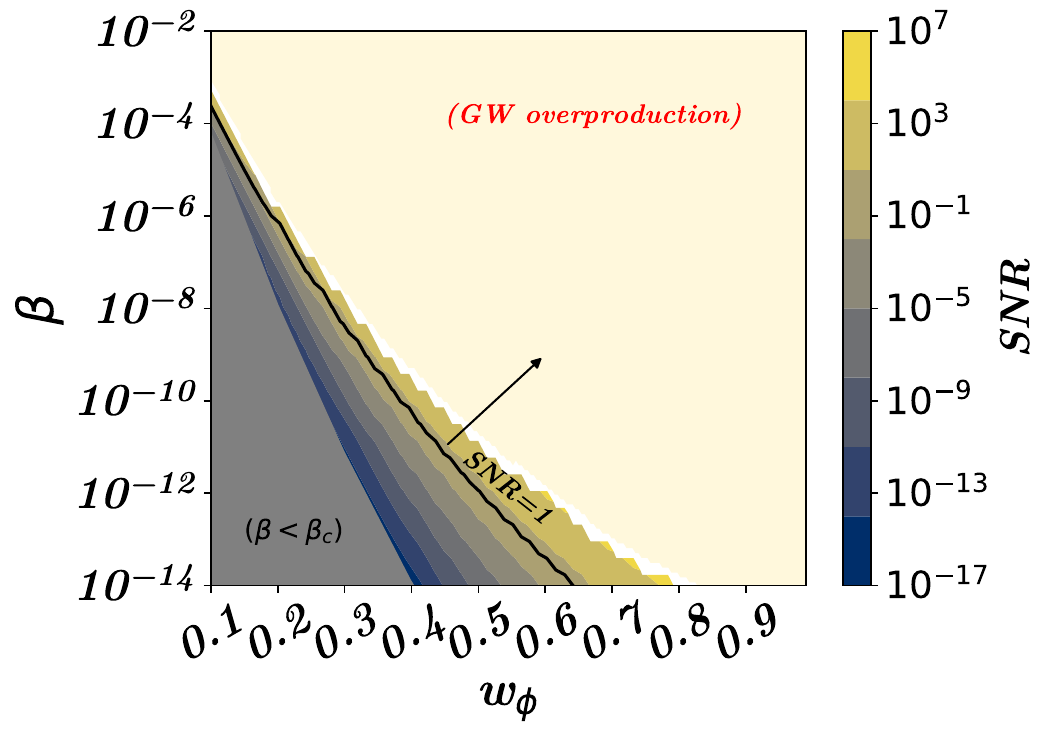}
    \caption{\it $M_{\rm in}=10^7$ g}
    \label{fig:SNR_sudden_LISA_m7}   
    \end{subfigure}
    \caption{\it Figure is same as Fig.~\ref{fig:SNR_iso_ET}, but for \textbf{LISA}, considering \textbf{adiabatic fluctuation}.}
    \label{fig:SNR_sudden_LISA}
\end{figure*}

Figs.~\ref{fig:SNR_sudden_ET} and \ref{fig:SNR_sudden_LISA} present the SNR plots for ET and LISA, respectively, for PBHs with masses of $10^5$ g and $10^7$ g. The plots follow the same structure as those for the isocurvature source. As in the isocurvature case, the SNR increases with $ \beta$ for a fixed $\w$ and vice versa, as evident from Eq.~\eqref{eq:OmegaGWres_Peak_adi}. This trend arises because a higher $\beta$ or $\w$ extends the PBH-dominated era, thereby increasing the GW spectral density, which directly depends on the duration of PBH domination.

\begin{figure*}[!ht]
    \centering
    \begin{subfigure}{.49\textwidth}
    \includegraphics[width=\textwidth]{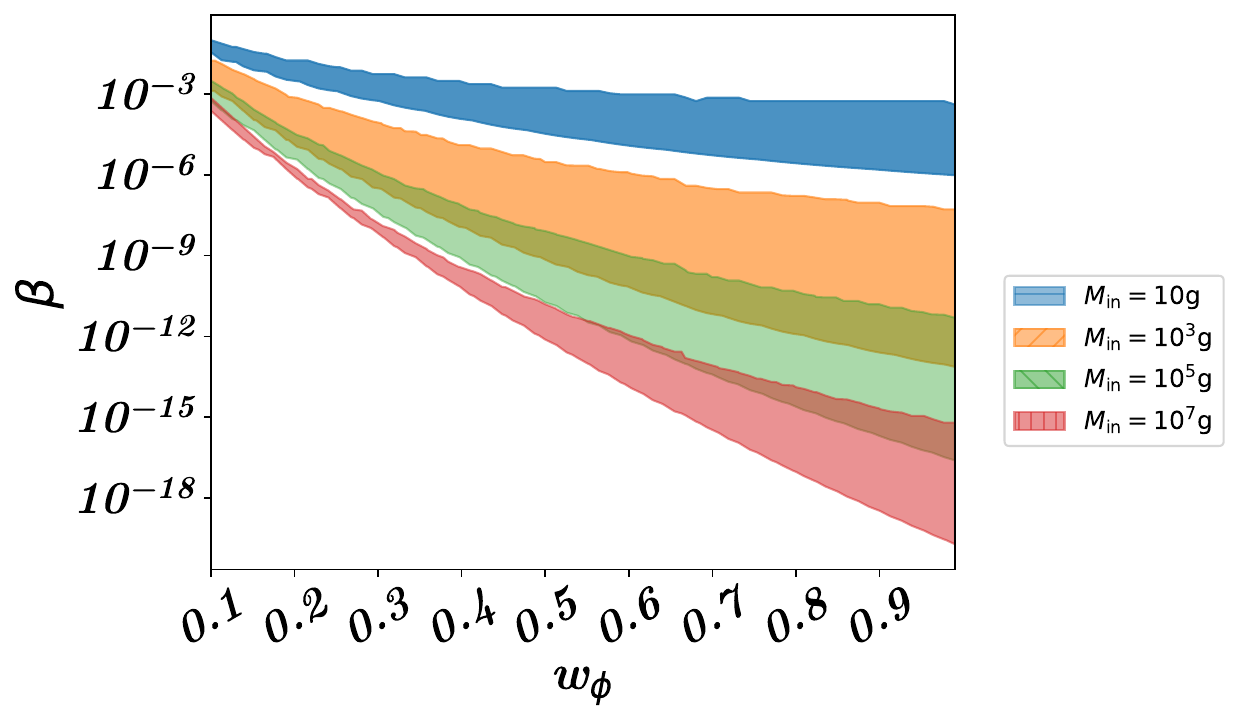}
    \caption{\it ET}
    \label{fig:SNR_sudden_all_masses_ET}
    \end{subfigure}
    \hfill
    \begin{subfigure}{.49\textwidth}
    \includegraphics[width=\textwidth]{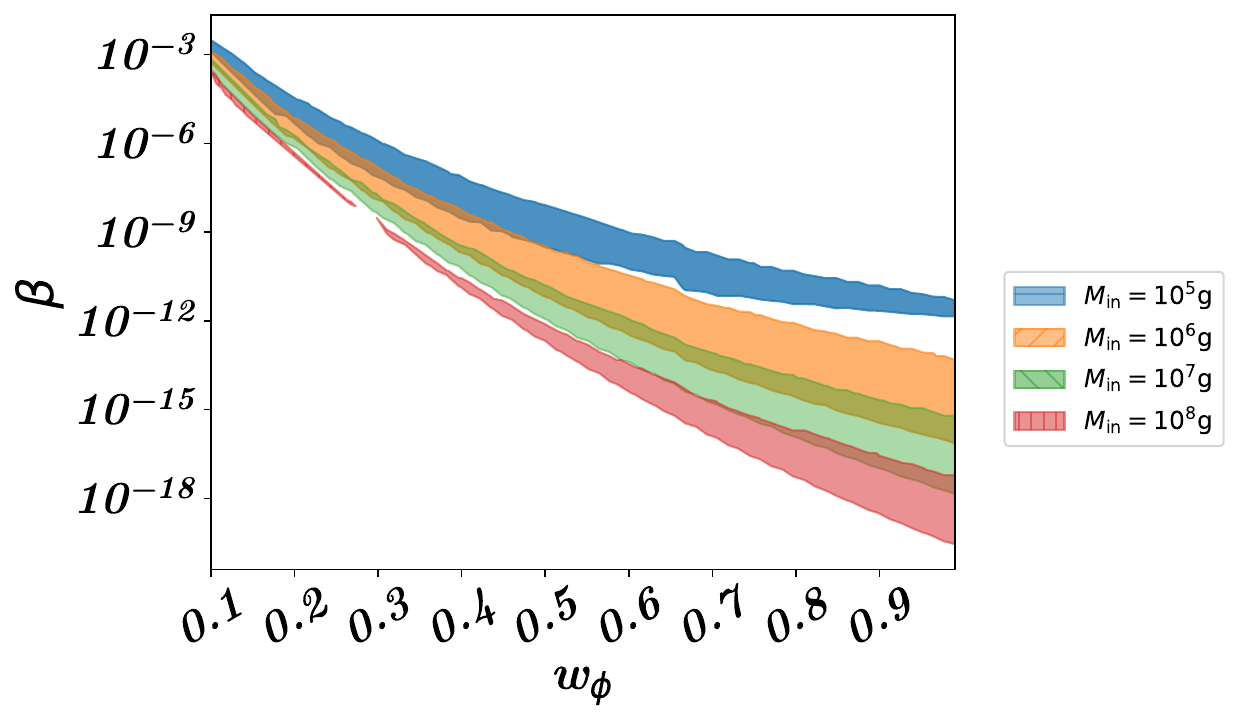}
    \caption{\it LISA}
    \label{fig:SNR_sudden_all_masses_LISA}   
    \end{subfigure}
    \caption{\it Illustration of extent of parameter space for various $\Min$ where ${\rm SNR}\geq1$, considering GWs due to \textbf{adiabatic fluctuation}, for ET (\textbf{left}) and LISA (\textbf{right}). The upper limit comes from the GW overproduction, derived from $\Delta N_{\rm eff}$ constraints at BBN.}
    \label{fig:SNR_sudden_all_masses}
\end{figure*}

The SNR analysis suggests that both ET and LISA show promise in detecting the adiabatic source, unlike the isocurvature one, which is only sensitive to ET. For ET (see, for instance, Fig.~\ref{fig:SNR_sudden_all_masses_ET}), similar to the isocurvature source, we observe that the parameter span in the $\w - \beta$ plane is considerably large for formation masses in the range $\Min \sim (10^3 - 10^7)$ g. However, it starts to shrink drastically for PBH masses above $10^7$ g, with no available parameter space for high $\rm SNR > 1$ for PBH masses above $4 \times 10^7$ g.  

On the other hand, for LISA (see, for instance, Fig.~\ref{fig:SNR_sudden_all_masses_LISA} ), the parameter span starts from $\Min \sim 2 \times 10^4$ g. For PBH masses below this threshold, the SNR remains $<1$ for all possible ranges of $\w$ and $\beta$. Furthermore, for LISA, the formation mass consistent with $\rm SNR > 1$ lies in the range $\Min \sim (2 \times 10^4 - 5 \times 10^8)$ g, where the upper limit corresponds to the maximum mass value that evaporated just before BBN.

Thus, in terms of SNR, LISA is expected to be a promising mission for detecting induced GW spectra arising from adiabatic fluctuations, in contrast to the isocurvature scenario. However, for ET, we find that both sources are compatible with achieving high SNR values.

\subsubsection*{Fisher matrix analysis:}
\begin{figure*}[!ht]
    \centering
    \begin{subfigure}{.49\textwidth}
    \includegraphics[width=\textwidth]{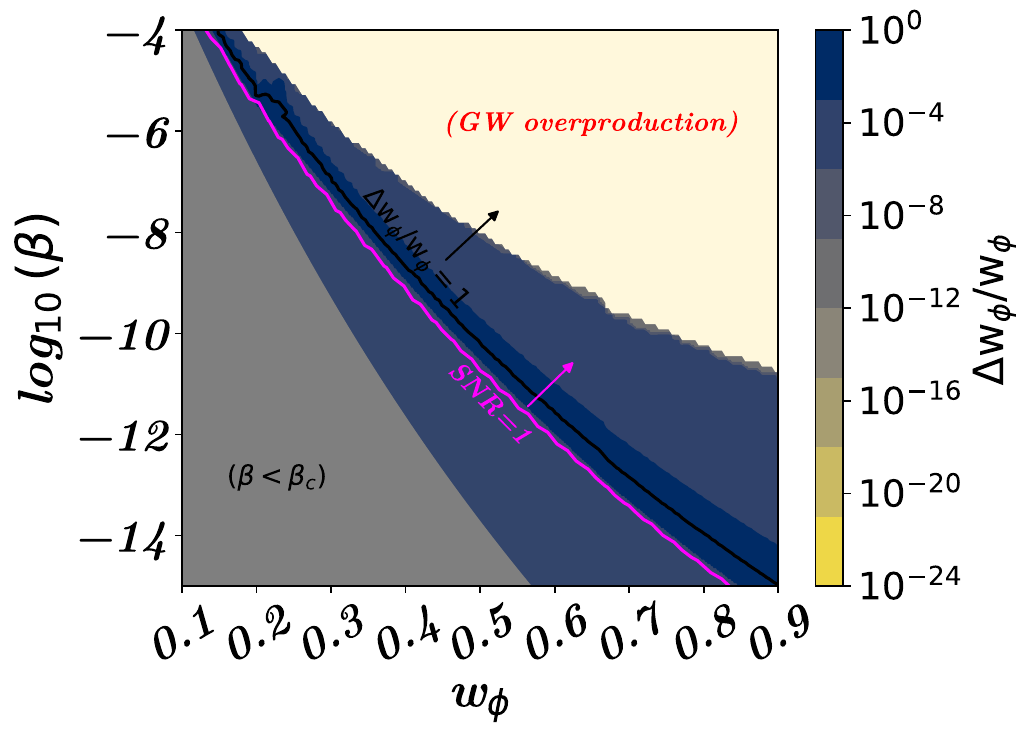}
    \caption{\it $\Min=10^5$ g}
    \label{fig:fisher_sudden_ET_deltaw_m5}
    \end{subfigure}
    \begin{subfigure}{.49\textwidth}
    \includegraphics[width=\textwidth]{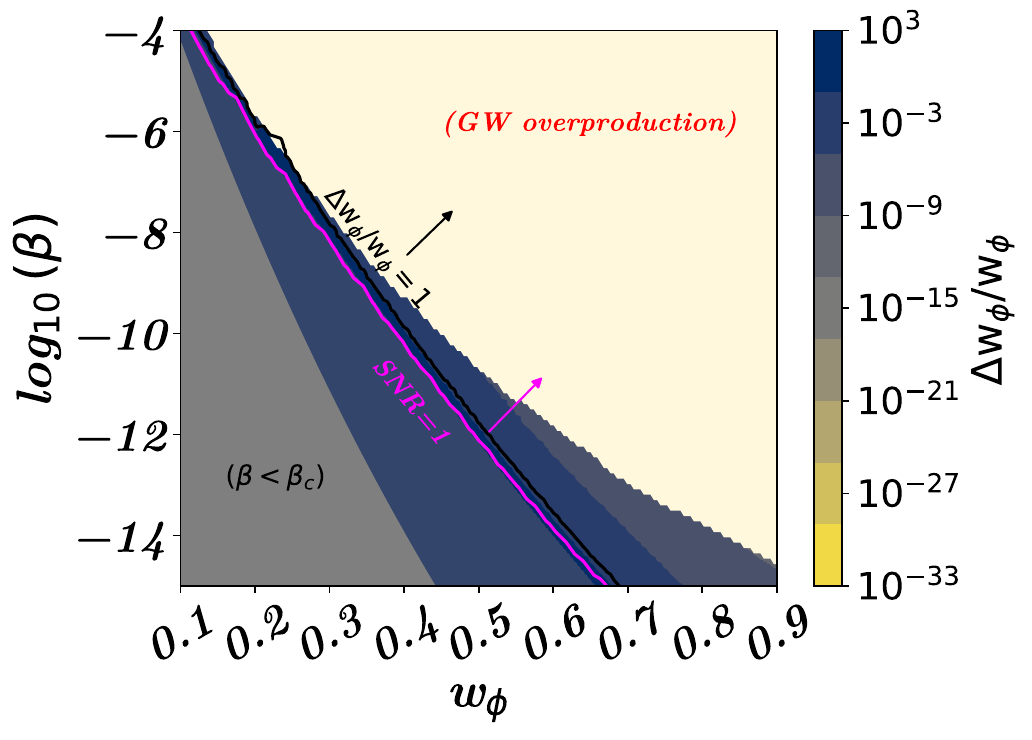}
    \caption{\it $\Min=10^7$ g}
    \label{fig:fisher_sudden_ET_deltaw_m7}   
    \end{subfigure}
    \caption{\it Illustration of relative uncertainties on $\w$, calculated by Fisher analysis, for the induced GWs from \textbf{adiabatic fluctuation} while considering \textbf{ET}. Black (magenta) solid line indicates ${\rm SNR}=1$ ($\Delta \w/\w=1$). The black (magenta) arrow indicates where ${\rm SNR}>1$ ($\Delta \w/\w<1$).}
    \label{fig:fisher_sudden_ET_deltaw}
\end{figure*}

\begin{figure*}[!ht]
    \centering
    \begin{subfigure}{.49\textwidth}
    \includegraphics[width=\textwidth]{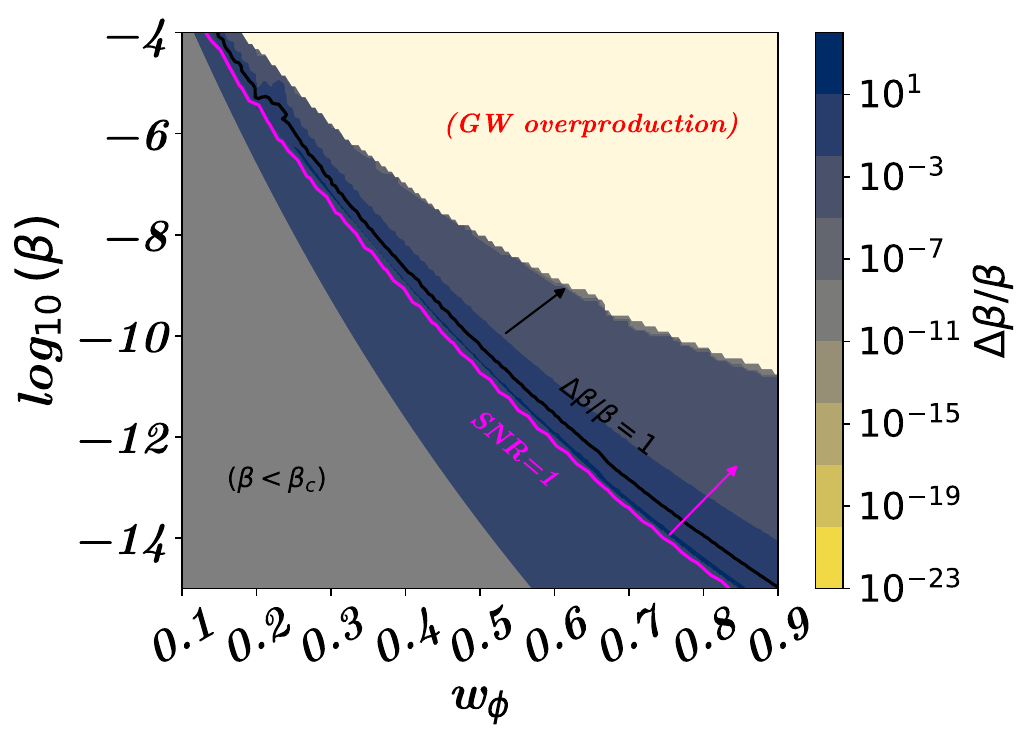}
    \caption{\it $\Min=10^5$ g}
    \label{fig:fisher_sudden_ET_deltabeta_m5}
    \end{subfigure}
    \begin{subfigure}{.49\textwidth}
    \includegraphics[width=\textwidth]{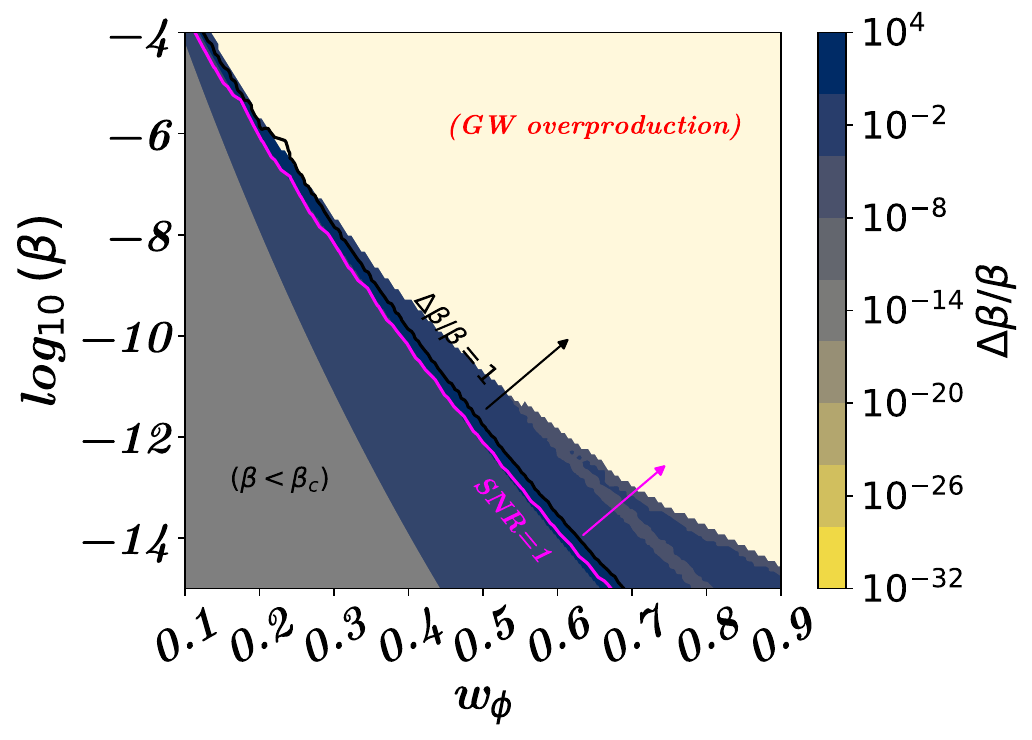}
    \caption{\it $\Min=10^7$ g}
    \label{fig:fisher_sudden_ET_deltabeta_m7}   
    \end{subfigure}
    \caption{\it Representation is same as Fig.~\ref{fig:fisher_sudden_ET_deltaw}. However, it is the illustration of relative uncertainties on $\beta$, for \textbf{ET}.}
    \label{fig:fisher_sudden_ET_deltabeta}
\end{figure*}

\begin{figure*}[!ht]
    \centering
    \begin{subfigure}{.49\textwidth}
    \includegraphics[width=\textwidth]{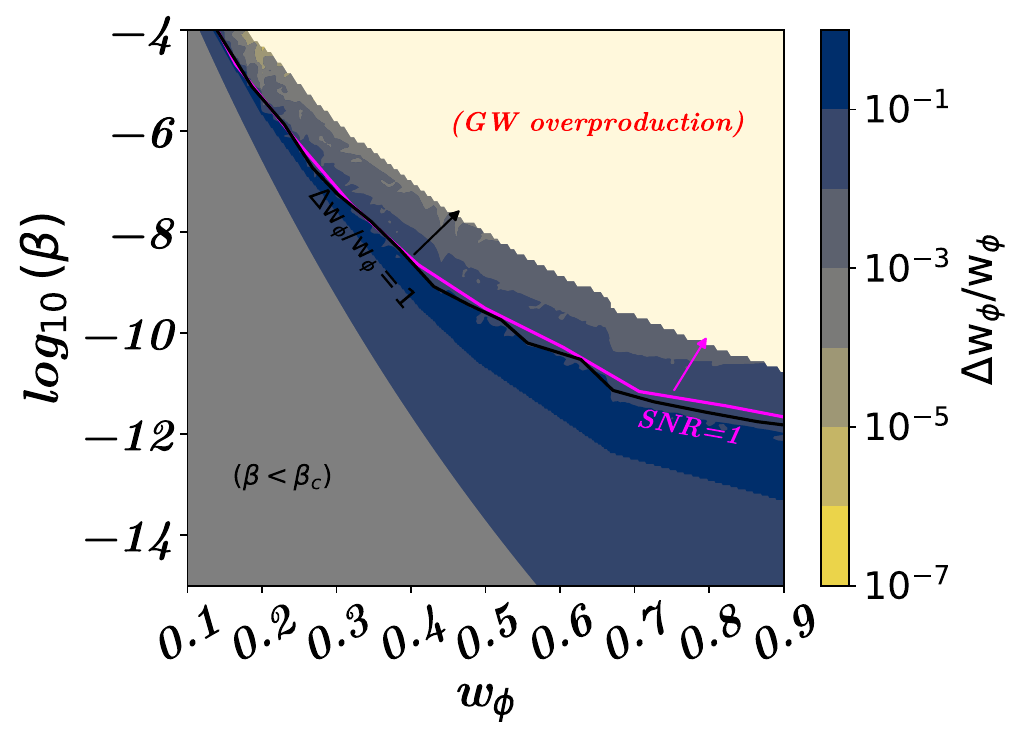}
    \caption{\it $\Min=10^5$ g}
    \label{fig:fisher_sudden_LISA_deltaw_m5}
    \end{subfigure}
    \begin{subfigure}{.49\textwidth}
    \includegraphics[width=\textwidth]{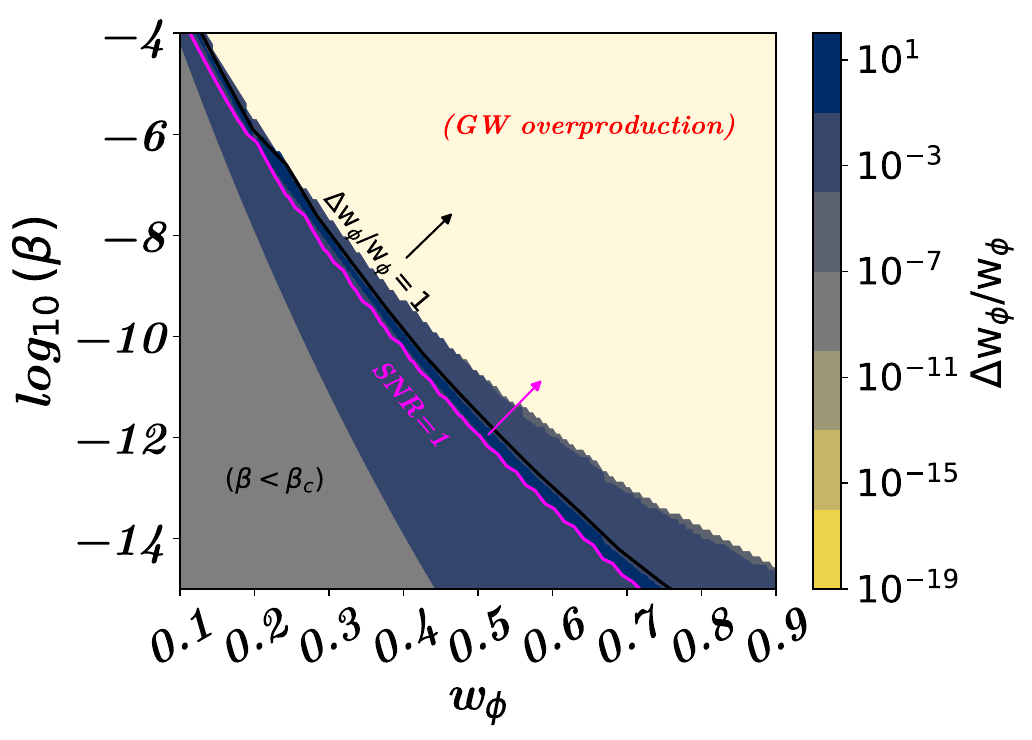}
    \caption{\it $\Min=10^7$ g}
    \label{fig:fisher_sudden_LISA_deltaw_m7}   
    \end{subfigure}
    \caption{\it Figure is same as Fig.~\ref{fig:fisher_sudden_ET_deltaw}, but for \textbf{LISA}.}
    \label{fig:fisher_sudden_LISA_deltaw}
\end{figure*}

\begin{figure*}[!ht]
    \centering
    \begin{subfigure}{.49\textwidth}
    \includegraphics[width=\textwidth]{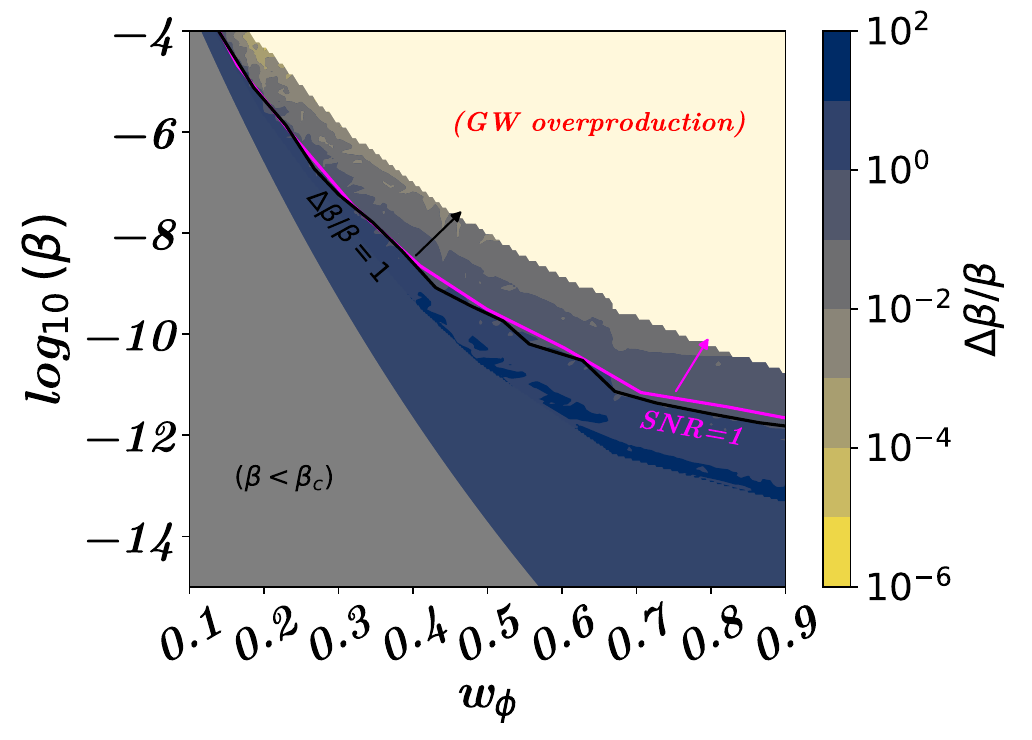}
    \caption{\it $\Min=10^5$ g}
    \label{fig:fisher_sudden_LISA_deltabeta_m5}
    \end{subfigure}
    \begin{subfigure}{.49\textwidth}
    \includegraphics[width=\textwidth]{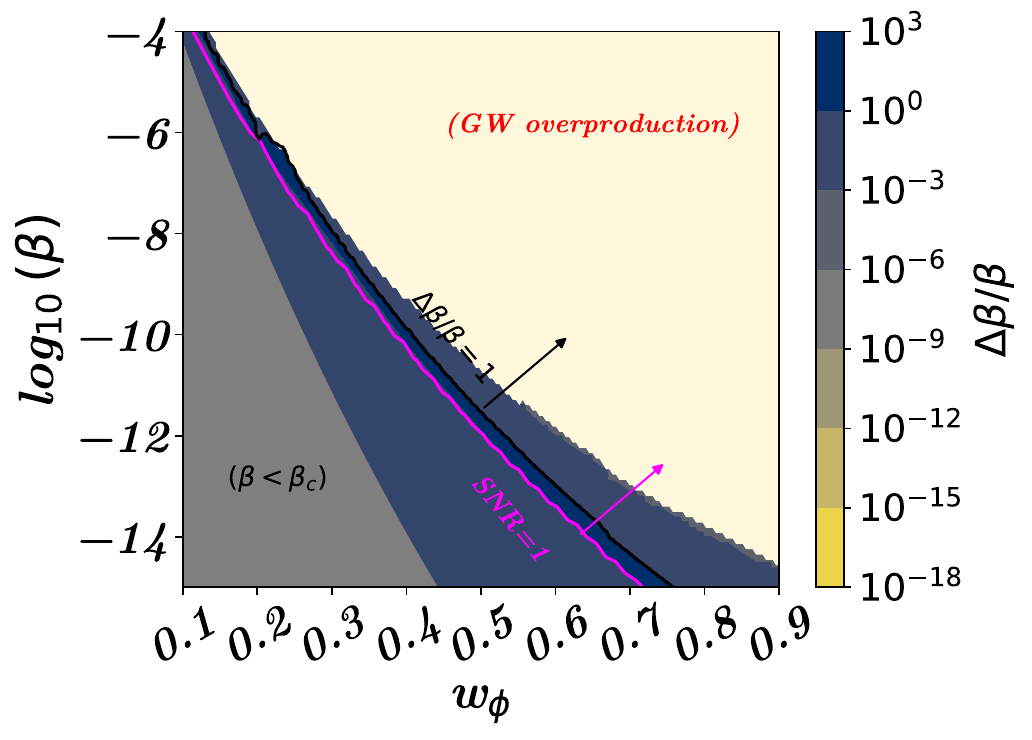}
    \caption{\it $\Min=10^7$ g}
    \label{fig:fisher_sudden_LISA_deltabeta_m7}   
    \end{subfigure}
    \caption{\it Figure is same as Fig.~\ref{fig:fisher_sudden_ET_deltaw}, but for relative uncertainties on $\beta$, considering \textbf{LISA}.}
    \label{fig:fisher_sudden_LISA_deltabeta}
\end{figure*}

In Figs.~\ref{fig:fisher_sudden_ET_deltaw} and \ref{fig:fisher_sudden_ET_deltabeta}, we present the relative uncertainties in $\w$ and $\beta$, respectively, for ET, considering $\Min = 10^5$ g and $10^7$ g. Similarly, Figs.~\ref{fig:fisher_sudden_LISA_deltaw} and \ref{fig:fisher_sudden_LISA_deltabeta} illustrate the same analysis for LISA. 
A key finding of this Fisher analysis is that, for the adiabatic source, the relative uncertainties in $\w$ and $\beta$, given by $\Delta \w/\w$ and $\Delta\beta/\beta$, are significantly smaller than $1$. In contrast, for the isocurvature source, achieving $\Delta \w/\w$ or $\Delta \beta/\beta < 1$ is not feasible, which makes the detection of isocurvature source with higher uncertainty. 

Notably, for both ET and LISA, in the high SNR region (\textit{i.e.} SNR $>1$), we find consistently low uncertainties (\textit{i.e.} $\Delta \w/\w, \Delta\beta/\beta < 1$), reinforcing the idea that induced GWs from the adiabatic source can be measured with high precision. Thus, in agreement with the SNR analysis, the Fisher analysis further confirms that both LISA and ET are well-suited for detecting induced GWs from the adiabatic source with minimal uncertainties.

\section{Combined analysis with primordial source and PBH domination}
\label{sec:full_spec}
The two sources of induced GWs are independent and contribute to the spectrum of the induced GWs simultaneously in the PBH reheating scenario. Consequently, the most realistic approach is to assess the detectability of the combined spectrum at the detectors. The resulting present-day energy density spectrum of the induced GWs \footnote{Note that our analysis is carried out within the framework of linear cosmological perturbation theory.} can be expressed as  

\begin{eqnarray}  
\label{eq:GW_combined}  
\Omega_{\rm GW,com}^{(0)}(f) h^2 = \Omega_{\rm GW,iso}^{(0)} (f) h^2 + \Omega_{\rm GW,ad}^{(0)} (f) h^2.  
\end{eqnarray}  

This linear superposition arises from the statistical independence of the two sources, meaning their contributions add up without interference. Evaluating the detectability of $\Omega_{\rm GW,com}^{(0)}(f)$ provides a more comprehensive understanding of the observational prospects for early Universe physics, particularly the PBH reheating scenario, in the light of future GW detectors.
\begin{figure*}[!ht]
    \centering
    \begin{subfigure}{.32\textwidth}
    \includegraphics[width=\textwidth]{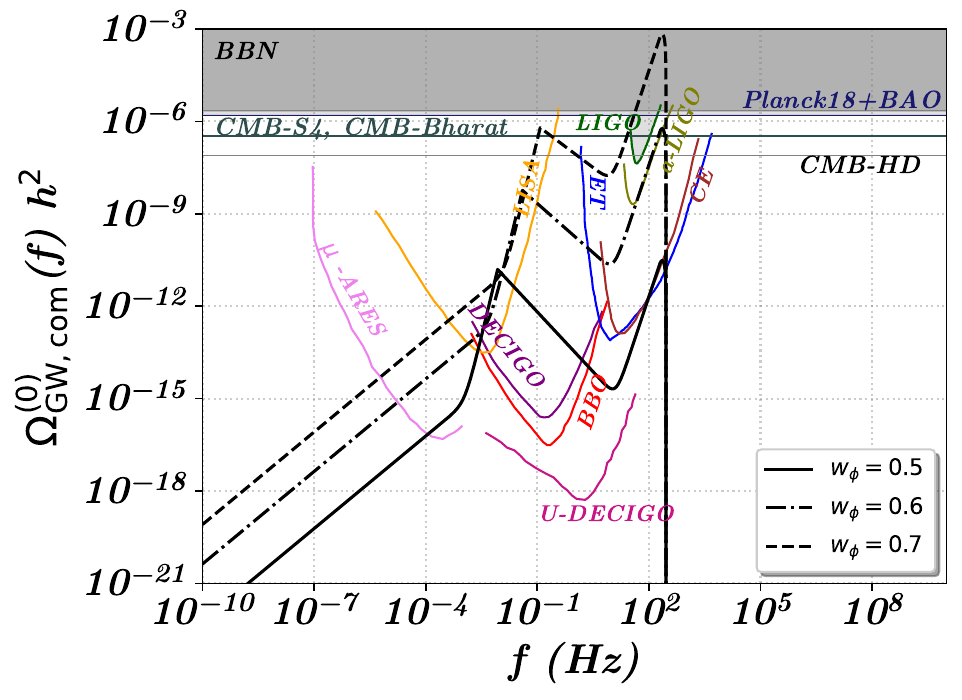}
    \caption{\it $\beta=10^{-10}$ and $\Min = 10^5$ g}
    \label{fig:GWfull_wvarry}
    \end{subfigure}
    \hfill
    \begin{subfigure}{.32\textwidth}
    \includegraphics[width=\textwidth]{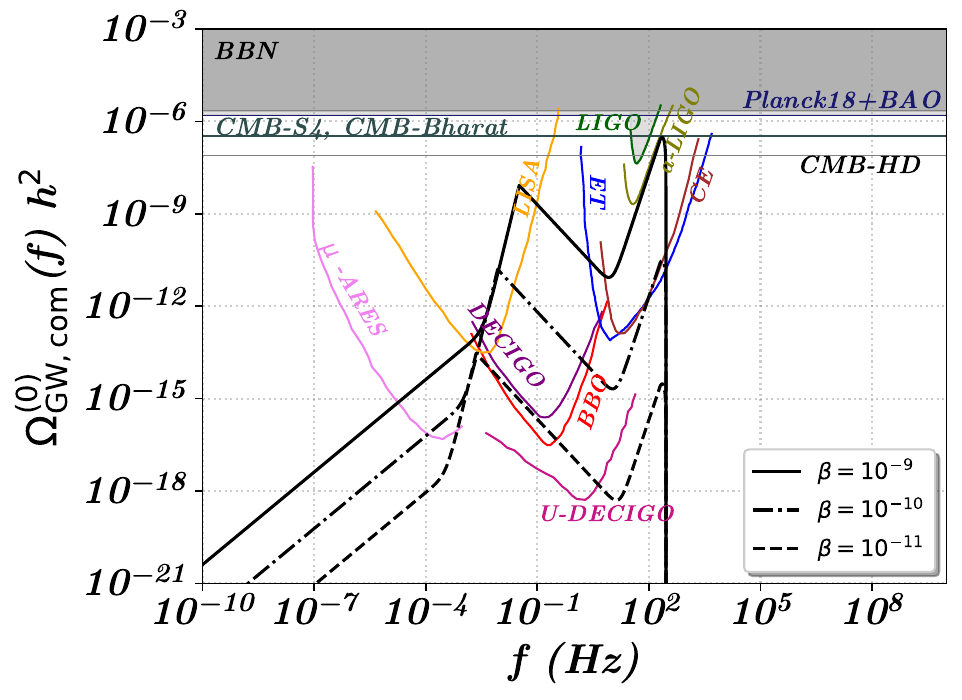}
    \caption{\it $\w=0.5$ and $\Min = 10^5$ g}
    \label{fig:GWfull_betavarry}   
    \end{subfigure}
    \hfill
    \begin{subfigure}{.32\textwidth}
    \includegraphics[width=\textwidth]{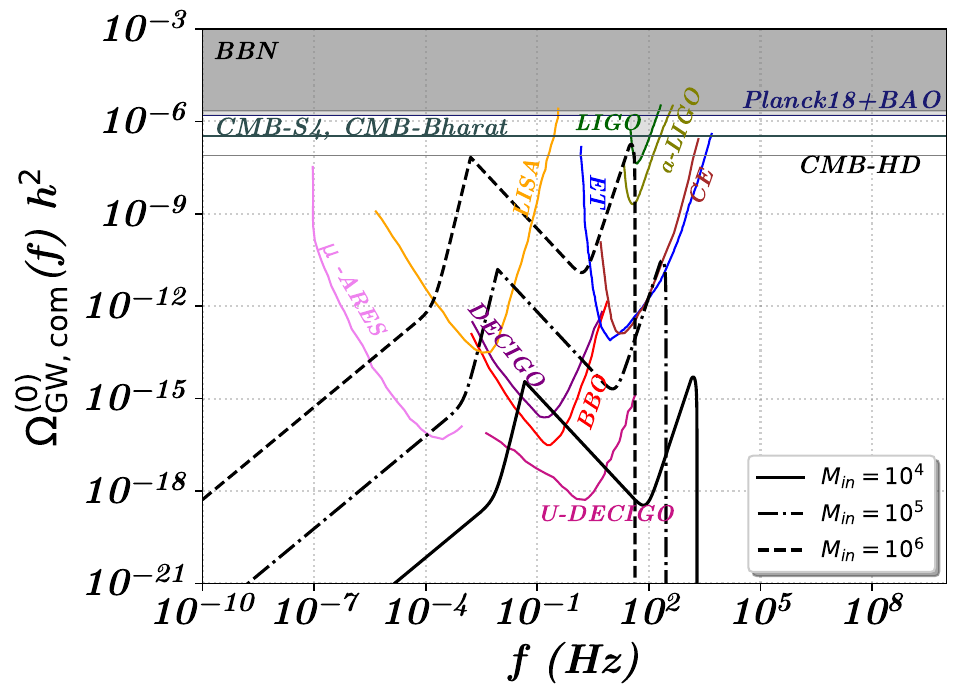}
    \caption{\it $\w=0.5$ and $\beta = 10^{-10}$}
    \label{fig:GWfull_mvarry}   
    \end{subfigure}
    \caption{\it Induced gravitational wave spectrum at the present considering \textbf{combined sources}. Each of the plots are the presentation for the variation of the three parameters while the other two are kept fixed.}
    \label{fig:GWfull}
\end{figure*}

The combined spectrum is illustrated in Fig.~\ref{fig:GWfull}, highlighting the distinct contributions of the two sources. The induced GWs spectrum for adiabatic fluctuations peaks around $k_{\rm UV}$ (Fig.~\ref{fig:GW_sudden_PBH}), whereas for the isocurvature scenario, it peaks at  $k_{\rm BH}$ (Fig.~\ref{fig:GWiso}). This results in two distinct peaks at characteristic frequencies, as shown in Fig.~\ref{fig:GWfull}, enhancing the overall prospects for detection across a broader frequency range. This complimentary coverage may increases the likelihood that the combined spectrum could be probed by future GW observatories.
\subsection{Statistical analysis in the light of the detectors}
\label{subsubsec:parameter_estimation_combined}
Since both the sources exhibit interesting prospects on their own merits, a combined statistical analysis of the combined spectrum is inevitable. As an example, for one source, the induced GWs may fall within the detection range, while for the other, an overproduction of GWs could make the scenario incompatible—an issue we will explore in detail in the next section. Therefore, a rigorous statistical analysis is essential for accurately estimating the parameters, which is presented in this section.
\subsubsection*{Evaluation of signal-to-noise ratio:}
\begin{figure*}[!ht]
    \centering
    \begin{subfigure}{.47\textwidth}
    \includegraphics[width=\textwidth]{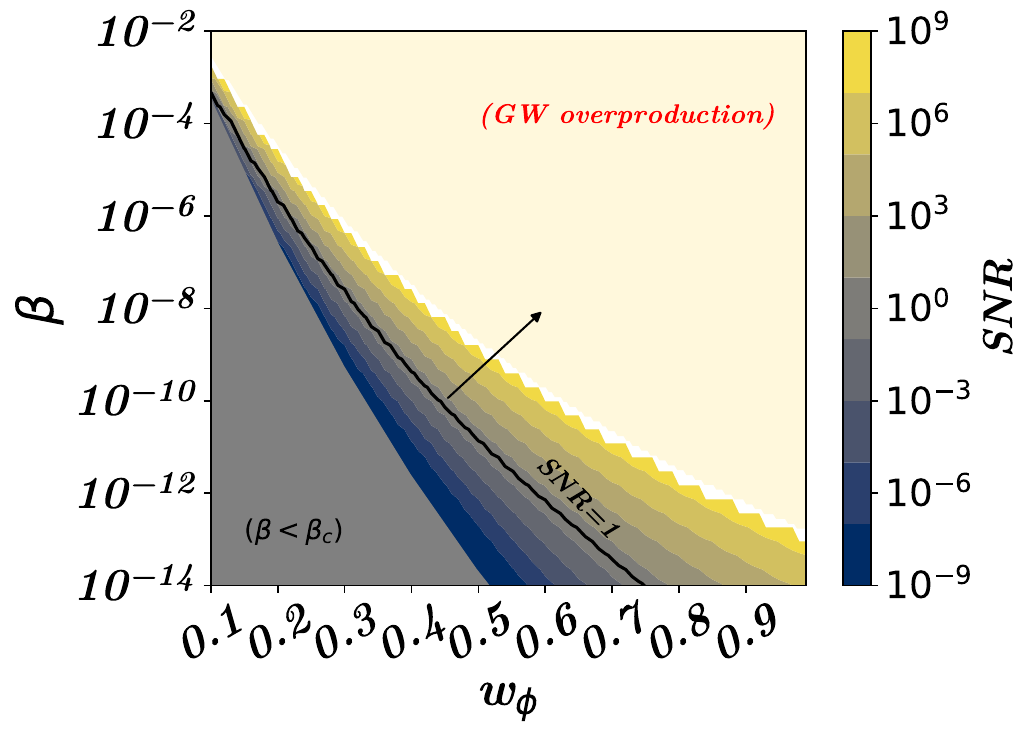}
    \caption{\it $\Min=10^5$ g}
    \label{fig:SNR_ET_combined_m5}
    \end{subfigure}
    \hfill
    \begin{subfigure}{.47\textwidth}
    \includegraphics[width=\textwidth]{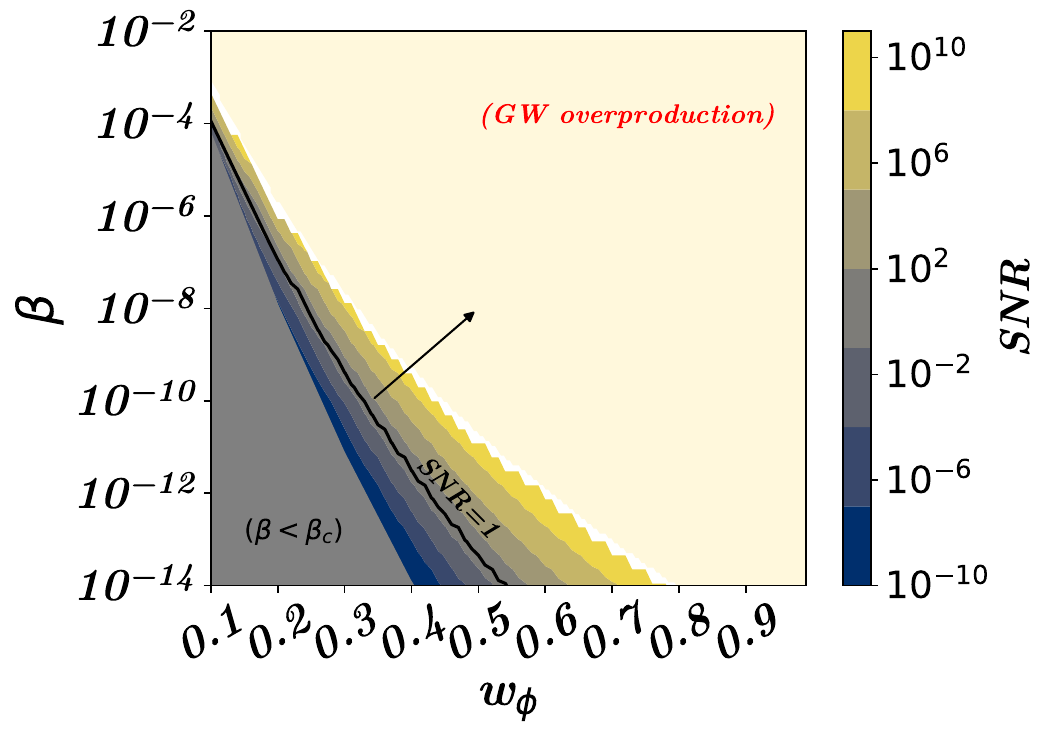}
    \caption{\it $\Min=10^7$ g}
    \label{fig:SNR_ET_combined_m7}   
    \end{subfigure}
    \caption{\it SNR plots for \textbf{ET} considering two different values of $\Min$ considering \textbf{the combined sources}. Rest of the illustration is same as Fig.~\ref{fig:SNR_iso_ET}.}
    \label{fig:SNR_ET_combined}
\end{figure*}

\begin{figure*}[!ht]
    \centering
    \begin{subfigure}{.47\textwidth}
    \includegraphics[width=\textwidth]{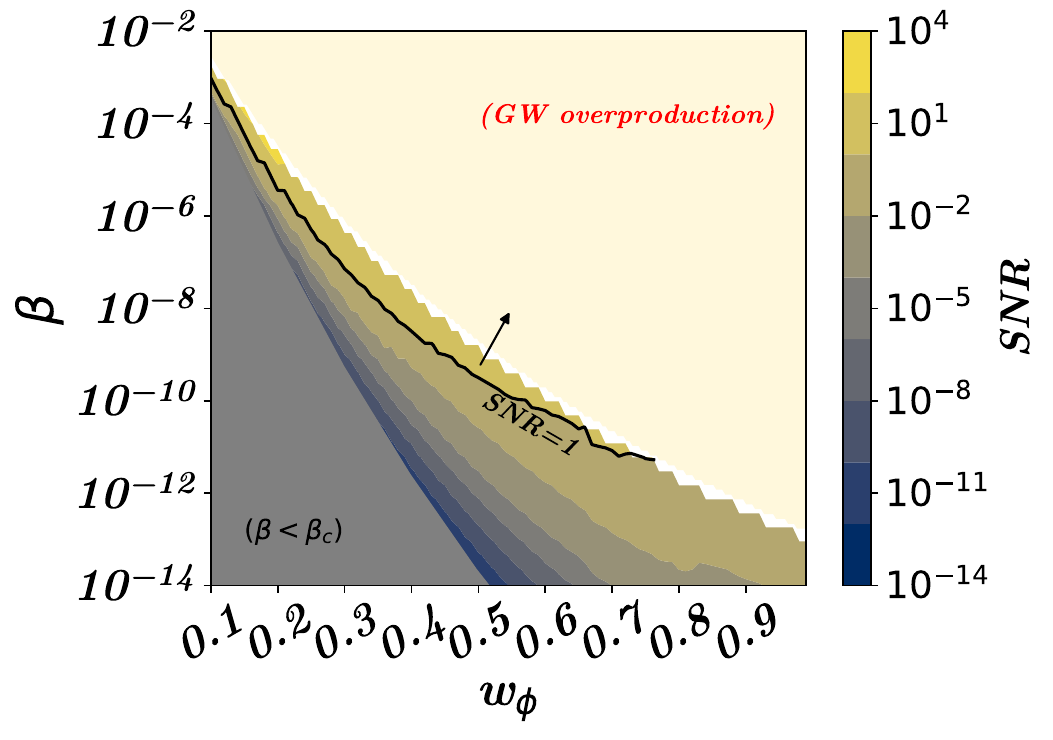}
    \caption{\it $\Min=10^5$ g}
    \label{fig:SNR_LISA_combined_m5}
    \end{subfigure}
    \hfill
    \begin{subfigure}{.47\textwidth}
    \includegraphics[width=\textwidth]{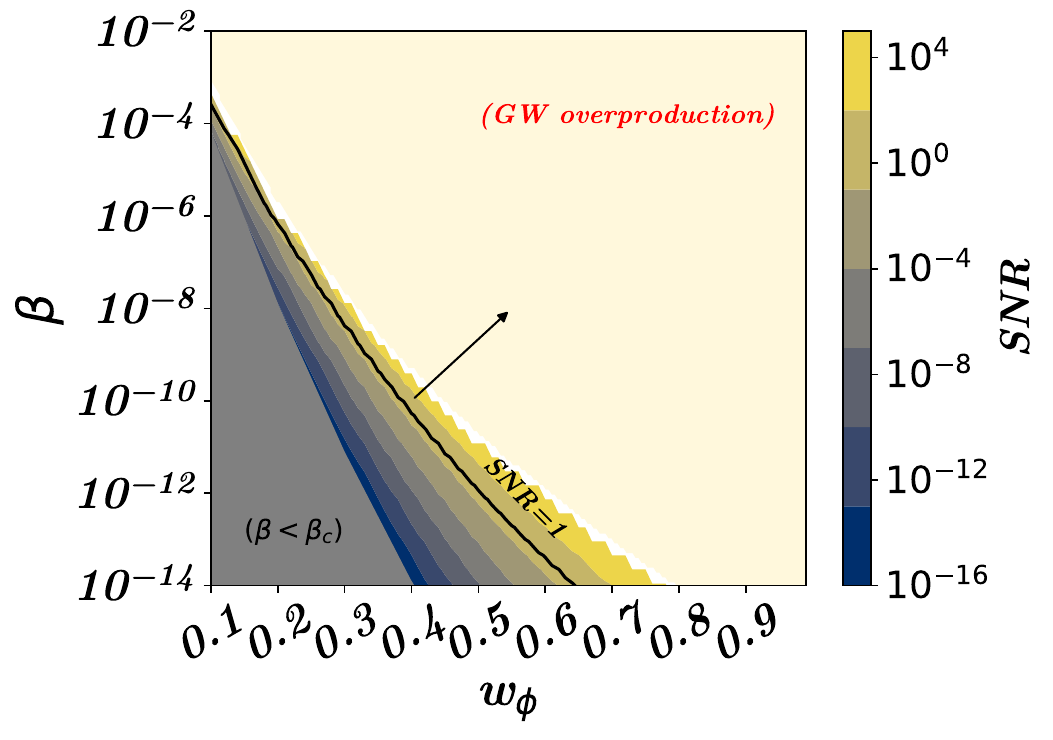}
    \caption{\it $\Min=10^7$ g}
    \label{fig:SNR_LISA_combined_m7}   
    \end{subfigure}
    \caption{\it Figure is same as Fig.~\ref{fig:SNR_iso_ET}, but for \textbf{the combined sources}, cosidering \textbf{LISA}.}
    \label{fig:SNR_LISA_combined}
\end{figure*}

\begin{figure*}[!ht]
    \centering
    \begin{subfigure}{.50\textwidth}
    \includegraphics[width=\textwidth]{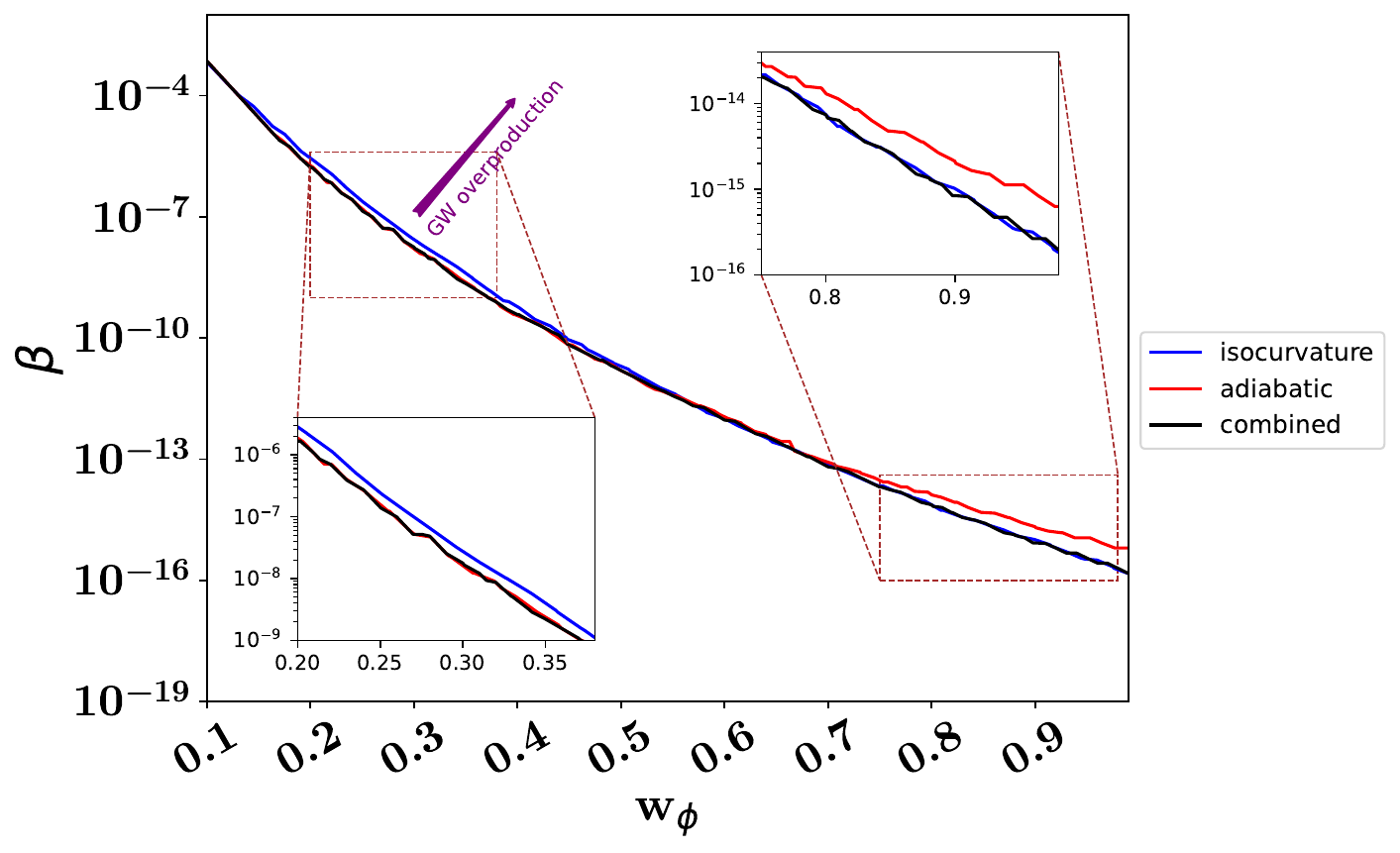}
    \caption{\it $\Min=10^7$ g.}
    \label{fig:max_GW_m7}   
    \end{subfigure}
    \hfill
    \begin{subfigure}{.42\textwidth}
    \includegraphics[width=\textwidth]{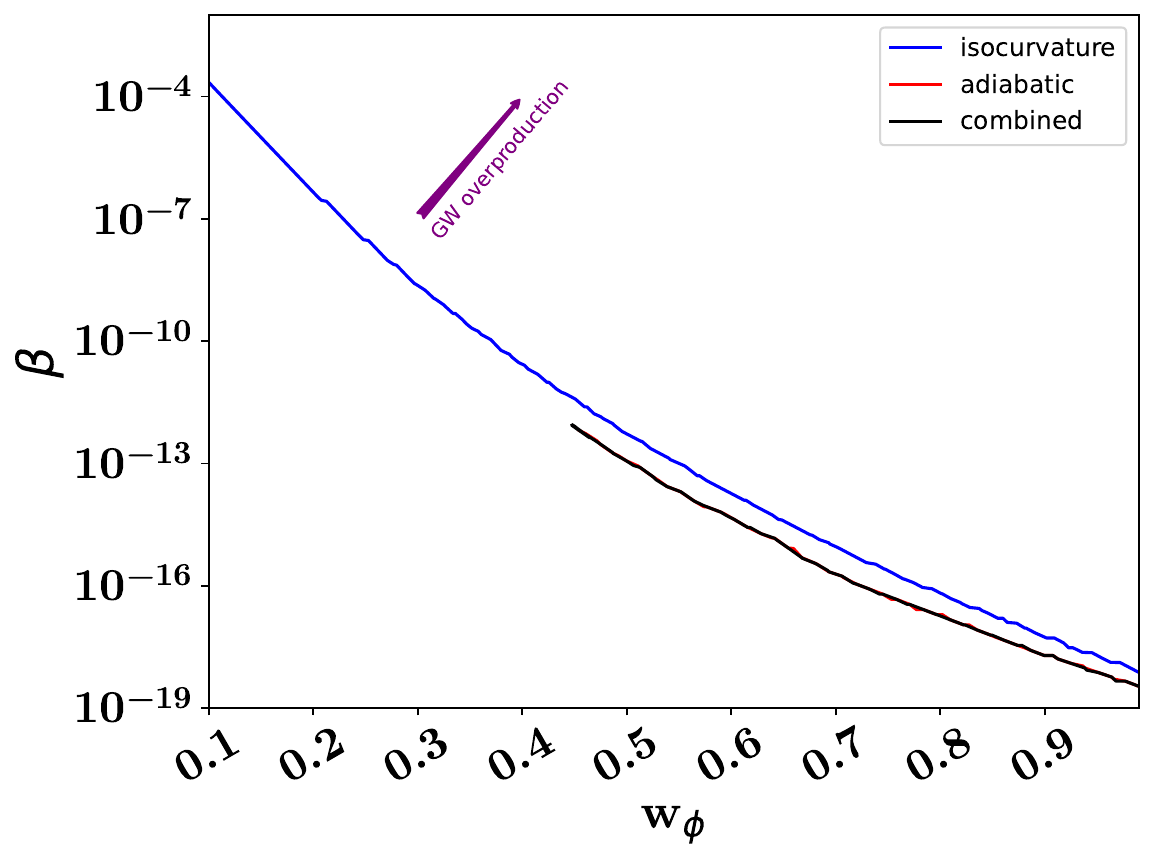}
    \caption{\it $\Min=4.8\times10^8$ g.}
    \label{fig:max_GW_m4e8}   
    \end{subfigure}
    \caption{\it Comparison of the maximum limit on the $\beta$ parameter imposed by $\Delta N_{\rm eff}$ constraints at BBN, considering induced GWs from three cases: isocurvature, adiabatic, and combined scenarios, for two distinct PBH formation masses.}
    \label{fig:max_limit_GW}
\end{figure*}

Similarly to the earlier study, we have estimated the SNR for ET and LISA in Figs.~\ref{fig:SNR_ET_combined} and \ref{fig:SNR_LISA_combined}, respectively, for two representative values of $\Min$ ($10^5$ g and $10^7$ g). The other annotations and shaded regions have the same physical interpretation as before. Fig.~\ref{fig:GWfull} suggests that the peaks in the spectrum, originating from two different sources, dominate each other for different PBH parameters. Consequently, some of the parameter space may be reduced due to the overproduction of GWs compared to individual sources. This effect is evident by comparing the adiabatic scenario (Fig.~\ref{fig:SNR_sudden_LISA_m5}) with the combined scenario (Fig.~\ref{fig:SNR_LISA_combined_m5}), where the SNR $>1$ region in the $\w-\beta$ plane shrinks in the combined case. To illustrate this effect, we have explicitly compared the GW overproduction limit of the combined spectrum with those from individual sources for different masses in Fig.~\ref{fig:max_limit_GW}, showing that the combined scenario follows a tighter limit than the individual sources. 

\begin{figure*}[!ht]
    \centering
    \begin{subfigure}{.49\textwidth}
    \includegraphics[width=\textwidth]{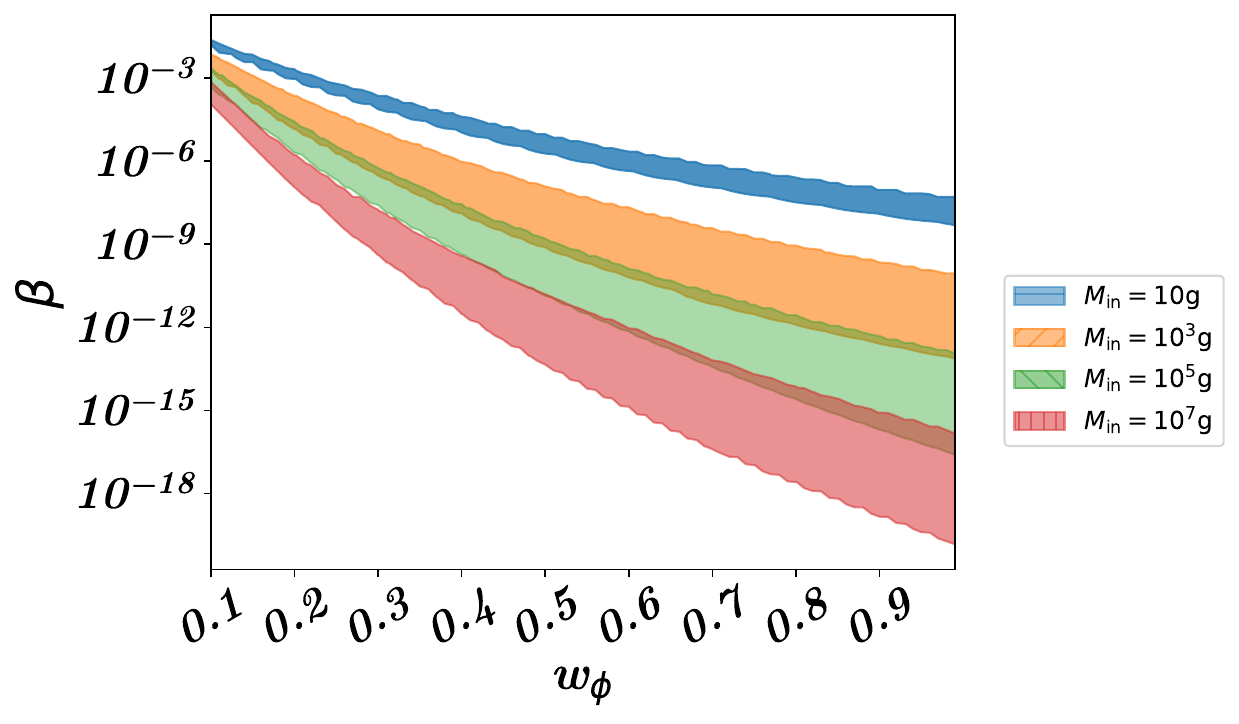}
    \caption{\it ET}
    \label{fig:SNR_full_all_masses_ET}
    \end{subfigure}
    \hfill
    \begin{subfigure}{.49\textwidth}
    \includegraphics[width=\textwidth]{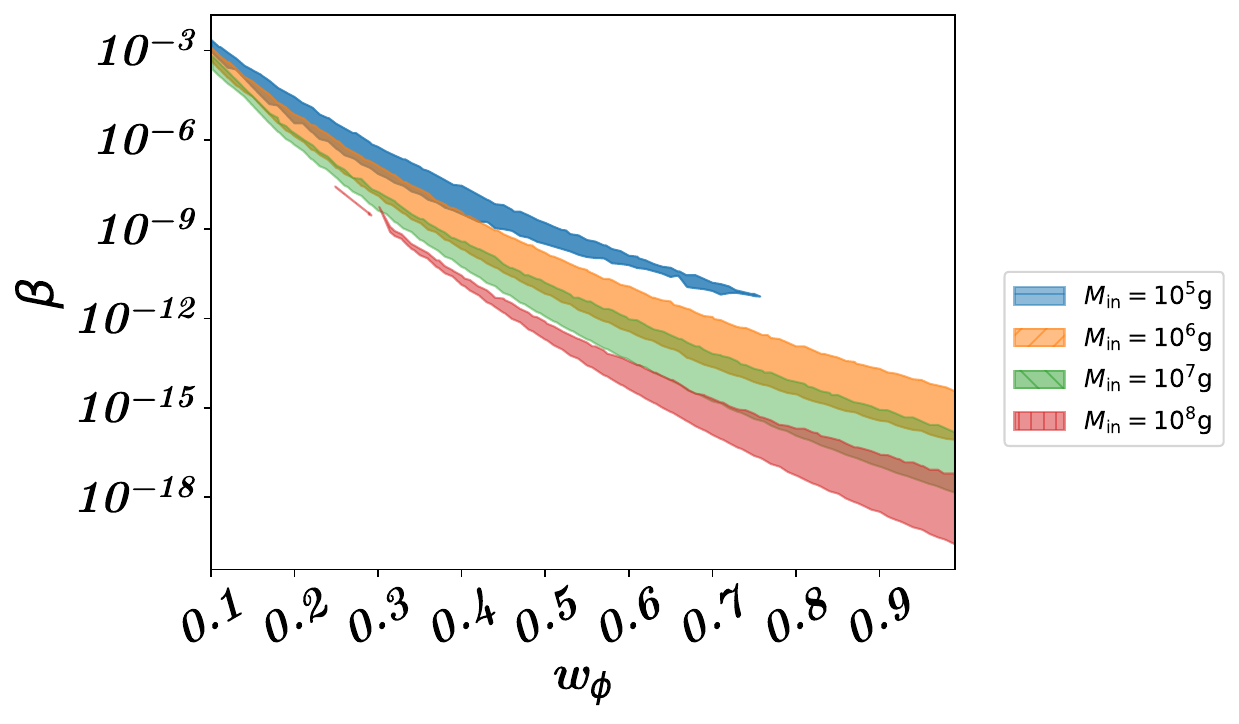}
    \caption{\it LISA}
    \label{fig:SNR_full_all_masses_LISA}   
    \end{subfigure}
    \caption{\it Illustration of extent of parameter space for various $\Min$ where ${\rm SNR}\geq1$, considering \textbf{the combined sources of GWs}, for ET (\textbf{left}) and LISA (\textbf{right}). The upper limit comes from the GW overproduction, derived from $\Delta N_{\rm eff}$ constraints at BBN.}
    \label{fig:SNR_full_all_masses}
\end{figure*}

Moreover, we have also represented the possible detectable range of parameters ($\rm SNR \geq 1$) in the $\w-\beta$ plane for distinct values of $M_{\rm in}$ in Fig.~\ref{fig:SNR_full_all_masses}, considering two interferometric missions, ET and LISA, analogous to the earlier analysis. Notably, this is not simply a direct sum of the parameter spaces from the two individual sources; rather, we observe a reduction in the compatible parameter space in the $\w-\beta$ plane. This effect is evident when comparing Fig.~\ref{fig:SNR_full_all_masses} with Figs.~\ref{fig:SNR_sudden_all_masses} and \ref{fig:snr_iso_et_all_mass}. This comparison demonstrates that, the combined scenario imposes tighter constraints than the individual sources, which is due to GW overproduction as we mentioned earlier.  

An important point to note is the inherent degeneracy between the two sources, particularly in terms of SNR. This implies that, based on SNR alone, it is challenging to determine whether the stochastic gravitational wave signal originates from the adiabatic or isocurvature source. Resolving this degeneracy requires further investigation. In particular, once observational data becomes available, the spectral shape of the detected signal will be crucial in distinguishing between the two sources. The distinct peak structures and frequency dependence of the adiabatic and isocurvature contributions, as discussed earlier, will play a key role in identifying the dominant mechanism responsible for the observed stochastic GWs background.

\subsubsection*{Fisher matrix analysis:}
\begin{figure*}[!ht]
    \centering
    \begin{subfigure}{.49\textwidth}
    \includegraphics[width=\textwidth]{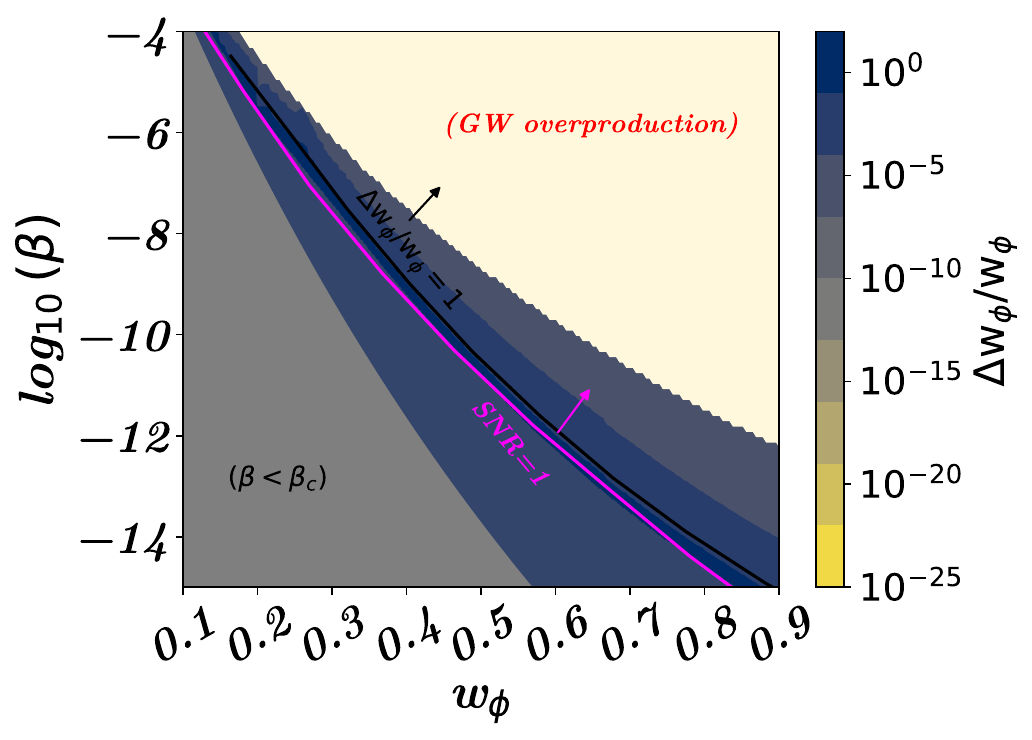}
    \caption{\it $\Min=10^5$ g}
    \label{fig:fisher_full_ET_deltaw_m5}
    \end{subfigure}
    \begin{subfigure}{.49\textwidth}
    \includegraphics[width=\textwidth]{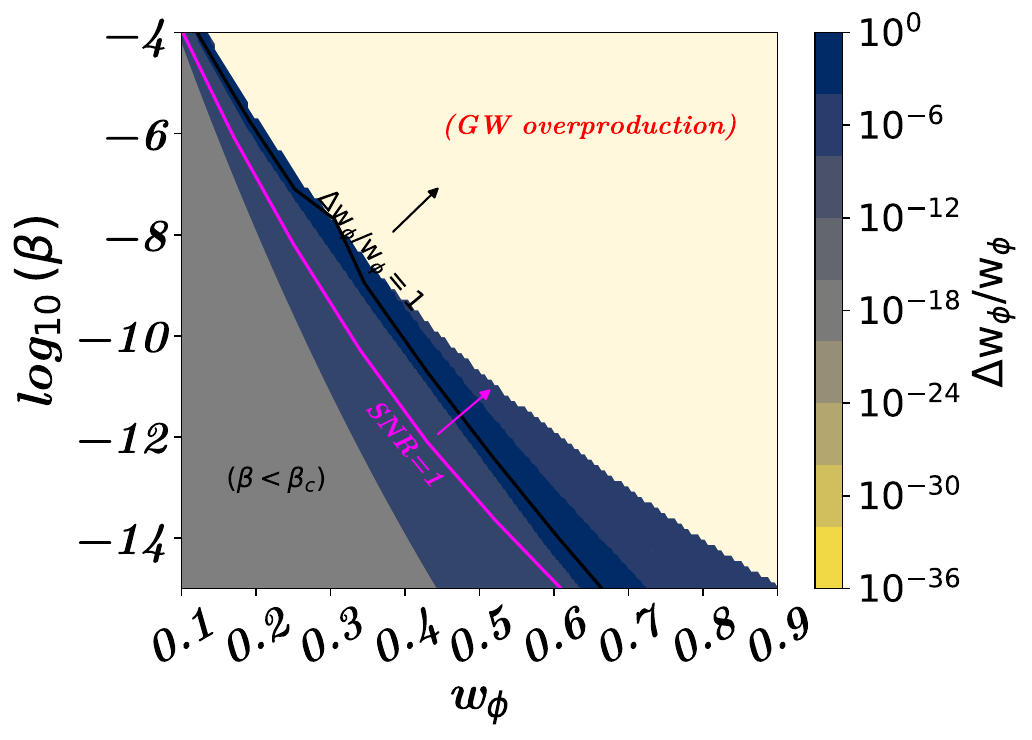}
    \caption{\it $\Min=10^7$ g}
    \label{fig:fisher_full_ET_deltaw_m7}   
    \end{subfigure}
    \caption{\it Illustration of relative uncertainties on $\w$, calculated by Fisher analysis, considering \textbf{combined sources of GWs} for \textbf{ET}. Black (magenta) solid line indicates ${\rm SNR}=1$ ($\Delta \w/\w=1$). The black (magenta) arrow indicates where ${\rm SNR}>1$ ($\Delta \w/\w<1$).}
    \label{fig:fisher_full_ET_deltaw}
\end{figure*}

\begin{figure*}[!ht]
    \centering
    \begin{subfigure}{.49\textwidth}
    \includegraphics[width=\textwidth]{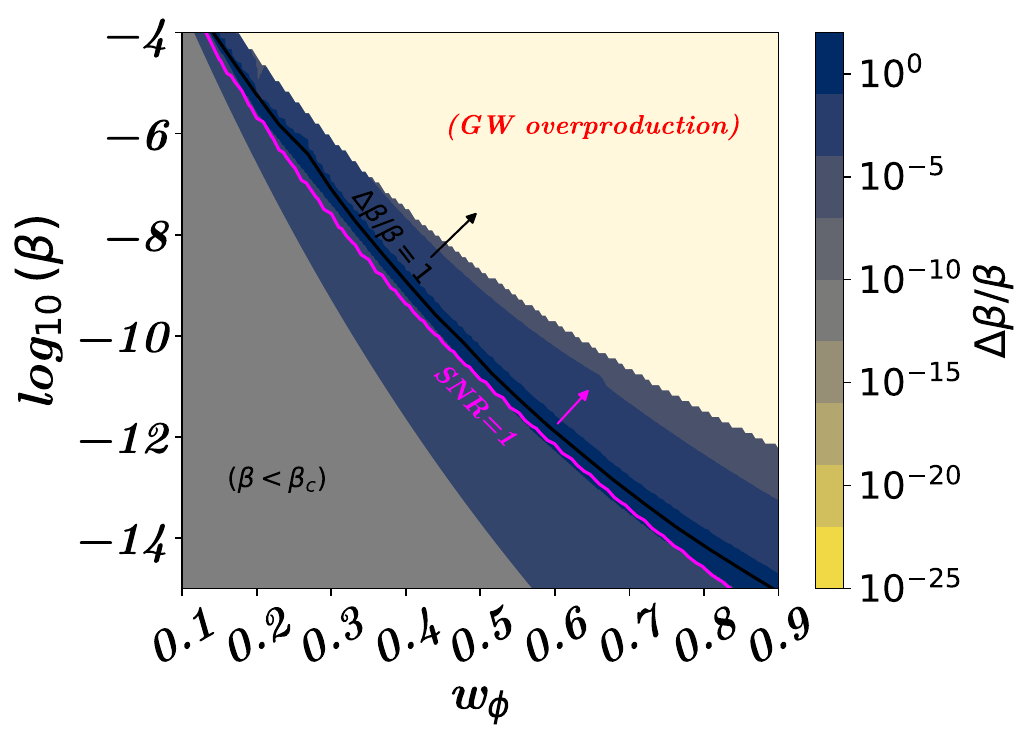}
    \caption{\it $\Min=10^5$ g}
    \label{fig:fisher_full_ET_deltabeta_m5}
    \end{subfigure}
    \begin{subfigure}{.49\textwidth}
    \includegraphics[width=\textwidth]{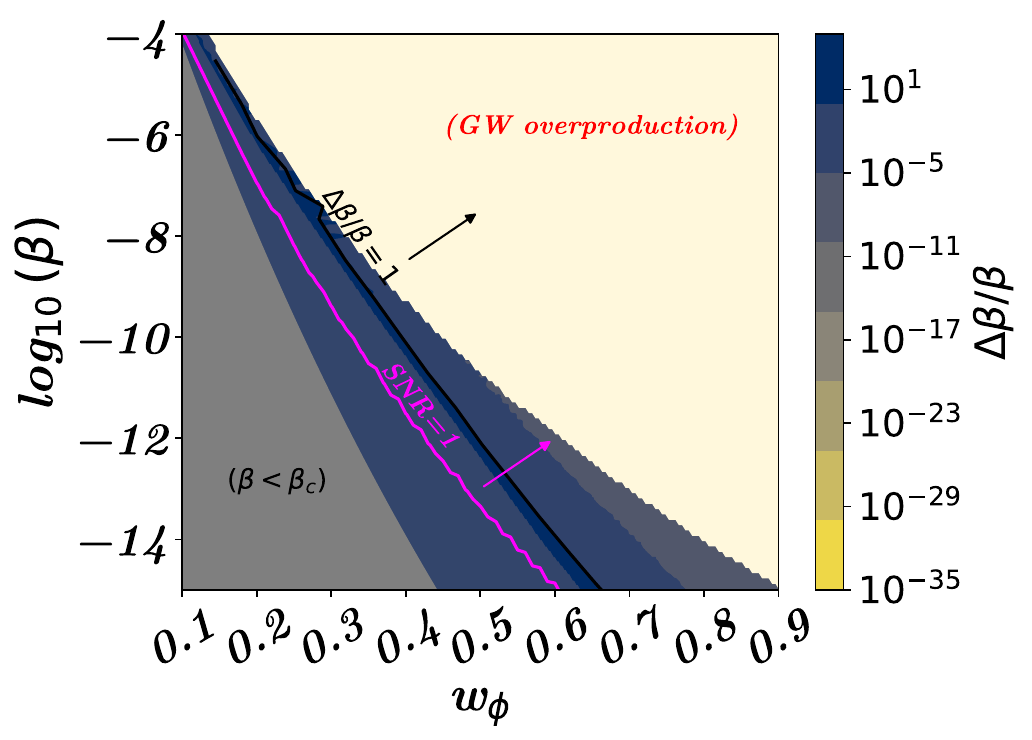}
    \caption{\it $\Min=10^7$ g}
    \label{fig:fisher_full_ET_deltabeta_m7}   
    \end{subfigure}
    \caption{\it Representation is same as Fig.~\ref{fig:fisher_full_ET_deltaw} but for relative uncertainties on $\beta$, cosidering \textbf{ET}.}
    \label{fig:fisher_full_ET_deltabeta}
\end{figure*}

\begin{figure*}[!ht]
    \centering
    \begin{subfigure}{.49\textwidth}
    \includegraphics[width=\textwidth]{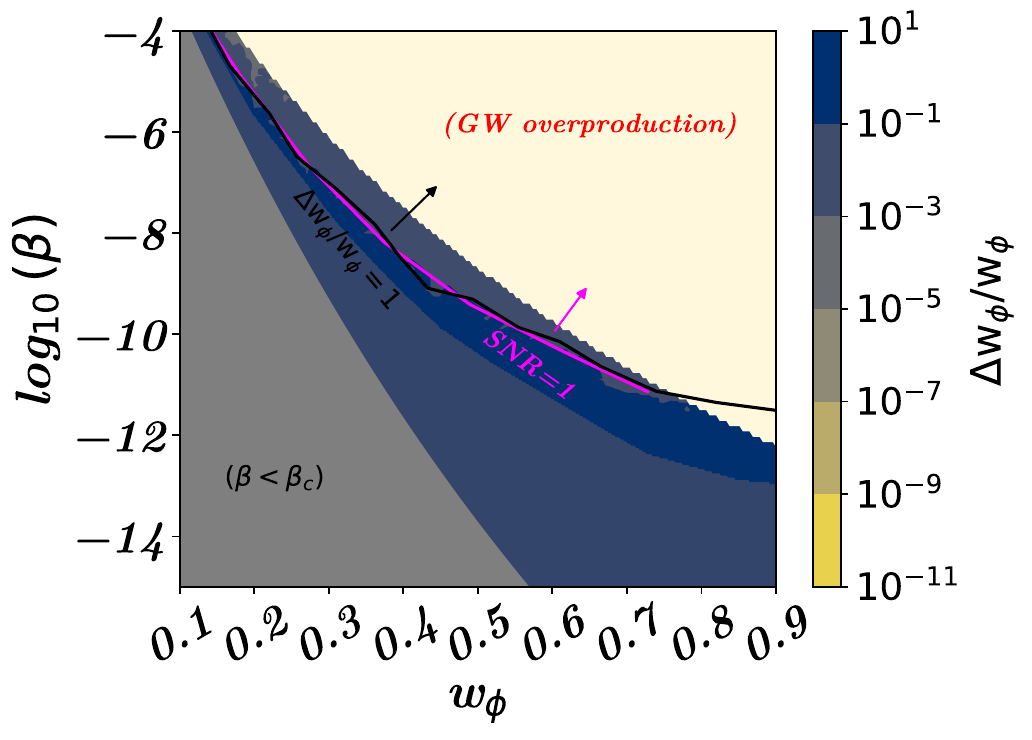}
    \caption{\it $\Min=10^5$ g}
    \label{fig:fisher_full_LISA_deltaw_m5}
    \end{subfigure}
    \begin{subfigure}{.49\textwidth}
    \includegraphics[width=\textwidth]{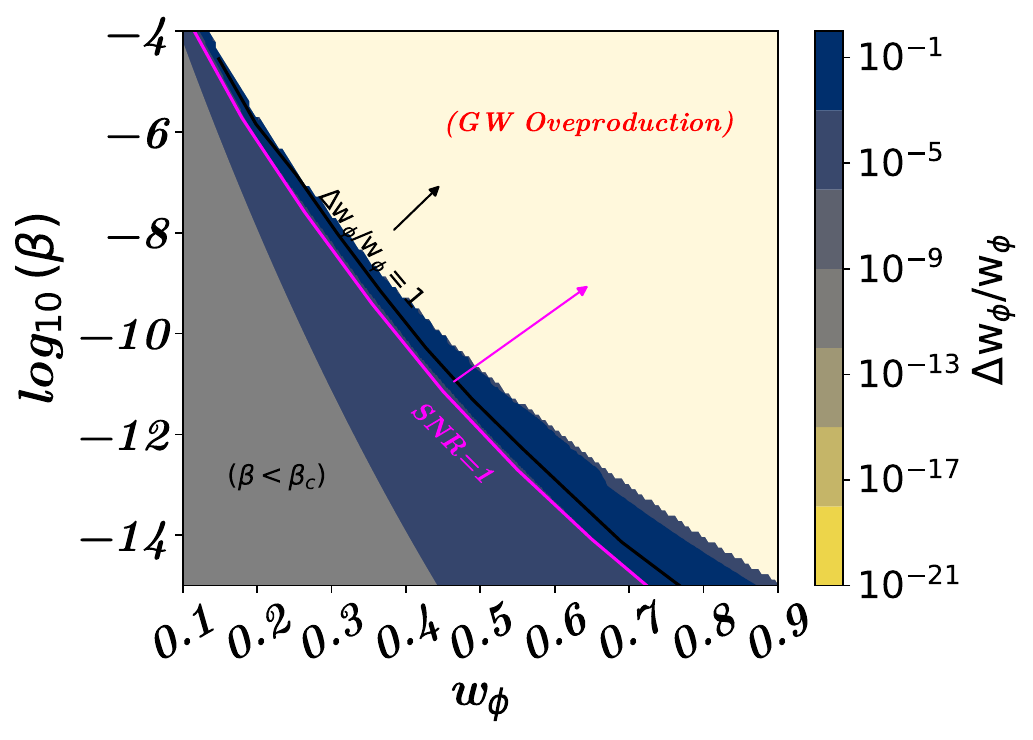}
    \caption{\it $\Min=10^7$ g}
    \label{fig:fisher_full_LISA_deltaw_m7}   
    \end{subfigure}
    \caption{\it Figure is same as Fig.~\ref{fig:fisher_full_ET_deltaw}, but for \textbf{LISA}.}
    \label{fig:fisher_full_LISA_deltaw}
\end{figure*}

\begin{figure*}[!ht]
    \centering
    \begin{subfigure}{.49\textwidth}
    \includegraphics[width=\textwidth]{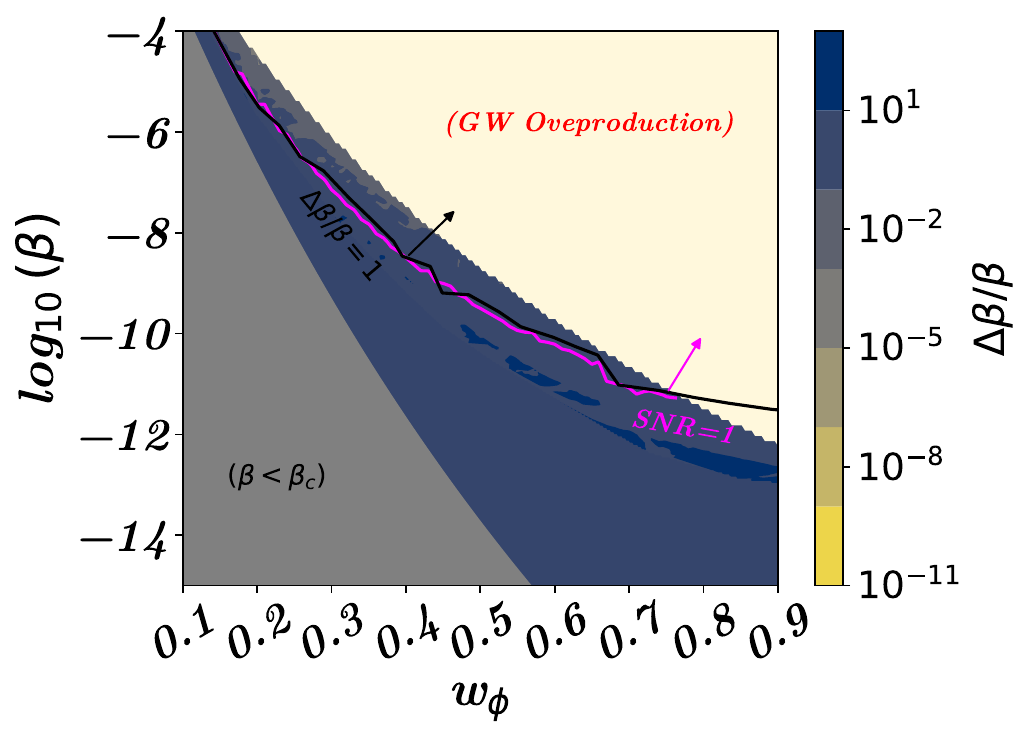}
    \caption{\it $\Min=10^5$ g}
    \label{fig:fisher_full_LISA_deltabeta_m5}
    \end{subfigure}
    \begin{subfigure}{.49\textwidth}
    \includegraphics[width=\textwidth]{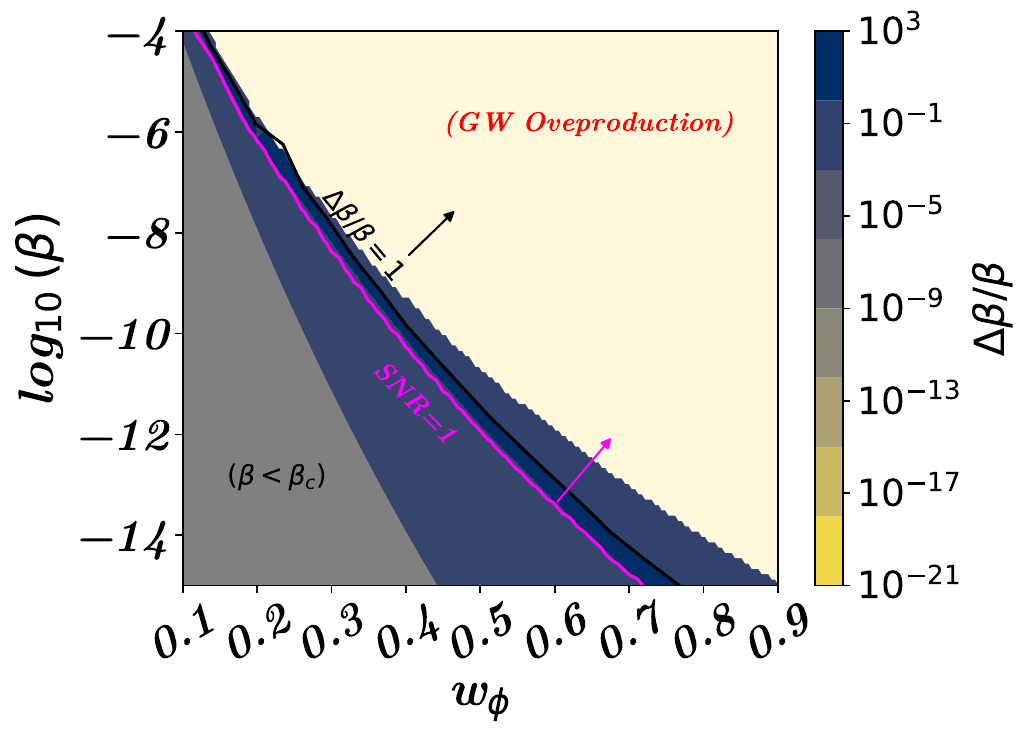}
    \caption{\it $\Min=10^7$ g}
    \label{fig:fisher_full_LISA_deltabeta_m7}   
    \end{subfigure}
    \caption{\it Figure is same as Fig.~\ref{fig:fisher_full_ET_deltaw}, but for relative uncertainties on $\beta$, cosidering \textbf{LISA}.}
    \label{fig:fisher_full_LISA_deltabeta}
\end{figure*}

Similar to the earlier sources, we have also performed the Fisher analysis to demonstrate the uncertainties on the parameters. Figs.~\ref{fig:fisher_full_ET_deltaw} and \ref{fig:fisher_full_ET_deltabeta} depict the forecasted uncertainties on $\w$ and $\beta$, respectively, for ET, considering two representative values of $\Min$. Similarly, Figs.~\ref{fig:fisher_full_LISA_deltaw} and \ref{fig:fisher_full_LISA_deltabeta} present the corresponding results for LISA. As the uncertainties associated with the parameters are smaller for the adiabatic scenario, combined analysis predominantly follows the adiabatic case. However, as observed in the SNR results, the reduced parameter space due to GW overproduction-limit in the combined scenario also shrinks the regions with lower uncertainties. This effect can be clearly observed by comparing Fig.~\ref{fig:fisher_full_LISA_deltaw_m5} with \ref{fig:fisher_sudden_LISA_deltaw_m5}, and \ref{fig:fisher_full_LISA_deltabeta_m5} with \ref{fig:fisher_sudden_LISA_deltabeta_m5}.

\begin{figure*}[!ht]
    \centering
    \begin{subfigure}{.49\textwidth}
    \includegraphics[width=\textwidth]{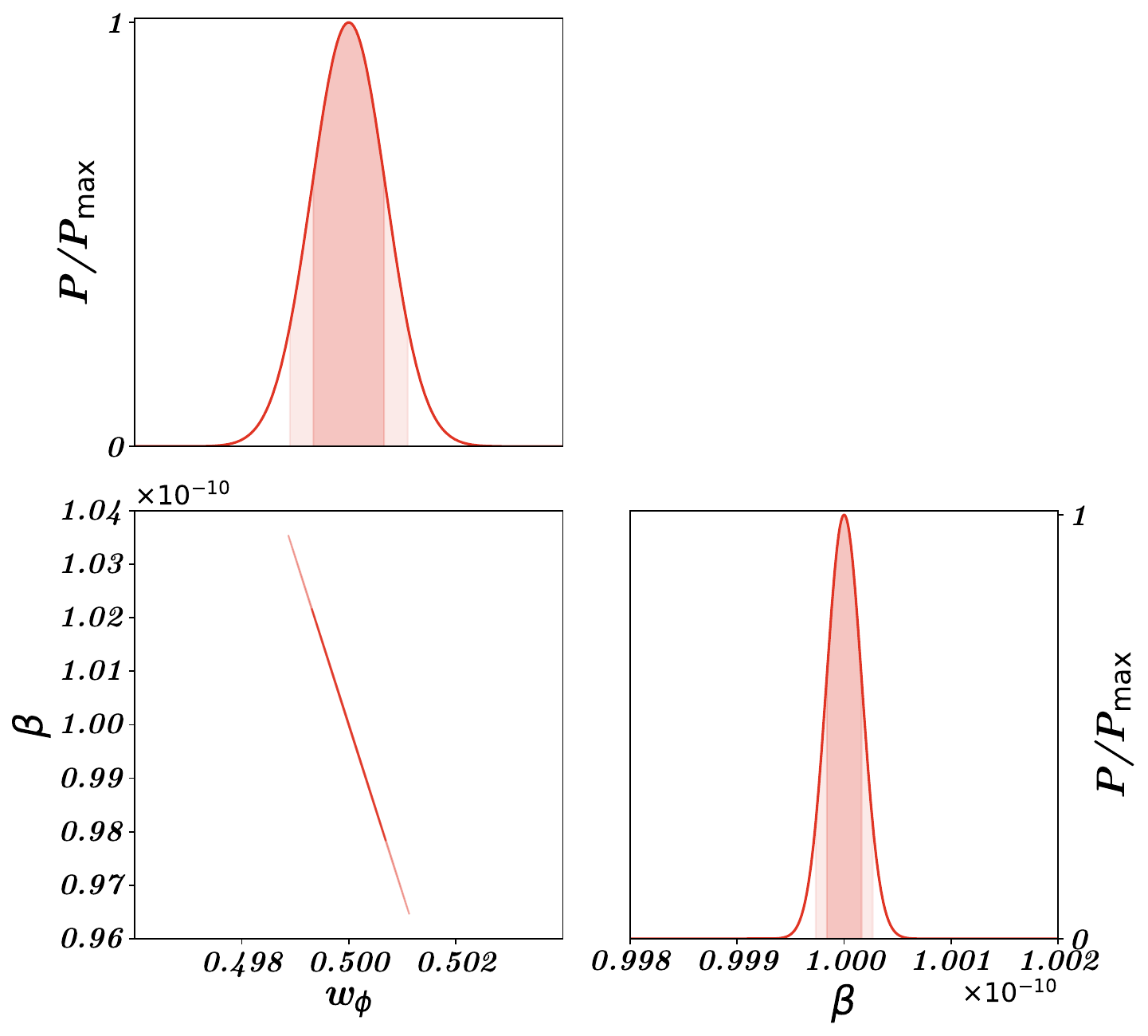}
    \caption{\it ET}
    \label{fig:fisher_full_et}
    \end{subfigure}
    \begin{subfigure}{.49\textwidth}
    \includegraphics[width=\textwidth]{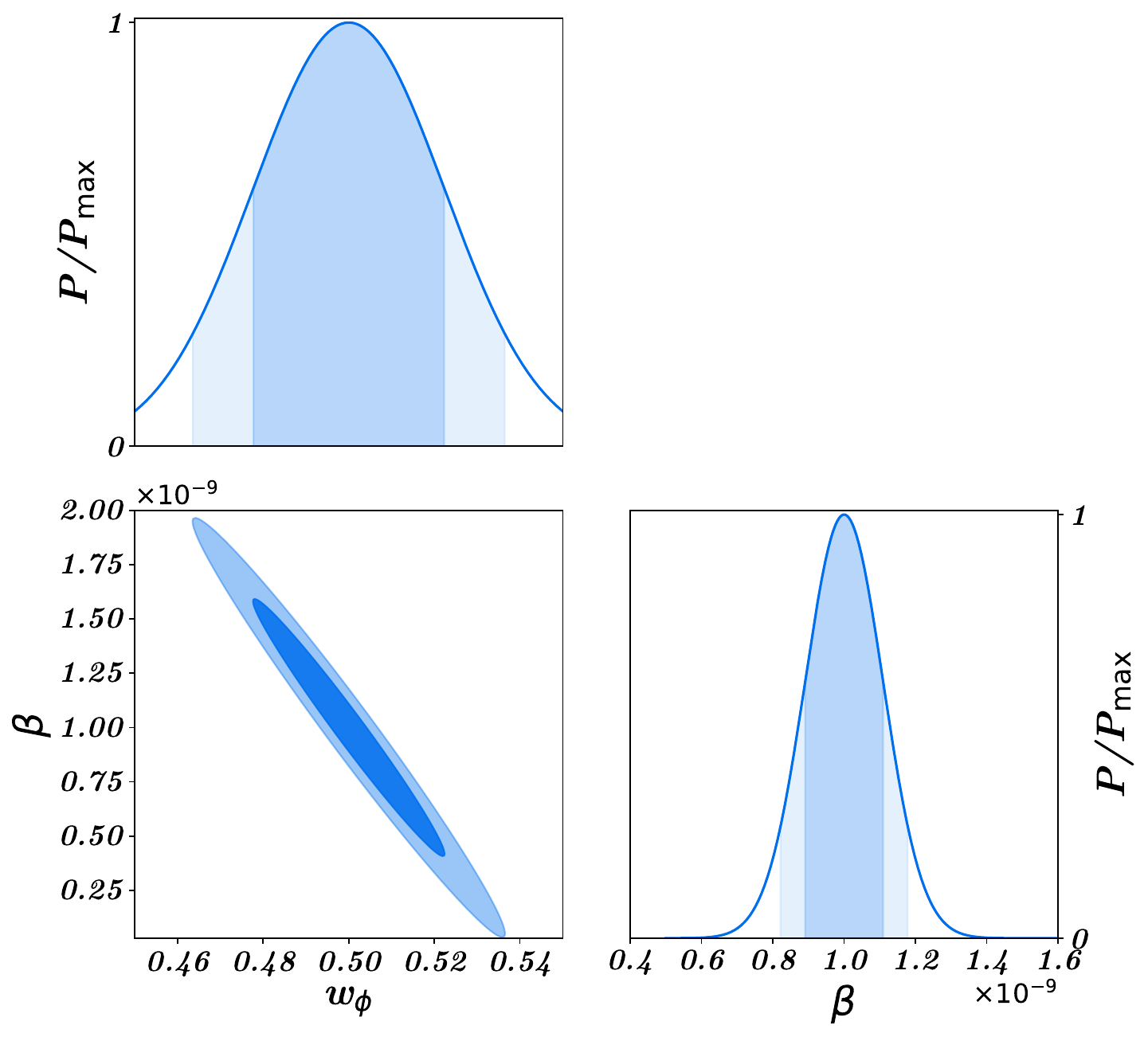}
    \caption{\it LISA}
    \label{fig:fisher_full_lisa}   
    \end{subfigure}
    \caption{\it 1-D and 2-D posterior distribution, derived from Fisher analysis, considering \textbf{the combined sources of GWs}. }
    \label{fig:fisher_full}
\end{figure*}

Accordingly, we have presented the posterior distribution of $\w$ and $\beta$, derived from Fisher analysis, for $\Min=10^5$ g in Fig.~\ref{fig:fisher_full}, for both detectors. Fiducials with corresponding uncertainties, for the other parameters, are tabulated in Table-\ref{tab:fisher_tab}. The 2D posterior distribution depicts anti-correlation between $\w$ and $\beta$. This trend of correlation is intuitive as both the parameters enhance the GW spectrum for the increasing values of them. Hence, for a given sensitivity threshold if $\w$ is larger, $\beta$ needs to be smaller. Furthermore, the results illustrate that the uncertainties are relatively smaller for ET compared to LISA. 
This may be an artefact of the following reasons: 
 The instrumental noise for LISA is larger compared to ET (as evident from Fig.~\ref{fig:noise}), and 
 the frequency binning, $N_b$ has been kept same for both detectors. However, as LISA operates over a broader frequency range than ET, $\Delta f$ (defined in Sec.~\ref{subsec:likelihood}) is comparatively larger for LISA. This leads to larger uncertainties for LISA, as a larger $\Delta f$ corresponds to smaller resolution of the detector.

\begin{table}[!ht]
    \centering
        \renewcommand{\arraystretch}{1.1}
        \begin{tabular}{c| c| c| c}
        \hline \hline
        \textit{Parameter} & \textit{Detectors} & \textit{Fiducials} & 1-$\sigma$ \\
        \hline
        \multirow{2}{*}{$\w$} & ET & 0.5 & 0.0010 \\
         & LISA & 0.5 & 0.0294 \\
        \hline
        \multirow{2}{*}{${\rm log}_{10}(\beta)$} & ET & $-10$ & 0.0160 \\
         & LISA & $-9$ & 0.3893 \\
        \hline
       \end{tabular}
       \caption{\it Fiducials and corresponding 1-$\sigma$ uncertainties associated with the parameters, derived from Fisher analysis, considering \textbf{combined scenario}. $\Min$ is set to $10^5$ g.}
       \label{tab:fisher_tab}
\end{table}

\subsubsection*{MCMC analysis with mock data:}

\begin{figure*}[!ht]
    \centering
    \begin{subfigure}{.49\textwidth}
    \includegraphics[width=\textwidth]{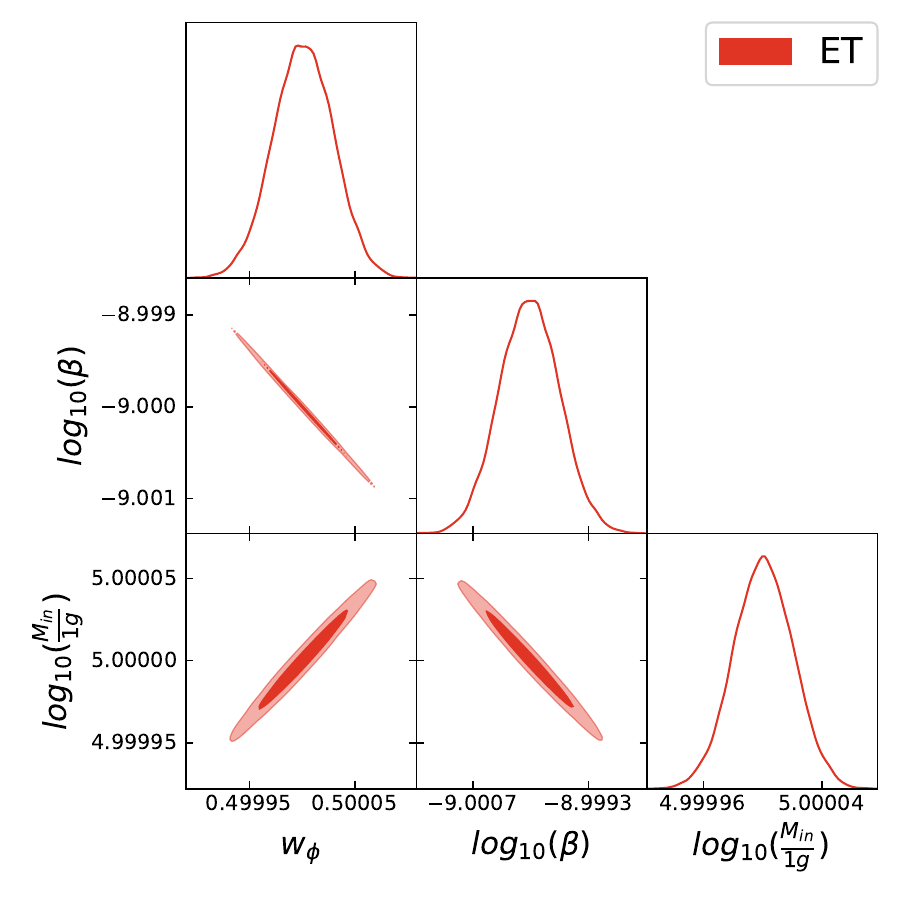}
    \caption{\it ET}
    \label{fig:mcmc_full_et}
    \end{subfigure}
    \begin{subfigure}{.49\textwidth}
    \includegraphics[width=\textwidth]{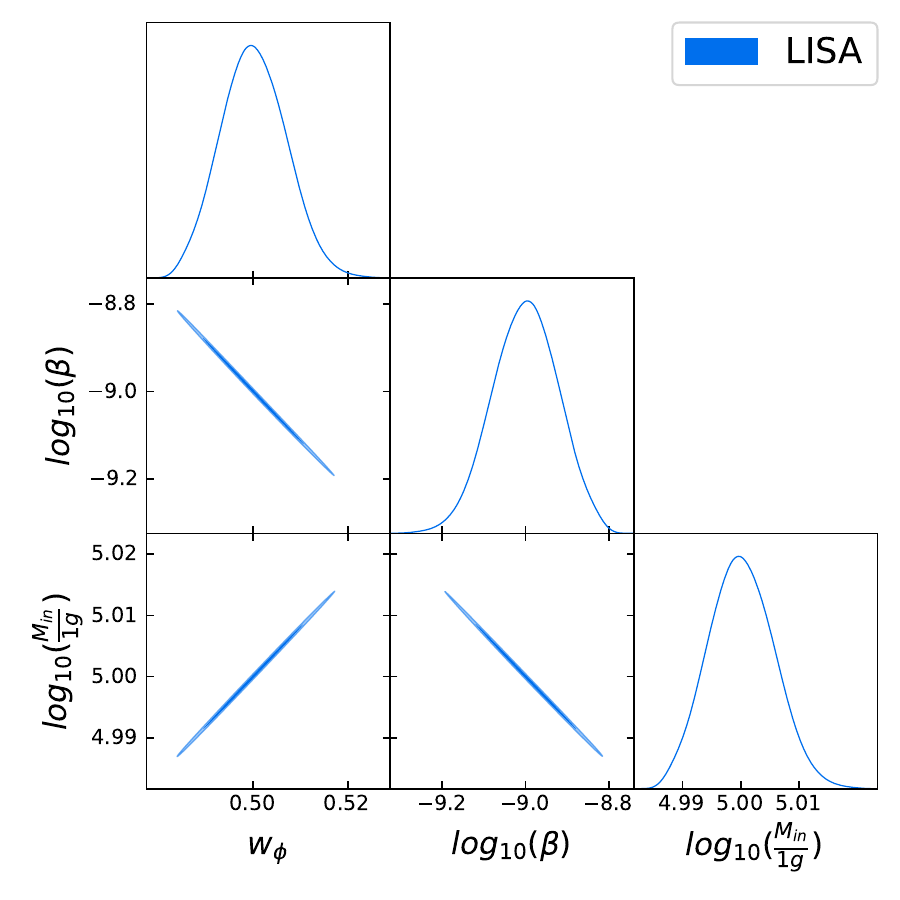}
    \caption{\it LISA}
    \label{fig:mcmc_full_lisa}   
    \end{subfigure}
    \caption{\it 1-D and 2-D posterior distribution for ET and LISA, considering three parameters associated with \textbf{combined scenario}.}
    \label{fig:mcmc_full}
\end{figure*}

Based on the basic setup described in Sec.~\ref{subsec:likelihood}, we have conducted the MCMC analysis for ET and LISA, considering the combined scenario, resulting in an estimation of the parameters under consideration using mock data generated from the instrumental specifications and possible sources of noise for both the missions. 
The priors adopted for the parameters are listed in Table-\ref{tab:prior}. To generate the mock data, we have considered  the fiducial parameter set by ($\w,{\rm log}_{10}(\beta),{\rm log}_{10}(\frac{\Min}{1{\rm g}})$) $\equiv$ ($0.5,-9,5$) for both detectors. This choice ensures that resulting GW spectrum satisfies SNR $>1$ and relative uncertainties $<1$, for both detectors. Finally, the mock catalogue itself is generated following methodology outlined in Sec.~\ref{subsec:likelihood}. Using on the mock catalogues, we have presented 1- and 2-dimensional posterior distributions for ET and LISA in Fig.~\ref{fig:mcmc_full}, with the corresponding statistical results summarized in Table-\ref{tab:mcmc_results}. 

\begin{table}[!ht]
    \centering
    \renewcommand{\arraystretch}{1.2}
    \begin{tabular}{c|c|c}
    \hline
    \hline
        \multirow{2}{*}{\textit{Parameter}} & \multicolumn{2}{|c}{\textit{mean$\pm\,\sigma$}}\\
        \cline{2-3}
        & ET & LISA\\
        \hline
        $\w$ & 0.5000$^{+0.00004}_{-0.00004}$ & 0.4995$^{+0.0084}_{-0.0081}$ \\
        ${\rm log}_{10}(\beta)$ &  -9.000$^{+0.0005}_{-0.0005}$ &  -8.995$^{+0.0917}_{-0.0967}$\\
        ${\rm log}_{10}(\frac{\Min}{1{\rm g}})$ &  5.000$^{+0.00003}_{-0.00003}$ & 5.000$^{+0.0066}_{-0.0068}$\\
    \hline
    \hline
    \end{tabular}
    \caption{\it Statistical results corresponding to the three parameters for each of the detectors, derived from MCMC analysis for \textbf{combined scenario}.}
    \label{tab:mcmc_results}
\end{table}
The posterior distributions reveal that ${\rm log}_{10}(\beta)$ exhibits a negative correlation with both $\w$ and ${\rm log}_{10}(\frac{\Min}{1{\rm g}})$, whereas $\w$ and ${\rm log}_{10}(\frac{\Min}{1{\rm g}})$ are positively correlated, for both detectors. This is fully consistent with the trends observed in the Fisher analysis, described earlier. Furthermore, Table-\ref{tab:mcmc_results} reveals that the uncertainties, associated with each of the parameters, are relatively less for ET compared to LISA, aligning with the findings from the Fisher analysis. As discussed previously, this is primarily due to the comparatively lower noise floor of ET and its lower $\Delta f$, representing finer resolution for ET compared to LISA. 

\section{Summary and outlook}
\label{sec:conclusion}
In this study, we explored the generation of stochastic GWs produced during the PBH reheating epoch, focusing on two distinct sources: \textit{isocurvature density fluctuations} from inhomogeneous spatial variations and \textit{adiabatic primordial curvature perturbation}. These sources generate complementary induced GWs spectra, which fall within the sensitivity range of future interferometry missions, providing a unique probe of early Universe cosmology through GW detection.
We have systematically analysed these two sources, first individually and then in combination, to explore the detection prospects of the relevant parameters $(\w, \beta, \Min$) at future interferometry missions such as LISA and ET. Note that we assume the tensor-to-scalar ratio to be sufficiently small, ensuring that the impact of primary GWs—originating from tensor perturbations generated by vacuum fluctuations during inflation—remains negligible (see, for instance, Ref.~\cite{Ghoshal:2024gai, Maity:2024cpq} for detection prospects of the early Universe in the context of primary GWs, considering future interferometry missions).

 Below we summarize the key findings from our analysis.

\begin{itemize}
    \item \textbf{Isocurvature fluctuations:} Our analysis reveals distinct detection capabilities for two different interferometry missions, ET and LISA. While ET demonstrates strong sensitivity across a broad PBH mass range, $ M_{\rm in} \in (0.5 - 3.34 \times 10^7)$ g, LISA is only sensitive to PBH masses above $10^8$ g. However, ET is not expected to detect the spectrum for PBHs with a formation mass exceeding $ 4 \times 10^7$ g, as the formation peak lies beyond its sensitivity range, highlighting a complementary detection window between LISA and ET. The parameter space in the $\beta - w_\phi$ plane is explored, constrained by the requirement $\text{SNR} > 1$ and limits from GW overproduction (\textit{i.e.}, the $ \Delta N_{\rm eff}$ bound at BBN). One notable outcome is that the detectable region $(\text{SNR} > 1$) in the $\beta - w_\phi $ plane for ET is significantly large for a PBH formation mass around $M_{\rm in} \sim 10^6 $ g. This is expected, as the peak frequency, associated with the UV cutoff scale $k_{\rm UV}$ for the masses around $10^{6}$ g, falls within the range where the noise spectrum $ \Omega_{\rm GW}^{\rm noise}$ for ET attains its minimum. In contrast, for LISA, achieving $ \text{SNR} > 1$ proves highly challenging, making it less favorable for detecting the isocurvature source of induced GWs.
    
    Although for ET we found high SNR values considering isocurvature source, the precision in constraining the parameters remains relatively poor. Fisher analysis indicates that achieving relative uncertainties below 1 for both parameters, $ w_\phi$ and $\beta$, is not possible for any PBH mass, with uncertainties exceeding 10 in most cases. The situation is even worse for LISA, where the SNR value is relatively low and the parameter uncertainties are significantly higher. Thus, while the GW spectrum induced by isocurvature fluctuations shows promising detection prospects, particularly for ET, the precision in constraining the underlying parameters remains low. As a result, although detection is feasible, accurately probing these parameters through either detector is unlikely for GWs induced by isocurvature fluctuations.

    \item \textbf{Adiabatic fluctuations:} This spectrum spans a broader frequency range, significantly enhancing detection prospects. As in the previous scenario, ET is sensitive to PBH masses in the range $ M_{\rm in} \in (0.5 - 3.97 \times 10^7)$ g. However, the detection prospects for LISA are notably improved in this case. SNR analysis suggests that for LISA, PBH masses within $ M_{\rm in} \in (2.3 \times 10^4 - 4.8 \times 10^8)$ g satisfy the condition $ \text{SNR} > 1$, in contrast to the isocurvature case, where achieving a significant SNR proved highly challenging. Additionally, the extent of the detectable parameter space in the $\beta - w_\phi$ plane reaches its maximum around $M_{\rm in} \sim 10^3$ g for ET, while for LISA, it is approximately $ 10^6$ g.  
    
    Thus, both detectors are well-placed in terms of  prospects of detection and constraining the parameters, with relative uncertainties often falling below 1 in regions where  $\text{SNR} > 1$. This makes adiabatic-induced GWs a more promising probe of PBH reheating physics compared to the isocurvature case for both detectors.
    \item \textbf{Combined scenario:} When both sources contribute simultaneously—which is generally the case—neither can be ignored, the combined GWs spectrum remains detectable across the same mass ranges expected from both the adiabatic and isocurvature cases, with LISA and ET exhibiting complementary sensitivity. Table~\ref{tab:Min_table} summarizes the detectable ranges of $M_{\rm in}$ for all three types of sources. In the combined scenario, the extent of the parameter space in the $\beta - w_\phi$ plane that satisfies $ \text{SNR} > 1$ is modified for both detectors (Fig.~\ref{fig:SNR_full_all_masses}), primarily due to the overproduction of GWs, as constrained by the $\Delta N_{\rm eff}$ bound from BBN (see, for instance, Fig.~\ref{fig:max_limit_GW}).

    \qquad Like the adiabatic case, ET and LISA seem to have prospects of detecting as well as constrainting the parameter with high precision. The posterior distributions for both Fisher analysis (Fig.~\ref{fig:fisher_full}) and MCMC analysis (Fig.~\ref{fig:mcmc_full}) reveal a negative correlation between $\w$ and $\beta$. Additionally, both statistical results infer that the uncertainties in probing the parameters are smaller for ET. 
\end{itemize}

\begin{table}[!ht]
    \centering
    \begin{tabular}{c|c|c|c}
        \hline
        \hline
        \textit{Detectors} & \textit{Isocurvature} & \textit{Adiabatic} & \textit{Combined} \\
        \hline
        ET & $(0.5-3.34\times10^7)$ & $(0.5-3.97\times10^7)$ & $(0.5-3.97\times10^7)$\\
        LISA & $(2\times 10^8-4.8\times10^8)$ & $(2.3\times10^4-4.8\times10^8)$ & $(2.3\times10^4-4.8\times10^8)$\\
        \hline
    \end{tabular}
    \caption{\it Detectable range of $\Min$ for both source types and the combined case, in grams.}
    \label{tab:Min_table}
\end{table}
This study highlights the synergy of next-generation GW observatories, with LISA and ET offering complementary sensitivity across a broad PBH mass range. It also sheds light on the distinct nature of the two sources in terms of both detection and constraint prospects. However, we currently lack a mechanism to break the degeneracy between them.  Once observational data becomes available, the spectral shape, distinct peak structures, and frequency dependence will be crucial in identifying the dominant source of the detected GW background. Detecting these induced GWs with LISA and ET could help distinguish between the two sources, providing valuable insights into the early Universe and the role of PBHs in GWs generation.  This work underscores the importance of multi-signal searches in future GW data, demonstrating how stochastic backgrounds from non-standard early-Universe processes could serve as key cosmological messengers in the upcoming era of gravitational wave astronomy.

The analysis presented in the article can be extended further in a number of directions.
This study investigates Schwarzschild primordial black holes with a monochromatic mass distribution. A significant extension would involve analyzing spinning PBHs and adopting an extended mass function, as these could profoundly influence the induced gravitational wave spectrum.
An extended mass distribution typically prolongs the transition from matter to radiation domination, leading to a pronounced suppression of the GW signal. This effect arises from a weakened resonance mechanism, and subtle cancellation further suppresses the signal. Prior studies, such as those in references \cite{Inomata:2019zqy,Pearce:2023kxp,Pearce:2025ywc,Inomata:2020lmk,Papanikolaou:2022chm}, demonstrate that extended mass distributions broaden evaporation timescales, resulting in a smoother transition and a suppression of the induced gravitational wave signal by an order of magnitude. Additionally, our model-independent approach to PBH parameters can be further developed by linking constraints to specific formation mechanisms. Connecting induced GWs to scenarios like inflationary dynamics, cosmic strings, or phase transitions could provide deeper insights into early Universe physics.

At this point, we emphasize that our analysis is based on linear cosmological perturbation theory, which offers a clear and tractable framework for predicting the gravitational wave spectrum, including the characteristic double-peak feature. However, if the early matter-dominated phase is sufficiently long, small-scale density perturbations can enter the non-linear regime, as their amplitude grows with the scale factor. As a result, adopting a fixed cutoff at $k_{\rm UV}$ may, in some cases, lead to an overestimation. To accurately capture the power spectrum in the non-linear regime, one must resort to detailed numerical simulations~\cite{Fernandez:2023ddy} or analytical techniques such as kinetic field theory~\cite{Konrad:2022tdu, Bartelmann:2019unp}.

On the observational front, next-generation detectors such as DECIGO and BBO, along with pulsar timing arrays (PTAs), could offer a multi-frequency probe of PBH-induced GWs. The synergy between space- and ground-based detectors will be essential for constraining PBH properties across different mass ranges. Furthermore, combining GW observations with multimessenger signals from the CMB, BBN, and high-energy astrophysics could provide a more comprehensive understanding of PBH reheating and early Universe dynamics.

\section*{Acknowledgements}
Authors sincerely thank Jan Tr\"ankle and Guillem Dom\'enech for their fruitful discussion, and the anonymous referee for valuable insights and suggestions in the manuscript. Authors also acknowledge the use of publicly available codes \texttt{emcee}~\cite{2013PASP..125..306F} and \texttt{GetDist}~\cite{Lewis:2019xzd}, and thank the computational facilities of Technology Innovation Hub, ISI Kolkata and of the Pegasus cluster of the high performance computing (HPC) facility at IUCAA, Pune. DP thanks ISI Kolkata for financial support through Senior Research Fellowship.
MRH acknowledges ISI Kolkata for providing financial support through Research Associateship. 
SP thanks the ANRF, Govt. of India for partial support through Project No. CRG/2023/003984. 

\appendix
\section{Noise modelling}
\label{app:noise}
\begin{figure}
    \centering
    \includegraphics[scale=0.6]{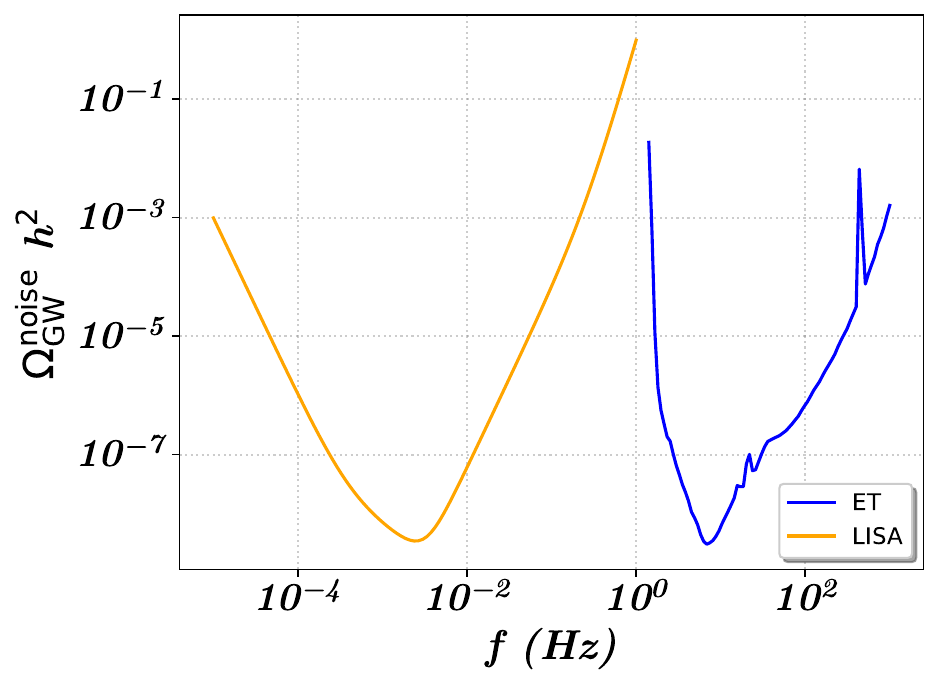}
    \caption{\it Noise spectra for LISA and ET.}
    \label{fig:noise}
\end{figure}

For any GW detectors, the effective noise power spectral density, $S(f)$, can be expressed as the ratio of detector's response function $R(f)$ and noise $N(f)$. As a result, the noise curves for any GW detectors can be described as~\cite{Gowling:2021gcy}:
\begin{eqnarray}\label{eq:omegagw}
    \Omega_{\rm GW}^{\rm noise} (f) \;=\; \left(\frac{4\pi^2}{3H_0^2}\right)f^3 S(f).
\end{eqnarray}
In the following, we discuss a summary of the $S(f)$ for the GW missions considered.
\begin{itemize}
\item
\textbf{LISA}

The key sources of instrumental noise in LISA arises mainly due to be optical metrology noise (omn) and acceleration noise of the test mass (acc), resulting from disturbances of the test mass~\cite{Gowling:2021gcy}. The contributions are expressed as following~\cite{Breitbach:2018ddu}: 
\begin{align}
S_{\rm acc}(f) &= 9 \times 10^{-30} \frac{1}{(2\pi f / 1{\rm Hz})^4} \left( 1 + \frac{10^{-4}}{f / 1{\rm Hz}} \right)~{\rm m^2Hz^{-1}},\\
S_{\rm oms}(f) &= \left(1.5\times 10^{-11} {\rm m}\right)^2\left[1+\left(\frac{2 {\rm mHz}}{f}\right)^4\right]\,{\rm Hz}^{-1}.
\end{align}
Hence, the instrumental noise for the LISA reads as 
\begin{align}
S_{\rm LISA}(f) &= \frac{10}{3L^2} \left(2S_{\rm acc}(f)\left(1+{\rm cos}^2 \left(\frac{f}{c/(2\pi L)}\right)\right)+ S_{\rm oms}(f)\right) \left[1 + \frac{6}{10}\left( \frac{f}{c/(2\pi L)} \right)^2\right],
\label{eq:SeffLISA}
\end{align}
with $L = 2.5 \times 10^9$~m being the constellation arm-length of the instrument. Now the GW energy density power spectrum can be calculated from Eq.~\eqref{eq:omegagw}.

\item
\textbf{ET}

For ET we have followed Ref.~\cite{Chowdhury:2022pnv} to generate the noise curve.
\end{itemize}
Fig.~\ref{fig:noise} represents the instrumental noise spectra for both detectors.

\section{Calculation of cutoff function}
\label{app:theta_UV}
For the isocurvautre case, UV cutoff function ($\Theta^{\rm (iso)}_{\rm UV} (k)$) for the resonant part, in Eq.~\eqref{eq:OGWres}, can be defined as~\cite{Domenech:2024wao}
\begin{eqnarray}
    \Theta^{\rm (iso)}_{\rm UV}(k) &\equiv& \int_{-s_0(k)}^{s_0(k)} ds
    \frac{\left(s^2-1\right)^2}{\left(1-c_s^2 s^2\right)^{5/3}} \label{eq:ThetaUV} \\
    &=&\frac{3 s_0 \left(5 c_s^4-2 c_s^2 \left(2 s_0^2+5\right)+9\right)-\left(5 c_s^2 \left(c_s^2+6\right)-27\right) s_0 \left(c_s^2 s_0^2-1\right) \, _2F_1\left(\frac{5}{6},1;\frac{3}{2};c_s^2 s_0^2\right)}{10 c_s^4 \left(1-c_s^2 s_0^2\right)^{2/3}},\nonumber
\end{eqnarray}
whereas for the adiabatic case, the cutoff function $\Theta^{\rm (ad)}_{\rm UV}(k)$ for the resonant part for the adiabatic scenario, in Eq.~\eqref{eq:OmegaGW_res_adi}, can be expressed as
\begin{align}
    \Theta^{\rm (ad)}_{\rm UV}(k) \equiv&  
    \int_{-s_0(k)}^{s_0(k)} ds\left(s^2-1\right)^2\left(1-c_s^2 s^2\right)^{n_{\rm eff}(\w)} \label{eq:ThetaUV_ad} \\
    =&\frac{2}{5} s_0^5 \, _2F_1\left(\frac{5}{2},-n_{\rm eff}(\w);\frac{7}{2};c_s^2 s_0^2\right)-\frac{4}{3} s_0^3 \, _2F_1\left(\frac{3}{2},-n_{\rm eff}(\w);\frac{5}{2};c_s^2 s_0^2\right) \nonumber \\
    &+2 s_0 \, _2F_1\left(\frac{1}{2},-n_{\rm eff}(\w);\frac{3}{2};c_s^2 s_0^2\right) \,. \nonumber
\end{align}
In the above two scenarios, $s_0(k)$ can be defined as 
\begin{eqnarray}
 s_0(k)\equiv
 \begin{cases}
        1  \quad & \frac{k_{\rm UV}}{k}\geq \frac{1+c_s^{-1}}{2}\\
        2\frac{k_{\rm UV}}{k}-c_s^{-1} \quad & \frac{1+c_s^{-1}}{2}\geq \frac{k_{\rm UV}}{k}\geq\frac{c_s^{-1}}{2}\\
        0 \quad & \frac{c_s^{-1}}{2}\geq\frac{k_{\rm UV}}{k}
    \end{cases} \,.
\end{eqnarray}

\bibliographystyle{bibi}
\bibliography{biblio.bib}

\end{document}